\documentclass{eethesis}

\usepackage{graphicx}

\renewcommand{\type}{Dissertation}

\renewcommand{\degree}{DOCTOR OF PHILOSOPHY}

\renewcommand{\major}{Physics}	

\let\a=\alpha
\let\b=\beta 
\let\c=\gamma
\def\C{\Gamma} 
 
\let\d=\delta

\let\e=\epsilon
\def\eb{{\bar\epsilon}}
\def\tb{{\bar\theta}}
\def\kb{{\bar\kappa}}
\def\etab{{\bar\eta}}
\def\ce{{\cal\epsilon}}
\def\tv{\tilde v}
\let\z=\zeta 
\def\h{\eta}  
\let\k=\kappa
\let\l=\lambda
\let\L=\Lambda
\let\m=\mu
\let\n=\nu 
\let\r=\rho
\let\s=\sigma
\let\S=\Sigma
\let\t=\theta

\let\th=\theta
\let\w=\omega

\let\P=\Pi

\let\T=\Theta
\let\Th=\Theta

\def\ha{\hat{\alpha}} 
\def\[{{[}}
\def\]{{]}}

\def\oneone{\rlap 1\mkern4mu{\rm l}}
\def\hm{{\hat\mu}} 
\def\hn{{\hat\nu}} 

\def\ha{{\hat A}}
\def\hb{{\hat B}}

\def\tb{\bar\theta}

\def\ve{\varepsilon}
\def\vt{{\vec\tau}}
\def\tr{{\rm tr}}
\def\det{{\rm det}}
\let\x=\xi

\def\ua{\underline{\alpha}}
\def\ub{\underline{\phantom{\alpha}}\!\!\!\beta}
\def\uc{\underline{\phantom{\alpha}}\!\!\!\gamma}

\def\una{\underline a}\def\unA{\underline A}
\def\unb{\underline b}\def\unB{\underline B}
\def\unc{\underline c}\def\unC{\underline C}

\def\unm{\underline m}\def\unM{\underline M}

\def\unG{\underline{G}}
\def\unT{\underline{T}}

\def\xz{\times}

\def\nab{\nabla}

\def\umu{{\underline \mu}}

\def\pl{ \: + }

\newfont{\bbbold}{msbm10 scaled \magstep1}

\def\cA{\cal A}

\def\cE{\cal E}
\def\cF{{\cal F}}

\def\cL{\cal L}

\newfont{\goth}{eufm10 scaled \magstep1}

\def\ge{\mbox{\goth e}}

\def\qq{\quad\quad}
\let\la=\label 
 
\def\p{\partial}

\def\bd{\begin{document}} 
\def\ed{\end{document}}
\def\be{\begin{equation}}
\def\ee{\end{equation}}
\def\ba{\begin{array}}
\def\ea{\end{array}}
\def\bea{\begin{eqnarray}}
\def\eea{\end{eqnarray}}
\def\nn{\nonumber}
\def\ni{\noindent} 
\let\bl=\bigl 
\let\br=\bigr
\let\Br=\Bigr 
\let\Bl=\Bigl
\let\bm=\bibitem
\def\pth{(\partial_1+\partial_2+\partial_3)}
\def\dt{d\tau_1\cdots d\tau_n}
\def\pn{(\partial_1+\cdots + \partial_n)}
\def\lra{\leftrightarrow}
\def\ra{\rightarrow}
\def\qq{\quad\quad}
\def\mod{{\rm mod}}

\def\4{\m_1\cdots \m_4}
\def\5{\m_1\cdots \m_5}
\def\6{\m_1\cdots \m_6}\def\h6{\hm_1\cdots \hm_6}
\def\Hat#1{\widehat{#1}}

\def\sh#1{\rlap{\hbox{$\mskip 1 mu /$}}#1}	

\def\ft#1#2{{\textstyle{{\scriptstyle #1}\over {\scriptstyle #2}}}}
\def\fft#1#2{{#1 \over #2}}
\def\sst#1{{\scriptscriptstyle #1}}
\newcommand{\eq}[1]{(\ref{#1})}
\def\eqs#1#2{(\ref{#1}-\ref{#2})}
\def\cites#1#2{\cite{#1}-\cite{#2}}
\def\Hat#1{\widehat{#1}}
\def\nosum{({\rm no\ sum\ over}\ i~)}

\def\oneone{\rlap 1\mkern4mu{\rm l}}

 \newcommand{\hoch}[1]{$\, ^{#1}$}

\newcommand{\ttbs}{\char'134}
\newcommand{\AmS}{{\protect\the\textfont2
  A\kern-.1667em\lower.5ex\hbox{M}\kern-.125emS}}

\begin{document}
\pagenumbering{roman}
\pagenumbering{arabic}
\setlength{\headheight}{12pt}
\pagestyle{myheadings}



\maketitlepage
{From Superalgebras to Superparticles and \\ Superbranes}   
{Igor V. Rudychev} 
{Doctor of Philosophy}                
{May 2001}
{Physics}         



\approvalone
{From Superalgebras to Superparticles and\\ Superbranes}
{Igor V. Rudychev}
{Ergin Sezgin}
{Richard Arnowitt}
{Christopher N. Pope}
{Stephen A. Fulling}
{Thomas W. Adair III}
{May 2001}



\absone
{From Superalgebras to Superparticles and \\Superbranes}
{May 2001}
{Igor V. Rudychev}
{B.S., Kharkov State University}
{Dr. E. Sezgin}
{In this work I investigate connections between superalgebras and their
realizations in terms of particles, branes and field theory models.
 I start from Poincar\'e superalgebras with brane charges and study its   
 representations. The existence
of new supermultiplets in different dimensions including 
an ultra short supermultiplet 
 in D=11 different
 from the supergravity multiplet is shown. Generalizations of
superalgebras containing brane charges, including those in 
D$>$11 are considered.
 The realization of these algebras at the level of relativistic particle
 models and, upon quantization, at the level of field theory is presented.
 Application of Hamiltonian/BRST methods of quantization of systems with mixture
 of first and second class constraints as well as a 
 conversion method are discussed for the
 models of interest. Using quantization of particle mechanics 
we obtain information on 
  the  spectrum and linearized equations of motion of the 
  perturbative, linearized M-theory.
   The generalization of particle models to p-branes is made using
 a geometrical formulation of superembedding approach to study 
  the example of
 L-branes which have a linear multiplet on their worldvolume. 
 The p-branes and
 strings in B-field are considered as well as the
 origin of noncommutativity and non-associativity in their low-energy limit.
  It is shown that the application of Hamiltonian/BRST methods for 
  those models leads to stringy
 version of Seiberg-Witten map and the 
  removal of the
 non-associativity/noncommutativity. }




\dedicate{Natalia L. Rudychev, my wife and friend.}



\acknow{I would like to thank E.Sezgin for his guidance, help and constant
           encouragement.}



\pagestyle{headings}
\setlength{\headheight}{36pt}
\tableofcontents
\clearpage



\chapter{\bf Introduction}

\body

 Supersymmetry plays an important role in modern physics.
Starting from the early 70's, the importance of supersymmetry became clear not
only in field theory but also in superstring models.
During the last century, physicists tried to answer the question 
of how to unify 
all possible interactions in Nature. The main problem is how to 
reconcile quantum
mechanics with gravity. 
A remarkable answer to this question comes from
string theory. In string theory, one does not have point-like
particles as fundamental objects but rather 
extended objects called strings. Strings could be open as well as closed
and the vibrational modes of the string correspond to particles. 

 The consistent string theory without a tachyon has to be supersymmetric
 and exists in ten dimensions.
The low energy limit of superstring theory gives ten-dimensional 
supergravities that are supersymmetric field theories.

 As far as unification is concerned,
 the problem with  superstring theory is that it is not unique.
There are five consistent superstring theories. 
The existence of their duality symmetries
 in different dimensions and the properties of their
brane solitons suggest  the existence of M-theory which unifies all five string
theories. In particular M-theory contains strongly coupled limit of Type IIA
string which lives in eleven dimensions.
Moreover, 
M-theory has the eleven dimensional supergravity as its low energy limit and 
it also
has  nonperturbative excitations in its spectrum such as membranes and
fivebranes.

      In this thesis we investigate how supersymmetry could be used to study
 properties of M-theory. In particular, we will start from 
 the properties of superalgebras
 with brane charges. We know that important information about
 M-theory is encoded
in the M-theory superalgebra, which is the 
eleven dimensional Poincar\'e superalgebra with membrane and fivebrane charges.
 In this work we investigate various aspects of this algebra
 including its representations. 
We also consider its realization 
 in terms of superparticle that
lives in an extended supermanifold. This supermanifold is
parametrized by ordinary superspace coordinates together with additional tensorial
coordinates which are in one-to-one correspondence with brane charges.
Upon
quantization of the superparticle model we obtain the 
spectrum and the linearized equations
of motion of supersymmetric field theory. Supersymetry transformations
of this field theory are generated by the M-theory superalgebra.

   In other part of this work
    we study superbranes using
 superembedding approach where both worldvolume and target space are
supermanifolds. Using this approach we investigate properties of a class of p-branes
called L-branes which have linear multiplets on their worldvolume.
Then we  discuss  various aspects of noncommutative geometry in
strings and p-branes in external fields.

    In the rest of the introduction we will 
outline the main ideas of our thesis in more detail.    
We begin with the discussion  of the relevant superalgebras.
  The simplest superalgebra in $D=4$ with $4N$ supercharges is
given by

\be
\{ Q^i_\a, Q^j_\b \} = (\c^m)_{\a\b} P_m \d^{ij},
\la{simple}
\ee

where $Q^i_\a$, $i=1,...,N$ is a supercharge and $P_m$ is momentum. 
It is possible to extend this superalgebra.
One of the possible extensions of \eq{simple} is obtained by 
adding
all symmetric in $(\a i), (\b j)$ 
combinations of $\c$-matrices together with $SO(N)$
invariant tensors. 
This leads to the appearance of central charges.
Central charges commute with every other generator of the superalgebra.
The superalgebra with central charges is given by 

\be
\{ Q^i_\a, Q^j_\b \} = (\c^m)_{\a\b} P_m \d^{ij} + C_{\a\b} V^{ij} + 
(\c^5)_{\a\b} U^{ij},
\la{simple-central}
\ee

where $V^{ij}$ and $U^{ij}$ are central charges. They are antisymmetric in $i,j$
indices and singlets in respect to the Lorentz group.
The representations of the superalgebra with central charges
\eq{simple-central}
are well studied. All of them are massive. 
This follows from the
positive energy theorem. 
The shortening of a multiplet occurs when central charges take 
particular values.
Those shortened
multiplets could be interpreted as BPS multiplets  \cite{c2}.
In \cite{M} and \cite{pkt1} higher-dimensional origin of central charges
 was discussed.

    Another possible extension of super Poincare superalgebra is given by 
 adding
tensorial central charges. The tensorial central charges are not singlets
under Lorentz rotation.
They commute with everything, except
generators of Lorentz rotations $M_{mn}$. 
This extension does not contradict 
the theorem by Haag, Lopuszanski and Sohnius because tensorial charges are
non-singlets under Lorentz rotations.
 To be precise, it is not quite correct to
call them central because they do not commute with $M_{mn}$.  
We would rather use the term brane-charges.  The four-dimensional $N=1$
superalgebra
with brane charges is

\be
\{ Q_\a, Q_\b \} = (\c^m)_{\a\b} P_m  + (\c^{mn})_{\a\b} Z_{mn},
\la{brane-charges}
\ee

\be
[Q_\a ,Z_{mn}] = 0, \qquad [Z_{mn},Z_{kp}] = 0,
\la{brane-charges1}
\ee

where $Z_{mn}$ are brane-charges and they commute with all generators except
$M_{mn}$.

 In  \cite{c9} it was shown that the brane charges could be interpreted
nonperturbatively as integrals over the brane worldvolume.
The brane charges carry an information about nonperturbative part of
the theory.
And they give a rise to existence of  p-branes. 

In four dimensional case $Z_{mn}$ corresponds to the integral
of the membrane current. Therefore we call it membrane charge.
The superalgebra already defines not only
perturbative but also nonperturbative extended objects.
The extended objects, such as p-branes, belong to the nonperturbative spectrum
of the theory. 
 Another way to explore the properties of the brane-charges is to start from
correspondent $Osp$ superalgebra.
Then, contraction with non-zero brane-charge generators gives extended
Poincar\'e
superalgebra.
 
 So far, we have discussed only nonperturbative interpretation of the
brane-charges.
In this work we will  consider  {\it perturbative} interpretation.
We will show existence of massless
representations of the superalgebra with brane-charges.
If central charges are singlets
superalgebra has only massive irreps. It follows from positive energy
theorem. All branes, we discussed so far, are massive extended object.
That means,  they can not be used for description of massless multiplets.
 We are going to discuss superparticle and field-theoretical
realization of massless supermultiplets.

  Superalgebra is a useful tool for  studying
properties of new theories. 
For example, properties of eleven dimensional
superalgebra with brane charges gives an information  about M-theory
\cite{M}.
 We have a few tools to study M-theory and the superalgebra is one of
them. 
 One more intriguing extension of Poincare superalgebra could be
 obtained by introduction of
tensorial charges with spinorial indices. It leads to the M-algebra \cite{ma3}.
In this case the superalgebra with new fermionic generators in ten
dimensions is \cite{Gstr}

\be
\{ Q_\a, Q_\b \} = (\c^m)_{\a\b} P_m, \quad [P_m, Q_\a] = -(\c^m)_{\a\b} Z^{\b},
\la{orig_of_M}
\ee

where $Z^{\b}$ is a new fermionic generator. It has an interpretation 
in terms of new string formulation given by Siegel in \cite{Sstr}. 
The superalgebra \eq{orig_of_M} could be lifted up to eleven dimensions.
In $D=11$ it forms the M-algebra constructed 
in \cite{ma3}.
The M-algebra contains the following generators.
$P_m, Q_\a, Z_m, Z_{mn}$, $Z_{m_1,...,m_5}$ which correspond to brane-charge 
extended Poincar\'e superalgebra in eleven dimensions, and 
$ Z_\a, Z_{m \a}, Z_{\a\b},$  $Z_{m_1,..., m_4 \a},$   $Z_{m_1,..., m_3 \a \b},
 Z_{m_1 m_2 \a \b \c},$

    $ Z_{m \a \b \c \d}, Z_{\a_1,..., \a_5}$
 which are new and have nontrivial commutation relations \cite{ma3}.

   There is another interesting problem that could be solved by 
using properties of
superalgebras. It is a unification of type IIA/B heterotic string theories
in the
context of M-theory. The web of $S$ and $T$ dualities allows 
to connect type IIA and type IIB theories in less then ten  dimensions.
One can call it a low-dimensional unification. On the other hand M-theory 
lives in eleven dimensions. It unifies five string theories indirectly
by going to low dimensions.
 Is it possible
to obtain connection between all string theories directly in ten dimensions?

   On the level of superalgebras it is possible  to obtain
type IIA/IIB, heterotic superalgebras  from the theory in higher
dimensions.  Then compactification gives all ten-dimensional 
superalgebras. The M-theory algebra does not do
this. The M-theory compactifications give type IIA but 
not IIB in ten dimensions.
Is it possible to go beyond eleven dimensions? 
 This question was discussed in \cite{b2}. In the next chapter
we will
show that in twelve dimensions such a unification is indeed possible.

  Let us return to the discussion of 
superalgebras with brane charges and their representations.
Supersymmetry algebras with central charges are playing one of the
central roles in string theory and supergravity. 
The very existence of the extended
objects such as p-branes is due to tensorial central charge extension of
the superalgebras in different dimensions. Even properties of M-theory
could be defined from eleven-dimensional 
 superalgebra with brane charges \cite{M}.
  There are two classes of central charges in supersymmetric theories. 
 First one includes the central charges which are Lorentz singlets.
Massive representations of those central extended superalgebras
were studied in \cite{c1}. They could be interpreted through presence of
the soliton-like objects in the theory \cite{c2}. They could be also used in
some supersymmetric Lagrangians \cite{c3} and in some particular
realizations of supermultiplets \cite{c4},\cite{c5}. 
Central charges play
crucial role in spontaneously broken Yang-Mills theory \cite{c3}
and also in building of shell supergravities
\cite{c6},\cite{c7}.

 The second class includes charges that are not Lorentz singlets
but rather transform as tensors under Lorentz
rotation. They commute with the rest of the generators. Tensorial
central charges
do not arise in superalgebras considered by 
Haag,Lopuszanski and Sohnius \cite{c8}.
They were introduced in \cite{c9} as p-brane charges. Representations of
the superalgebra with tensorial central charges were studied much less
intensively then the ones with singlet central charges. 
Recently, representations were
studied in \cite{Hull} and the $N=1$, $D=4$ case was considered in
\cite{3/4}. Other applications of brane charges as well as extensions of
different superalgebras involving brane charges
were investigated in \cite{Ferr}, \cite{dW1}, 
\cite{Shif}.

   In  this work we consider a different interpretation of the tensorial
central charges. We associate them  perturbatively 
 with extra coordinates of a particle in an extended
superspace. 
Superparticle with central charge coordinates was considered before in
\cite{RS1}. The quantization in the massless case for $D=4$ $N=1$ was
presented in \cite{BLS}. In this
models the particle lives on a supermanifold parametrized by usual
coordinates and by twistor variables associated with p-form charges. 
The quantization in
the massless case is claimed to give rise to
 an infinite number of states of arbitrary
spin \cite{BLS}. The
ordinary N-extended version of the spinning particle \cite{Pi}
describes higher spin particle but the number of such states is finite.

     So far our analysis covered only particles in Minkowski space-time.
Particle models and their quantization in $AdS$ spaces have attracted
a lot of attention recently, especially in the context of AdS/CFT
correspondence.
The bosonic massive particle in $AdS_5$ was 
quantized in
\cite{K1}. Only recently, generalization of the light cone approach to
$AdS$ spaces was applied to the quantization of superparticle in $AdS$
background \cite{MT}. 

 There is one more interesting observation: if one considers massless
multiplets of Poincare superalgebra with tensorial central charges, then 
it is
possible to obtain multiplet shortening.
This could be explained as follows. When 
tensorial central charges take some particular value the number of creation
and annihilation supercharges is reduced. It means
that in four dimensions there is a massless multiplet shorter then the 
$N=1$ Wess-Zumino
multiplet. In eleven dimensions there are supermultiplets different from
supergravity multiplet. 
This result will be explained in Chapter 2. 

    Superalgebra with brane charges has massless BPS representations.
 They could be associated with massive non-perturbative objects.
The example of massive case and  application of brane-charges
in construction of eleven-dimensional 
preons is given in \cite{BAIL}. 

The superalgebra with brane charges also has massless supermultiplets.
  How do we interpret massless multiplets from field theory point of view?
One way is to use the analogy with the ordinary superparticle/field theory
correspondence.
Let us consider an example in eleven dimensions.
The spectrum and the linearized
equations of motion of eleven dimensional supergravity could be obtained
from  quantization
of the superparticle model. 
The superparticle model  realizes  eleven-dimensional 
Poincare superalgebra
without brane-charges. 
The quantization of ordinary Brink-Schwarz superparticle in eleven dimensions
gives 
linearized equations of motion and spectrum of an eleven-dimensional 
supergravity.
For the case of superparticle with brane charge coordinates, the quantization
should give supersymmetric field theory.  
The supersymmetry of this field theory is generated by  M-theory superalgebra.
One can argue that this field theory could be associated with linearized 
perturbative M-theory.

     In this work we mostly concentrate on massless representations.
If one is interested in BPS states \cite{BAIL} the situation is quite 
different.
 Here we will briefly describe the approach of \cite{BAIL},
which considered BPS states, 
  and compare it with our interpretation. 
From BPS-states point of view,  states are interpreted as
extended objects, i.e. p-branes. Moreover 
if one has
zero in the right-hand
side of the anticommutator of two supercharges, then supersymmetry is
not broken, and if the right-hand side is not zero, then supersymmetry is broken.
For example if only one of 
all supercharges anticommutes non-trivially (i.e. gives zero), 
then  only one supersymmetry is broken. In \cite{BAIL} those
states are called preons.
It is also possible to show at the level of  superalgebra that
all other BPS states in eleven dimensions including membrane and five-brane
could be obtained from preons. 
It is not clear yet
what these preons are, and whether they correspond to extended 
objects or particles.
It is important to understand how bound states of preons can 
create membranes and fivebranes.
 In our approach, we interpret the massless as well as massive 
 states 
 from pure field-theoretical/particle point of view. 
Therefore, if one has only one 
non-trivially anticommuting supercharge, then one uses it 
to built states starting from lowest
weight state. This supercharge is a creation operator.
We see the major difference in two approaches. In the case of BPS states 
one has only one
broken supersymmetry and the rest are unbroken. In the case of 
field-theoretical description,
we have only one nontrivial supercharge that could be interpreted 
as 
a creation generator
and the rest of the supercharges trivially anticommute and therefore
 their action on vacuum states
does not give new physical states.

     We conclude this introduction by briefly mentioning the other topics
covered in the dissertation.

   In  Chapter 2, we study representations of Poincar\'e superalgebra with
brane charges. We also discuss  superalgebras with brane charges in D$>$11.
On the level of superalgebras, we investigate the higher dimensional
origin of Type IIA/B, heterotic unification.

    In  Chapter 3, we investigate the realizations of the superalgebras with brane
charges in terms of superparticles and superstrings. We also present the realizations
of superalgebras in D$>$11 in terms of multi-particle and string models.

      In  Chapter 4, we study the quantization of superparticles with
brane-charge coordinates in twistorial form. We start from applying 
BRST/Hamiltonian methods for the bosonic case.In the process of 
quantization BRST/Hamiltonian methods are
playing important role. In this work we show that problem of 
mixture of first and second class constraints 
could be resolved using conversion procedure even for complicated systems
of massless/massive superparticles with brane charge coordinates.

     In  Chapter 5, we consider higher p-branes models. We start from
applying superembedding approach to construct equations of motion and action
of new class of branes, called L-branes. L-branes have linear multiplet on
world-volume which includes higher p-forms. We present Born-Infeld type action
for the higher p-forms. This example could be also important from point of
view of nonlinear dualization of higher p-forms.
In the end of this chapter we show that BRST/Hamiltonian methods are 
 useful for studying properties of
strings and branes in background fields.
 We apply Hamiltonian formalism for the string in  
constant as well as variable B-field. It leads to new methods to study
noncommutativity for strings in  B-field and membrane in 
constant C-field. We also consider higher rank p-branes.
BRST/conversion methods help us to study stringy origin of Seiberg-Witten
map between noncommutative/ordinary theory that appears in the low energy limit
of open string/membrane.
We also show that starting from string/membrane Hamiltonian analysis 
it is possible to remove noncommutativity/non-associativity.

\bigskip
\newpage



\chapter{ \bf  
 Superalgebras
 }

 
 
\section{\bf  New massless representations of Poincare superalgebra
  with brane charges }

In this section we study massless representations of the
supersymmetry algebra with brane charges. We start from  $D=4$, $N=1$
and then generalize our result to eleven dimensions \cite{RSS-new}.
 
 In arbitrary dimensions, superalgebra with brane charges is given by
 
\be
\{ Q_\a, Q_\b \} = Z_{\a\b},
\la{alg-arb}
\ee
 
 where $\a, \b  = 1,...,n$ and  $Z_{\a\b}$ is given by decomposition in terms of all symmetric
$\c$-matrices.
Using $GL(n,R)$ rotations,  it is possible to bring $Z_{\a\b}$ to
diagonal form
(see for example \cite{BAIL},\cite{W1})

\be
Z_{\a\b} = diag(c_1, ... c_s,0, ... ,0).
\la{Zdiag-s}
\ee
 
For different values of $s$ one has different number of the nontrivial
supercharges.  We incorporate  $GL(n,R)$ rotations in generalization of
Wigner method for the
superalgebras with brane charges. In generalized Wigner method we choose 
generalized Little group not as subgroup of $SO(n,1)$ but as subgroup of
$GL(n,R)$.
 The nontrivial part of the superalgebra could be represented as Clifford
algebra.
 Then, one decomposes the remained supercharge operators into creation and
annihilation operators.
 Surprisingly, it is possible, by fixing some particular $GL(n,R)$
 rotation, to have only one nontrivial creation operator. This analysis is
aplyed for massive as well as massless representations.
The difference with analysis of \cite{BAIL} is that they considered only
massive
BPS states. Those BPS states break fraction of the
supersymmetry and the number
of non trivially commuting supercharges 
corresponds to the number of broken supersymmetries.
In applying Wigner method to 
massless representations the situation is different. 
The number of non trivially commuting supercharges corresponds to the
number of
annihilation and creation operators, therefore, connected to the length
 of the supermultiplet.
 It is more convenient to consider some particular examples.

 In four dimensions $N=1$ superalgebra is 

\be
\{ Q_\a, Q_\b \} = Z_{\a\b}.
\la{alg4}
\ee

The decomposition of $Z_{\a\b}$ in the basis of $\c$ matrices 
is taking the following form  

\be
Z_{\a\b} = (\c^\m)_{\a\b}P_\m + (\c^{\m\n})_{\a\b}Z_{\m\n}.
\la{P4}
\ee
As it was mentioned before, 
we use a generalization of the Wigner method. 
Instead of defining little group generators, we use
matrices $T_{\a\b}$, that are the
elements of $GL(4,R)$. This group is the automorphysm group of the
\eq{alg4}.
 Then, $Z_{\a\b}$ could be diagonalize using subgroup of $GL(4,R)$
rotations. It could be done in the similar
 way, comparing to the situation when one chooses  form of momentum
invariant under transformations of
the little group.
In this case $Z_{\a\b}$ is taking the form 

\be
Z_{\a\b} = diag(c_1,c_2,0,0).
\la{Zdiag}
\ee

Superalgebra \eq{alg4} is 

\be
\{ Q_i,Q_j \} = diag(c_1, c_2),
\la{QQ4}
\ee

where $i,j = 1,2$ and the rest of the supersymmetry 
generators anticommute. Moreover.
it is possible to have $GL(4,R)$ rotation  that gives $c_2 = 0$.
 In this case there is additional multiplet shortenning. 

Applying Wigner method to derive massless representations of
superalgebra {\it without}
brane-charges, one ends up with two nontrivial
supersymmetry generators.  
One of them is interpreted
as
creation and another one as annihilation operators

\be
Q_1 |>_0 = |>_{1/2}.
\la{Q012}
\ee

If brane charges are present and $c_2 = 0$ then the number of
generators is reduced by two. The similar effect happens in eleven
domensions.

To study representations of this superalgebra let 
us first investigate Casimir operators.
It is possible to impose covariant condition on $Z_{\a\b}$ such as

\be
Z_{\a\c} Z^{\c \b} = 0.
\la{Z24}
\ee

This particular choice will be explained later from point of view of the field
theory, 
which realizes this representation. It is necessary to mention, that even
without refering to particle model it is interesting to look for subclass
of all representations generated by the condition \eq{Z24}. 
The equation \eq{Z24} is 
generalized masslessness condition. The mass-shell condition $P^2 = 0$ is
also
imposed, because we are
interested in massless representations.

To find solution of the \eq{Z24} let us apply \eq{P4}   

\be
(P_\m P^\m) - 2(Z_{\m\n}Z^{\m\n}) = 0,
\la{PPsup}
\ee

\be
\e^{\m\n\l\r}Z_{\m\n}Z_{\l\r} = 0,
\la{ZZsup}
\ee

\be
\e^{\m\n\l\r}P_\n Z_{\l\r} = 0.
\la{PZ}
\ee

We see that in the general case 
the mass is proportional to central charge. Conditions \eq{PPsup},\eq{ZZsup} 
and \eq{PZ} could be solved as

\be
Z_{\m\n} = P_\m K_\n - P_\n K_\m,
\la{PK}
\ee

\be
(P \cdot K)^2 = P^2(K^2 - 1),
\la{PK2}
\ee

and in general case P and K do not have to be orthogonal.
For massless representations \eq{PK2} reduces to $P_\m K^\m = 0$.
It is possible to interpret equation \eq{PK} from two points of view.
In first one we represent $Z, P, K$ not as 
operators but rather as eigenvalues of those
operators on physical states.
The second point of view states that the decomposition \eq{PK} could be
understood literally. In this
case we have finite dimensional nonlinear deformed superalgebra 
with generators
$P$ and $K$. 
The interpretation in terms of non-linear deformed superalgebras is
very interesting possibility and needs further investigation.
The examples of finite-dimensional nonlinear deformed superalgebras were
considered before
in \cite{RS1},\cite{B1}, \cite{RSS} for the case 
of particle and string models.
They also we studied in context of
the unification of type IIA and type IIB superalgebras
in higher dimensions.

 Using the information given above the equation
\eq{QQ4} could be simplified. 
Choosing momentum, in the form $p_\m = (\omega,0,0,\omega)$ one has
 
\be
\{ Q_1, Q_1 \} = (\omega + c), \qquad \{ Q_2, Q_2 \} = (\omega - c),
\la{QQoc}
\ee

where $c$ is the value of $K_\m = (0,0,c,0)$. 
Even in massless case
if $\omega = c$ one has multiplet shortening.
This condition resemble BPS equation. The difference is that in BPS case
all multiplets are massive but here we are discussing massless
supermutliplets only.
 Using the example \eq{QQoc} 
 one can 
interpret different 
values of $s $ in equation \eq{Zdiag-s}.
 
 Now we want to study the representations of the superalgebra \eq{alg4}.
Usually, superalgebra with central charges does not have massless
representation
and the shortest massive multiplet is BPS.  But in presence of brane
charges,
it is possible to have
massless representations of superalgebra  
without breaking positive energy theorem \cite{BL}.

In addition to the Casimir $C_1 = P^2$ we have another one described by

\be
C = K^2.
\la{CK4}
\ee

Also one can have extra Casimir operator associated with Pauli-Lubanski
vector if one substitutes $K_m$ instead of $P_m$.
One can see that massless multiplet includes arbitrary parameter $C$ that is associated
with Casimir \eq{CK4}. In the next chapters we will
 interpret inclusion of this parameter, effectively,  as existence 
of additional "spinning"
degrees of freedom.
 
  Therefore we see, tha in $D=4$ one has a multiplet consisting of one
fermion and one boson. States of this supermultiplet are generated by
 $Q_1$ and parametrized by the arbitrary value of $c$ as well as
eigenvalues of
additional Casimir operators.

We see that even superalgebra with central charges can have massless
representations if those charges associated not with extended objects but
with ordinary particle states. Moreover, for some particular value of
brane
charges 
  the number of
nontrivial supersymmetries in $D=4$ reduces to
1/4, i.e. one has ultra short massless multiplet (comparing to BPS one).
The possibilities of new multiplets with different number of nontrivial
supersymmetries were considered in \cite{Hull} and \cite{3/4} for
massive BPS representations only.
In that approach, central charges were associated with extended objects
such as 
p-branes. $D=4$, $N=1$ case was considered previously in \cite{BLS}.

 For the case of eleven dimensions one can have different "little" groups under 
$GL(32,R)$. It is convenient to choose $T_{i \b}$ among generators which live invariant
diagonal form of $Z_{\a\b}$, i.e.

\be
Z_{\a\b} = diag (c_1, ... , c_r,0,...,0).
\la{Z114}
\ee

Then the superalgebra 

\be
\{ Q_\a, Q_\b \} = Z_{\a\b},
\la{alg5}
\ee

where 

\be
Z_{\a\b} = (\c^\m)_{\a\b}P_\m + (\c^{\m\n})_{\a\b}Z_{\m\n} + (\c^{\m_1 ... \m_5})_{\a\b}
Z_{\m_1 ... \m_5}.
\la{P5}
\ee

could be written in the form

\be
\{ Q_i, Q_j \} =  diag(c_1,...,c_r),
\la{QQ11}
\ee

where $i=1,...,r$. Here we have to be careful, because condition \eq{Z24}
assumes that $r < 16$ or $r=16$.  But in simplest case $ i=1 $ one sees that additional symmetries
of $GL(32,R)$ comparing to its  Poincar\'e subgroup allows us to fix values 
of central charges in such a way that massless multiplet is getting shortened up to one
nontrivial supercharge. In most general case one can have 8 and  16
supercharges. Nontrivial part of superalgebra giving supergravity
multiplet in
$D=11$ has 16 supercharges (8 creation and 8 annihilation charges).
 Similar shortening one can have starting from example of massless
representations 
of $OSp(1/32,R)$ \cite{Gun1}. 
Usually, in difference from singleton multiplets, massless multiplets of
$OSp(1/n,R)$ supergroup survive Ignonu-Wigner contraction, and in this case for some 
particular value of central charges massless multiplet of $OSp(1/32,R)$ which contains
scalar, spinor and three-form, but doesn't contain graviton, could be 
contracted to
Poincar\'e limit. It also shows existence of other multiplets apart from 
supergravity multiplet
in eleven dimensions.

 Now it is interesting to discuss what follows from condition \eq{Z24}
in eleven dimensions. It is more convenient for now to consider two different limits. When 
five form brane charge is equal to zero but two form is not and when two form is zero but
five form is not.
 In the first case five-form $ Z^{(5)} = 0$. In this case one can extract equations 
equivalent to the system \eq{PPsup} - \eq{PZ}.  Solution of those
equations could be written in the form
similar to \eq{PK}, \eq{PK2}.  
 But in the case when $Z^{(2)} = 0$ situation is slightly different. Equations that follow
from \eq{Z24} are

\be
Z_{(5)}^2 = 0, \qquad Z^{(5)}_{[\m_1,..., \m_4}{}^\n Z_{\n_1,...,\n_4] \n } = 0
\qquad (P Z^{(5)} )_{\n_1,...\n_4} + ( Z^{(5)} Z^{(5)} )_{\n1,...,\n_4} = 0.
\la{PZZ5}
\ee

Those equations also could be solved in manner analogous to \eq{PK}, \eq{PK2}.

  Now let us discuss possible massless irreps in eleven dimensions coming
from \eq{QQ11}
for the cases of different $r$. If $r=1$ the situation is similar to four-dimensions and
one has just one creation operator and therefore only two states: with spin zero and one-half.
For the case of $r=2$ we still have spin zero and one-half but in addition to previous case
we have to add CPT conjugated states into the consideration. $r=8$ corresponds to
the state of spin $(0, 1/2, 1)$, in the final case of $r=16$ one has multiplet with highest
spin two. The important observation  here that one should not naively
interpret states 
with highest spin one as super Yang-Mills multiplet. The difference appears that in addition to
ordinary mass and spin those states are parametrized by the value of additional Casimir
operators made out of $Z^{(2)}$ and $Z^{(5)}$. In most general case it is
not quite clear how to
interpret them in most general case.  In the following chapters using
connection between
particle model and
field theory, we will describe them as pure "twistorial" degrees of
freedom. This will be demostrated on
the examples of four-dimensions and  in eleven-dimensional
space-time. In these examples, the wave functions (or fields) will
depend on additional spinorial 
degrees of freedom. To make a connection with ordinary field theory and to
get rid of extra
spinorial dependence we will decompose fields/wave functions in powers of
those spinors. Interestinly,
the coefficients of that decomposition will be symmetric tensors in their
spinorial indices.
The imposing additional constraints, that appear from the particle 
quantization, will give
us fields describing
 linearized field theory. This field theory posess supersymmetry generated
by the super Poincar\'e algebra with branme charges.

Now we have to interpret the representations of Poincar\'e superalgebra with brane charges
in eleven dimensions
 from the point of view of field theory, and then, discuss a
connection, if any, with spectrum of linearized  M-theory. This will be
discussed in the
following chapters.

\bigskip



\section{\bf Supersymmetry in dimensions beyond eleven}


 Attempts to extend the supersymmetry beyond eleven dimensions have a long history.
The major obstacle was due to appearance of spins higher then two in the
supermultiplet. Surprisingly, 
in twelve dimensions this problem could be resolved.
 In this section we will follow \cite{RSS}.
 For a first sight one can not have highest spin 2 in the supermultiplet
if $D=12$, because
of the 64 Majorana Supercharges. But it is true if
signature of
space time is $(11,1)$. It is still possible to have 32 Supercharges if they are
Majorana-Weyl spinors. This interesting result imposes restriction on
signature of space-time. It should be $(10,2)$ now. In twelve dimensional
space-time of this signature it is possible to have Majorana-Weyl
spinors.
In this case highest spin in the supermultiplet is 2 and everything is
consistent.
There is one problem that appears here. In $D=(10,2)$,
$(\c^m)_{\a\b}$ is antisymmetric and one can not have
a momentum generator in the right hand side of the 
 superalgebra.  One can have the following $(10,2)$ superalgebra

\be
\{ Q_\a, Q_\b \} = (\c^{mn})_{\a\b} Z_{mn}.
\la{s-algebra-12}
\ee

There are many different ways to interpret the generator $Z_{mn}$, but if
one
wants to have momentum generator in the theory there are two distinct
possibilities. One of them is

\be
\{ Q_\a, Q_\b \} = (\c^{mn})_{\a\b} P_m n_n,
\la{s-algebra-12-1}
\ee

where $n_n$ is constant vector and another one

\be
\{ Q_\a, Q_\b \} = (\c^{mn})_{\a\b} P^1_m P^2_n,
\la{s-algebra-12-2}
\ee

where $ P^1_m$ and $P^2_n$ are two independent commuting generators and each one
can play role of momenta.
The first case assumes that there is one chosen direction in twelve-dimensional
space-time. It means that twelve-dimensional theory is effectively lower
dimensional. In some sense it could be dangerous, because it could 
mean that we managed to rewrite ten-dimensional theory using new variables in
twelve-dimensional fashion. Therefore, theory effectively is still
ten-dimensional.
Nevertheless, this approach could be extremely useful in unification
of type IIA/B and heterotic algebras from higher dimensions. We will 
discuss this unification later in this chapter. 
If higher dimensional theory has one 
chosen direction it does not necessary mean that it gives only one theory in lower
dimensions. We will see that even with constant vector in tact the twelve dimensional
theory is able to produce type IIA/B and heterotic superalgebras in $D=10$.
 One can ask: why should we start from the theory with broken Lorentz 
symmetry? There are few possible answers. First of all, this broken
symmetry could be
sign of more extended theory, which is Lorentz invariant. The other, is that if initial
symmetry has one chosen direction, it could give us the right theories
upon compactification of that fixed direction.
From this point of view important question arises: should the underlined theory of
everything be completely Lorentz invariant?
 So far there are no complete evidence that this must be the
case. Moreover, if one starts from theory of everything with one fixed
direction, and this
theory is beautiful and easy to produce low dimensional results with, then 
upon restoring
Lorentz invariance the result could be much more complicated and theory
could have no other
uses. The different theories with fixed directions in dimensions beyond 
eleven were 
considered in \cite{ns} - \cite{Nishino:1998sw}.

Another possibility described by \eq{s-algebra-12-2} assumes that theory
intrinsically is twelve-dimensional but we have to give an interpretation
to new generator
$P^2_n$. It will be done in next chapter.

It is by now well known that the type IIA string in ten dimensions is
related to M-theory on $S_1$, and the $E_8\times E_8$ heterotic string
is related to M-theory on $S^1/Z_2$ (See \cite{pkt1} for a review).
However, the connection between M-theory and type IIB theory, $SO(32)$
heterotic string and the type I string theory is less direct. One needs
to consider at least a dimensional reduction to nine dimensions to see
a connection. 

One may envisage a unification of the type IIA and type IIB strings in
the framework of a higher than eleven dimensional theory. The simplest
test for such an idea is to show that the Poincar\'e superalgebras of the
IIA/B theories are both contained in a spacetime superalgebra in 
$D\ge 10$ dimensions. The downside of this reasoning is an old result due to
Nahm \cite{nahm}, who showed that, with certain assumptions made,
supergravity theories are impossible in more than $(10,1)$ dimensions
(and supersymmetric Yang-Mills theories in more than $(9,1)$
dimensions)
%
\footnote {A candidate supermultiplet in $(11,1)$ dimensions was
considered in \cite{pvn}, but it was shown that no corresponding
supergravity model exists. }.
%
He assumed Lorentzian signature, and required that no spin higher than
two occurs. Much later, an analysis of super p-brane scan allowing
spacetimes with non-Lorentzian signature, the possibility of a $(2,2)$
brane in $(10,2)$ dimensions was suggested \cite{duff}. 

More recently, various studies in M-theory have also indicated the
possibility of higher than eleven dimensions
\cite{B1,hull,vafa,km1,at,pkt2}. (See also,
\cite{ b2,RS1,B1,ns}, \cite{jr, b3, n1, b5}, \cite{RS, b6, b7}
and  \cite{n2, b8, n3}  ).

In most of these
approaches, however, one needs to introduce constant null vectors into
the superalgebra which break the higher dimensional Poincar\'e symmetry.
Accordingly, one does not expect the usual kind of supergravity theory
in higher than eleven dimensions \cite{hull,vafa}. 

An approach which maintains higher dimensional Poincar\'e symmetry has
been proposed \cite{b2}. However, much remains to be done to determine
the physical consequences of this approach \cite{B1, b3}, since it
requires a nonlinear version of the finite dimensional super-Poincar\'e
algebra, in which the anticommutator of two supercharges is proportional
to a product of two or more translation generators \cite{b2} (see also
\cite{b5,es1})
\footnote {
While the occurrence of nonlinear terms in a {\it finite}
dimensional algebra is unusual, we refer to \cite{db} for a study of
finite W-algebras where such nonlinearities occur (see also
\cite{w1,w2,w3}). It would be interesting to generalize this study to
finite super W-algebras that contain nonlinear spacetime superalgebras
of the kind mentioned here. 
}.
Simplest realizations of these types of algebras involves 
multi-particles, as was shown first for bosonic systems 
in \cite{B1}, and later for superparticles in \cite{RS1,RS,b7,b8}.
Putting all particles but one on-shell yields an action for a
superparticle in which the constant momenta of the other particles
appear as null vectors.

In what follows, we shall focus our attention on the superalgebraic
structures in $D>11$ that may suggest a IIA/B unification and their
field theoretic realizations which involves null vectors, or certain
tensorial structures, explicitly. 


\subsection{\bf Unification of IIA/B algebras in $D > 10$}


To begin with, let us recall the properties of spinors and Dirac
$\c$-matrices in $(s,t)$ dimensions where $s(t)$ are the number of
space(time) coordinates. The possible reality conditions are listed in
Table 1,
Appendix A, where $M, PM, SM, PSM$ stand for Majorana, pseudo Majorana,
symplectic Majorana and pseudo symplectic Majorana, respectively
\cite{kt}. An additional chirality condition can be imposed for $s-n=
0~{\rm mod}~4$.
 
The symmetry properties of the charge conjugation matrix
$C$ and the $\c$-matrix $(\c^\m C)_{\a\b}$ are listed in Table 2,
Appendix A. The parameters $\e_0$ and $\e_1$ arise in the relation
\bea
&& C^T=\e_0 C\ ,\nn\\
&& (\c^\m C)^T = \e_1 (\c^\m C)\ . \la{sp}
\eea
This information is sufficient to deduce the 
symmetry of $( \c^{\m_1\cdots \m_r} C)_{\a\b}$ for any $r$, since 
the symmetry property alternates for $r~{\rm mod}~2$.

Using Tables 1 and 2 from Appendix A, it is straightforward 
to deduce the structure of
the type IIA/B superalgebras in $(9,1)$ dimensions. The $N=(1,1)$ super
Poincar\'e algebra (i.e. type IIA) contains a single $32$ component
Majorana-Weyl spinor generator $Q^\a$, with $\a=1,...,32$ and a set of
$528$ bosonic generators, including the translation generator $P^\m$,
that span a symmetric $32\times 32$ dimensional symmetric matrix. The
non-trivial part of the algebra reads
\be
{\bf D=(9,1)\ ,   IIA:}  \la{2a}
\ee
\bea
\{ Q_\a, Q_\b\} &=& \c^\mu_{\a\b}\ P_\mu + (\c_{11})_{\a\b}\ Z
        		+(\c_{11}\c^\mu)_{\a\b}Z_\mu \nn\\
&& +\c^{\mu\nu}_{\a\b}\ Z_{\mu\nu}
       + (\c_{11}\c^{\mu_1...\mu_4})_{\a\b}\ Z_{\mu_1...\mu_4}\nn\\
&& +\c^{\mu_1...\mu_5}_{\a\b}\ Z_{\mu_1...\mu_5}\ . \nn
\eea

All the generators labelled by $Z$ in this algebra, and all the
algebras below, commute with each other. The generators of the Lorentz
group can be added to all these algebras, and the $Z$-generators
transform as tensors under Lorentz group, as indicated by their indices.

It is clear that the algebra \eq{2a} can be written in a $(10,1)$
dimensional covariant form
\be
 {\bf D=(10,1)\ ,   N=1:} \la{11d}
\ee
$$
\{ Q_\a, Q_\b\}=\c^\mu_{\a\b}\ P_\mu +\c^{\mu\nu}_{\a\b}\ 
Z_{\mu\nu}\ +\c^{\mu_1...\mu_5}_{\a\b}\ Z_{\mu_1...\mu_5}\ . 
$$



 
Next, we consider the $N=(2,0)$ algebra in $(9,1)$ dimensions which
contains two Majorana-Weyl spinor generators $Q_\a^i\, (\a=1,...,16, \,
i=1,2)$ and $528$ bosonic generators. The nontrivial part of this algebra
takes the form 
\be
{\bf D=(9,1)\ ,   IIB: } \label{2b}
\ee
\bea
\{ Q_\a^i, Q_\b^j\} &=&\tau^{ij}_a\left(\c^\mu_{\a\b}\ Z_\mu^a +
                \c^{\mu_1...\mu_5}_{\a\b}\ Z_{\mu_1...\mu_5}^{a+}\right) \nn\\
             && +\e^{ij}\c^{\mu\nu\rho}_{\a\b}\  Z_{\mu\nu\rho}\ , \nn
\eea
where $\tau_a$ are the $2\times 2$ matrices $\tau^a= (\s_3,\s_1, 1)$. We
can make the identification $Z_\mu^3\equiv P^\mu$. 

Various aspects of the algebras discussed above have been treated in
\cite{pkt3} in the context of brane charges, and in \cite{b5} in the
context of higher dimensional unification of IIA/B superalgebras. 

In order to unify the superalgebras \eq{11d} and \eq{2b}, we consider
superalgebras in
\be
D=(10+m,1+n)\ , \qquad m,n=0,1,...\la{dims}
\ee
dimensions. To simplify matters, we shall restrict the number real
supercharges to be
\be
{\rm dim}~Q \le 64\ .
\ee
Using Table 1 and Table 2 from Appendix A, we learn that this 
restriction requires
dimensions \eq{dims} with
\be
m+n \le 3\ . 
\ee
Examining all dimensions \eq{dims} for which $(m,n)$ obey this
condition, we find that there are two distinct possibilities: 
\begin{itemize}
\item{$N=(1,0)$ algebra in $D=(11,3)$}    
\item{$N=(2,0)$ algebra in $D=(10,2)$}  
\end{itemize}

Both of these algebras contain $64$ real supercharges, and the second
one is not contained in the first.

The $N=(2,0)$ algebra in $(10,2)$ dimensions has two Majorana-Weyl spinor
generators $Q_\a^i\ (\a=1,...,32,\ i=1,2)$ obeying the anticommutator
\be 
{\bf D=(10,2)\ , N=(2,0): } \la{12-12}
\ee
\bea
\{Q_\a^i,Q_\b^j\} =& \tau_a^{ij}\left( \c^{\mu\nu}_{\a\b}\, Z_{\mu\nu}^a +
\c^{\mu_1...\mu_6}_{\a\b}\, Z_{\mu_1...\mu_6}^{a+}\right)\nn\\
& +\e^{ij}\left( C_{\a\b}\, Z + \c^{\mu_1...\mu_4}_{\a\b}\, 
Z_{\mu_1...\mu_4}\right) \ .\nn 
\eea 

The $N=(1,0)$ algebra in $D=(11,3)$ dimensions, on the other hand, takes
the form

\be
{\bf D=(11,3)\ , N=(1,0): } \la{14}
\ee
$$
\{Q_\a, Q_\b\} = (\c^{\m\n\r})_{\a\b}~Z_{\m\n\r}
+ (\c^{\m_1\cdots \m_7})_{\a\b}~Z^+_{\m_1\cdots \m_7}\ .
$$ 
Here and in \eq{12-12}, the $\c$-matrices are chirally projected. Factors
of $C$ used to raise or lower indices of $\c$-matrices are suppressed
for notational simplicity. Note also that both \eq{12-12} and \eq{14} have
$2080$ generators on their right hand sides, spanning $64\times 64$
dimensional symmetric matrices.

Various dimensional reductions of the algebra \eq{14} yield:
\begin{itemize}
\item[--]{ $N=1$ algebras in $D=(11,2),\ (10,3)$ }
\item[--]{ $N=1$ algebra in $D=(11,1)$ }
\item[--]{ $N=(1,1)$ algebra in $D=(10,2)$ }
\item[--]{ $N=2$ algebra in $D=(10,1)$ }
\end{itemize}
all of which have 64 real supercharges, and contain the $(9,1)$
dimensional IIB and $(10,1)$ dimensional $N=1$ algebras. The algebra \eq{12-12}
reduces to the last one in the above list. 

The master algebras \eq{12-12} and \eq{14} also give the $N=1$ algebra in
$D=(9,3)$, the $N=2$ algebra in $D=(9,2)$, the $N=(2,2)$ and $N=(2,1)$
algebras in $D=(9,1)$, all of which contain the IIA/B algebras of
$D=(9,1)$, but not the $N=1$ algebra of $D=(10,1)$.

Now we would like to show the embedding of the IIA/B and
heterotic algebras of $D=(9,1)$ in the master algebras \eq{12-12} and
\eq{14}. 

In the case of \eq{12-12}, the spinor of $SO(10,2)$ decomposes
under $SO(10,2)\rightarrow SO(9,1)\times SO(2)$ into two left-handed
spinors $Q_{\a}^{i}$ and two right-handed spinors $Q^{\a i}$. We are
using chiral notation in which lower and upper spinor indices refer to
opposite chiralities and there can be no raising or lowering of these
indices. We keep only $Z_{\hm \hn}^{a}$ ($\hm=0,1,...,11$) for
simplicity, and make the ansatz
$Z_{\hm\hn}^{a}\s_{a}^{ij}=P_{[\hm}v_{\hn]}^{ij}$. In the reduction to
$(9,1)$ dimensions we set $P_{\hm}=(P_{\m};0,0)$ and
$v_{\hm}^{ij}=(\vec{0};v_{r}^{ij})$, where $r=10,11$. Now the non-vanishing part of 
the algebra
\eq{12-12} reads:
\bea
\{Q_{\a}^{i},Q_{\b}^{j}\}&=&\c^{\m}_{\a\b}P_{\m}v_{+}^{ij} \ ,\\
\{Q^{\a i},Q^{\b j}\}&=&(\c^{\m})^{\a\b}P_{\m}v_{-}^{ij} \ , 
\eea
where $v_{\pm}={1\over 2}(v_{10}\pm v_{11})$. The desired embeddings are
then obtained by setting 
\begin{itemize}
\item[IIB:] { $\quad v_{+}^{ij}=\delta^{ij} \ ,\quad v_{-}^{ij}=0 \ .$}
\item[IIA:] { $\quad v_{+}^{ij}=v_{-}^{ij}={1\over 2}(1+\s^{3})^{ij} \ ,$}
\item[Het:] { $\quad v_{+}^{ij}={1\over 2}(1+\s^{3})^{ij} \ , \quad v_{-}^{ij}=0 \ .$}
\end{itemize}

In the case of \eq{14}, the spinor of $SO(11,3)$ decomposes under
$SO(11,3)\rightarrow SO(9,1)\times SO(2,2)$ into two left-handed spinors
$Q_{\a A}$ and two right-handed spinors $Q^{\a}_{ \dot{A}}$, where
$A,\dot{A}=1,2$ label left- and right-handed spinors of $SO(4)$. We keep
only $Z^{\hm\hn\hat{\r}}$ ($\hm=0,1,...,13$) for simplicity, and make
the ansatz $Z_{\hm\hn\hat{\r}}=P_{[\hm}F_{\hn\hat{\r}]}$
%
\footnote {See \cite{b5} which achieves the embeddings of the full IIA/B
algebras \eq{2a} and \eq{2b}, by using multi $F$-tensors and taking into
account $Z^{\hm_1...\hm_7}$ }.
%
We now set $P_{\hm}=(P_{\m};0,0,0,0)$, $F_{\m\n}=0=F_{\m r}$ and
$F_{rs}=
(\s_{rs})^{AB}v_{AB}+(\s_{rs})^{\dot{A}\dot{B}}v_{\dot{A}\dot{B}}$,
where $r,s=10,11,12,13$ and $\s$-matrices are the van der Waerden
symbols of $SO(4)$. Now the non-vanishing part of the algebra \eq{14}
reads:
\bea
\{Q_{\a A},Q_{\b B}\}&=& \c^{\m}_{\a\b}P_{\m}v_{AB} \ ,\\
\{Q_{\a \dot{A}}, Q_{\b \dot{B}}\}&=&
\c^{\m}_{\a\b}P_{\m}v_{\dot{A}\dot{B}} \ .
\eea
The desired embeddings are then obtained by setting 
\begin{itemize}
\item[IIB:] { $\quad \det (v_{AB})\neq 0 \ , 
\quad v_{\dot{A}\dot{B}}=0 \ ,$}
\item[IIA:] { $\quad v_{AB}=u_{(A}u_{B)} \ , 
\quad v_{\dot{A}\dot{B}}=u_{(\dot{A}}u_{\dot{B})} \ ,$}
\item[Het:] { $\quad v_{AB}=u_{(A}u_{B)} \ , 
\quad v_{\dot{A}\dot{B}}=0 \ ,$}
\end{itemize}
where $u_{A}$, $u_{\dot{A}}$ are constant spinors. Note that in the case
of IIB, the symmetric matrix $v_{AB}$ has rank two and it can be chosen
to be $\d_{AB}$ in a suitable basis.

For the heterotic case, which we will focus on in the rest of this
paper, the embedding can be equally well realized by choosing (dropping
the hats)
\be
Z_{\m\n\r}=P_{[\m} v_{\n\r]}\ , \la{z}
\ee
where
\be
v_{\m\n} \equiv n_{[\m} m_{\n]}\ , \la{vmn}
\ee
and $n, m$ are constant and mutually orthogonal null vectors:
\be
m^\m m_\m=0\ , \quad n^\m n_\m=0\ ,\quad  m^\m n_\m=0\ . \la{nm}
\ee
With these choices, it is clear that the matrix $\{Q_\a, Q_\b\}$ in
\eq{14} has rank $16$, as appropriate for the heterotic algebra.


\subsection{\bf Properties of supersymmetry in $D>11$}


We started out by considering an algebraic unification of the $(9,1)$
dimensional IIA/B superalgebras in higher dimensions, with emphasis on
$(11,3)$ dimensional $N=(1,0)$ algebra \eq{14}. Having made the choices
\eq{z} and \eq{vmn}, however, we have restricted ourselves to the
embedding of a supersymmetric theory with only $16$ supercharges. While
this is useful in understanding how the null vectors arise in a field
theoretical realization, ideally one should seek a master field theoretic
realization in which both IIA and IIB (and hence heterotic) symmetries
are realized according to the suitable choices to be made for the
three-form charge occurring in \eq{14}. 

In the next chapter we will focus our attention on zero-brane and 
the super Yang-Mills system it couples to, but the considerations 
will apply to higher branes as well \cite{n1,b6,b7}. 

We have also restricted our attention to the IIA/B unification in a
maximal dimensional spacetime (with $64$ real supercharges), namely
$D=(11,3)$. Null reduction of our results for
yield corresponding results
for $N=(1,0)$ supersymmetric models in $(10,2)$ dimensions. However,
$N=(2,0)$ supersymmetric results in $(10,2)$ dimensions cannot be
obtained in this way. In fact, a IIA/B unification in the framework of
the $N=(2,0)$ algebra in $(10,2)$ dimensions does not seem to have
attracted attention previously, and it may be interesting to investigate
this case further. 

One of the dividends of a higher than eleven dimensional unification of
IIA/B systems should be a more manifest realization of various duality
symmetries among the ten dimensional strings/branes. As it has been
stressed in \cite{b2, bars1}, these symmetries are to be interpreted as the
similarity transformations of the $64\times 64$ symmetric matrix
$\{Q_\a, Q_\b\}$ which leaves the BPS condition ${\rm det}~\{Q_\a,
Q_\b\} =0$ invariant. An explicit realization of these symmetries at the
level of brane actions would be desirable.

The introduction of structures, e.g. null vectors which break the higher
dimensional Poincar\'e symmetry may give the impression that not much is
gained by a higher dimensional formulation, and that it may amount to a
rewriting of the original theory. This is not quite so, even if one
considers the embedding of a single type of algebra in higher
dimensions, when one considers the null vectors as the averages of
certain quantities, e.g. momenta, attributed to other branes
co-existing with the brane under consideration, as has been illustrated
in \cite{B1,b3}. Furthermore, as mentioned briefly before and 
will be showed in the next chapter,
there exists now a simple realization of $D>11$ superalgebras which
involve the momenta multi-superparticles \cite{RS1,b8}. These do not
involve constant null vectors and maintain manifest covariance in
$D>11$. The multi-brane extension of these results and the nature of
target space field theories they imply, are interesting open problems.

Another aspect of the theories considered here is that their reductions
to lower than ten dimensions give rise to new kinds of super Yang-Mills
theories which, together with supergravity sector which can be included,
are candidates to be the low energy limits of certain $N=(2,1)$ strings.
The utility of such strings lies in the fact that they provide a unified
picture of various branes, e.g. string and membranes \cite{km1,km2},
resulting from different choices for the null vectors. 
In the next chapter we will 
generalize the construction of \cite{km1} to higher than $(2,2)$
dimensional targets, indeed to $(n,n)$ dimensional ones.
The description of their effective target space models is an interesting
problem. We expect that the field equations in these models are related
to the generalized self-dual Yang-Mills systems studied in \cite{hull2}. 
   
In all the algebras considered here the $Z$-type generators commute with
each other. However, there exist interesting extensions of the
Poincar\'e superalgebra in $(10,1)$ dimensions that includes super
two-form \cite{ma1,ma2}, and super five- form \cite{ma3} generators. The
most general such algebra with $N=1$ supersymmetry in $(10,1)$
dimensions has been called the $M$-algebra. In this algebra, there are
non-vanishing (anti) commutators of super two-form generators. The role
of the charges, some of which are bosonic and some fermionic, has not
been understood yet in the context of $M$-theory. However, we expect
them to play an important role. It would be interesting, therefore, to
determine if the $M$-algebra generalizes to the $(10,2)$ and $(11,3)$
dimensional algebras reviewed here, and if such an algebraic structure
can help in arriving at a more unifying picture of a wealth of
$M$-theory phenomena.

\bigskip

\newpage




\chapter{\bf Realization of Superalgebras}


 In the previous chapter we discussed the superalgebras in different dimensions
as well as massless representation of Poincar\'e superalgebra with brane charges.
As we mentioned before, it is necessary, for physical applications, to
know the 
realizations
of superalgebras not only in terms of supersymmetric field theory but also on 
the level of
superparticles and superstrings. Even if supersymmetric field theory is not 
known yet, it is
possible to extract some information about its properties from
quantization of 0-brane model. This approach
will be
discussed in next chapter. In this chapter we will start from realization of 
Poincar\'e
superalgebra with brane charges in different dimensions using corresponding 
superparticle
model. Quantization of the superparticle with brane coordinates gives us
information about spectrum and equations 
of motion of
unknown field theory. We will see that using realization of 
eleven-dimensional
superalgebra with brane charges, i.e. M-theory superalgebra, 
it is possible to
identify properties of linearized M-theory from perturbative field theory 
point of view.
 First of all, it is important to mention what exactly we mean under 
 superparticle
realization of the superalgebra. Let us consider example of ordinary 
eleven-dimensional
superparticle. This model has global supersymmetry, generators of which 
satisfy commutation
relations of eleven dimensional superalgebra $without$ brane charges.

\be
\{ Q_\a, Q_\b \} = (\C^m)_{\a\b} P_m.
\la{QQ11ord}
\ee

 The superparticle model also possess
kappa-symmetry, i.e. local fermionic symmetry, that reduces number of unbroken 
supersymmetries by two. This model is described by action

\be
S_{particle} = \frac{1}{2} \int d\tau \frac{1}{e} \Pi^m \Pi_m ,
\la{SP11}
\ee

\be
\Pi^m = \dot X^m - i \bar \theta \C^m \dot\theta   ,
\la{sp11P}
\ee

where $X^m$ and $\theta_\a$ are coordinates of eleven-dimensional target 
superspace.
The action is invariant under global Poincar\'e symmetry and supersymmetry:

\be
\d_\e \theta = \e , \qquad \d_\e X^m = i \e \C^m \theta ,
\la{p11susyord}
\ee

as well as mentioned above $\k$-symmetry

\be
\d_\k \theta = i \C^m \Pi_m \k , \quad \d_\k X^m = i \bar\theta \C^m \d_k \theta ,
\quad  \d_\k e = 4 e \dot{\bar \t} \k ,
\la{11ordkappa}
\ee

where the  $\k_\a$ is kappa-symmetry parameter.
 We say that this model gives us superparticle realization of
eleven-dimensional superalgebra. The quantization of 0-brane action
\eq{SP11} gives spectrum and linearized equations of motion of
eleven-dimensional supergravity \cite{green} (and references therein). 
Supergravity realizes superalgebra in terms of
field theory. There is also supermembrane in eleven dimensions, that
realizes superalgebra on the level of extended objects. 
If we are interesting in only  particle and
field-theoretical realization of the supersymmetry we need to consider
superparticle action and field-theory equations of motion.
 If one introduces brane-charges into algebra \eq{QQ11ord} and still follows
later realization, then it is necessary to interpret variables associated
with brane charges as coordinates of target superspace of some extended
version of the superparticle that will be main topic of the next section.

Superparticle with central charge coordinates was considered before in
\cite{RS1} and quantization in the massless case for $D=4$ $N=1$ was
presented in \cite{BLS}. In this
models, particle live on the supermanifold parametrized by usual
coordinates and variables associated with p-form charges. Quantization in
the massless case gives infinite number of states. The
ordinary N-extended version of the spinning particle \cite{Pi} can
also give higher then two spin spectrum,
but number of states is finite and there
is no interacting field theory interpretation of such states.
The generalization of the superparticle with central charges to AdS
background was conducted in \cite{BLPS} but because of complicated
structure of the fermionic part of the action quantization of this model
in AdS is not known yet. Although the bosonic massive particle in $AdS_5$ 
was quantized in
\cite{K1} and  supersymmetric case was considered in \cite{MT}
using light-cone formulation of particle on $AdS$ recently.\\

\bigskip


\section{\bf The Universal model}


The  generalized model of superparticle in any dimension can be 
realized if one starts from superalgebra \cite{RS1}, \cite{RSS-new}

\be
\{ Q_{\a},Q_{\b} \} = Z_{\a\b}.
\la{alg}
\ee

Then the generalized superspace in
which coordinates are associated with $Q_\a$ and $Z_{\a\b}$:
$$
(Q_\a\ ,\  Z_{\a\b}) \ \ \ra \ \ (\t^\a\ , \ X^{\a\b} ) \ . \la{ma}
$$
In addition, we introduce the momenta $P_{\a\b}$ (symmetric) and
the Lagrange multipliers $e_{\a\b}$ with the same symmetry property of
the charge conjugation matrix :
\be
e_{\a\b}=\e_0\,e_{\b\a}\ ,\qq   C^T=\e_0\,C\ . \la{ecs}
\ee

Here coordinates of particle are not only vector $X_\m$ but also antisymmetric tensors
appearing from decomposition of $X_{\a\b}$ in the basis of $\c$-matrices.

Having defined the basic fields of the model, the action is following \cite{RS1}
\be
I= \int d\tau \left( P_{\a\b}\, \Pi^{\a\b} 
+\ft12 e_{\a\b}\, (P^2)^{\a\b} \right) \ ,
\la{na}
\ee
where
\be
\Pi^{\a\b} = dX^{\a\b} - \t^{(\a} d\t^{\b)} \ ,
\la{Pi}
\ee
and $(P^2)^{\a\b}\equiv P^{\a\c} P_\c{}^\b$. The raising and lowering of
spinor indices is with charge conjugation matrix $C$ which has 
symmetry property \eq{ecs}.

The bosonic symmetry of the action takes the form

\be
\d_\L e_{\a\b} = d\L_{\a\b}\ , \qq
\d_\L X^{\a\b} = -\L^{(\a}{}_\c\, P^{\b)\c}\ ,   \qq
\d_\L P_{\a\b} = 0 \ , \qq
\d_\L \t = 0 \ .  \la{nl}
\ee 

 The action \eq{na} has also fermionic
symmetries. Firstly, it is invariant under the global supersymmetry
transformations

\be
\d_\e \t^\a = \e^\a\ , \qq \d_\e X^{\a\b} = \e^{(\a} \t^{\b)}\ , \qq
\d_\e P_{\a\b} = 0\ ,\qq \d_\e e_{\a\b} = 0\ , 
\ee

which clearly realize the algebra \eq{alg}. The action \eq{na} also has
local $\k$-symmetry given by
\bea
&& 
\d_\k \t^\a = P^{\a\b}\, \k_\b\ ,\nn\\
&&
\d_\k X^{\a\b} = \t^{(\a}\, \d_\k \t^{\b)}\ , \nn\\
&&
\d_\k e_{\a\b}  = 2 ( \k_\b\,d\t_\a + \e_0\,\k_\a\,d\t_\b)\ , \nn\\
&&
\d_\k P_{\a\b} = 0\ . \la{nk}
\eea
Recall that $ C^T=\e_0\,C$. These transformations close on-shell on the
bosonic $\L$-transformations and trivial bosonic transformations $\Sigma$, which are proportional 
to equations on motion:
\be
[\d_{\k_{(1)}},\d_{\k_{(2)}}]= \d_\L + \d_\Sigma \ , \la{closen}
\ee
where the composite gauge transformation parameters are
\bea
&&
\L_{\a\b} = 
\left( \k^{(2)}_\b~P_\a{}^\c~\k^{(1)}_\c
  + \e_0\,\k^{(2)}_\a~P_\b{}^\c~\k^{(1)}_\c \right)-(1\lra 2)\ , \nn\\
&&
\Sigma_{\a\b} = 4\e_0\,\k^{(1)}_{[\a}~d \k^{(2)}_{\b]}-(1\lra 2) \ .
\eea
We need the $e_{\a\b}$ equation of motion in showing the closure on
$X^{\a\b}$, and vice versa.

First of all the number of $\k$ symmetries depends on rank of $P_{\a\b}$.
For example, if one considers ordinary Brink-Schwarz superparticle , which
corresponds to 

\be 
P_{\a\b} = (\c^\m)_{\a\b}P_\m, \qquad e_{\a\b} = eC_{\a\b},
\ee

then it can be showed that rank of P is N/2 if $\a = 1,...,N$.
If we identify number of $\k$ symmetries with part of the target-space
supersymmetries preserved by particle (brane) configuration then we can say that
one-half of supersymmetries conserved. For the most general case number of the
conserved target-space supersymmetries (i.e. number of $\k$ symmetries) is equal 
to the $(N-rank(P))$. That could be seen from constraint analysis and from having 
$(N-rank(P))$ first class constraints corresponding to the same number of $\k$-symmetries.
It is not only case when it is possible to have different number of $\k$ symmetries
for the same generalized model. The other examples are given in \cite{RS1} and \cite{RS}.

The field equations that follow from the action \eq{na} are
\bea
&&
P_{\a\b}\,P^{\b\c}=0\ ,\la{e1}\\
&&
dP_{\a\b}=0\ , \qq  P_{\a\b}\ d\t^\b =0\ ,  \la{e23}\\
&& 
dX^{\a\b}-\t^{(\a} d\t^{\b)} +e^{\c(\a}\, P^{\b)}{}_\c =0 \ . \la{e4}
\la{neom}
\eea
In particular \eq{e1} implies 
\be
\det~\, P=0\ ,
\ee
which is the familiar BPS condition. 
Action \eq{na} is written in fist-order form. It is possible to rewrite action
in the second order form only by using infinite expansions:
the last equation can be solved for $P$ yielding the result
\be
P= -2\e_0 e^{-1} \left(E\,\wedge\,\Pi\right)\ , \la{solp}
\ee
where
\be
E \equiv \left({1\over 1+ e^{-1}}\right)\ ,
\ee
and the definition
$$
e^n  \wedge \Pi\,\equiv\,e^n\,\Pi\,e^{-n}\ ,
$$
is to be applied to every term that results from the expansion of $E$ in
$e$. Substituting \eq{solp} into the action \eq{na}, we find 
\be
I =-2 \int d\tau\ \tr\, \left[e^{-1}\left(E\wedge\Pi \right)\right]^2\ .
\la{2-order-part}
\ee
That is why it is much more convenient to work with first-order form of the action
\eq{na}.

\bigskip


\subsection{\bf Superparticle with central-charge
 coordinates and twistors}

It is also possible to rewrite Lagrangian \eq{na} using bosonic spinors
 $\l_\a$ which can be associated with first component of the four-dimensional
 twistor $Z$ \cite{BLS}, \cite{RSS-new}. To describe general model which includes massive as well as
 massless particles one can  use not one but few sets of such spinors. For
 example, ordinary massless particles in $D=4$ Minkowski space can be
described by the momentum which
 could be represented as $P_{\a\b} = \l_\a \l_\b$, where we used Penrose
  decomposition. For massive case it is not working that could be seen by
even by counting
 degrees of freedom. Instead, one could use pair of spinors $\l_\a^i$ where
 $i=1,2$. But in space-time of an arbitrary dimension one have to use different
 number of spinors. In our case  $P_{\a\b}$ can be chosen directly in 
 the following form if $i=1,..,s$ , then

\be
P_{\a\b} = \l^i_\a \l^i_\b,
\la{Pl1}
\ee

where $\l^i_\a \l^{j \a}=0$.
In this case rank of P is equal to s.  Different value of $s$ leads to
either
massless or massive superparticle. It is also possible to connect $\l$ to
 spinorial Lorentz harmonics.

Using \eq{Pl1} the Lagrangian of superparticle is taking the form:

\be
L=\l^i_\a \l^i_\b \Pi^{\a\b},
\la{L}
\ee

where $\Pi^{\a\b} = dX^{\a\b} - \t^{(\a} d\t^{\b)}$.
 
 This Lagrangian, as will be showed later,  could be quantized in more
convenient
 way as one in \eq{na}, but they are classically equivalent  on the
 level of equations of motion.

It is also possible to consider model with use of spinors slightly different 
from \eq{L}.  One can add to Lagrangian \eq{L} term with constraints 
$\l^{\a i} \l_\a^j = 0$. In previous consideration we did not do 
that because it was
possible to express $\l_\a$ in terms of Lorentz harmonics as it will be 
described later.
It means that it is possible to find most general solution of $\l$ in
terms of spinorial Lorentz
harmonics and condition $\l^{\a i} \l_\a^j = 0$ appears naturally. 
Here, for a moment, we assume that we can not do that anymore.

The action \eq{na} is given in first order form where $P_{\a\b}$ is
the momentum associated with $X^{\a\b}$. For many purposes, it is
convenient to pass to a second order formulation. This can be done
formally by solving for $P_{\a\b}$ in equation of motion of
\eq{na}. However the
result is not illuminating. Instead, a replacement of $P_{\a\b}$ with a
bilinear expression in commuting spinors is more interesting and
promising. In doing so, it should be first noted that 
 symmetric matrix $P_{\a\b}$ has
a maximum rank of $k/2$, if $\a=1,...,k$. Thus, introducing $k/2$
commuting spinors $e_\a^i$, the momentum matrix can be expressed as

\be
P_{\a\b}= e_\a^i\,e_{\b i}\ ,
\la{tr}
\ee

where $\a=1,...,k$ and $i=1,...,k/2$. The contraction is with
Kronecker delta $d_i^j$ and hence gives the manifest local $SO(k/2)$
symmetry of this relation. Introducing the notation

\be
\Sigma_{ij} = e_i^\a\,e_{\a j}\ ,
\la{ds}
\ee

the action \eq{na} can now be written as

\be
I=\int d\tau \left( e_\a^i\,e_{\b i}\, \Pi^{\a\b} +\ft12
e_\a^i\,e_\b^j\,\Sigma_{ij} \l^{\a\b} \right) \ .
\la{na1}
\ee

An immediate consequence of this action comes from the $\l$-equation
of motion

\be
e_\a^i\,e_\b^j\,\Sigma_{ij}=0\ .
\ee

Contracting this equation with $C^{\a\b}$ gives tr\ $\Sigma^2=0$,
which in turn implies that $\Sigma_{ij}=0$, as can be seen by
bringing $\Sigma_{ij}$ into the standard canonical form by means of
a unitary similarity transformation. Thus we have the condition

\be
e^\a_i\,e_{\b j} = 0\ .
\la{ps}
\ee

This is a generalized version of what is known as the pure spinor
conditions which have been considered previously, for example the
condition $\bar u\, \c^\m u=0$ in $D=10$, and the additional
condition $\bar u\, \c^{\m\n} u=0$ in $D=11$ on the spinors $u^\a$
see for example \cite{H1}. Also it appears that pure spinors are useful
in covariant quantization of Green-Schwarz superstring as was shown in
\cite{B1}.

Another consequence of the action \eq{na1} is that the generalized
momentum $P_{\a\b}$ associated with $X^{\a\b}$ is

\be
P_{\a\b}= e_\a^i\,e_{\b i}\ .
\ee

A special case of the action \eq{na} was considered in \cite{BLS}  where a
single spinor is used and consequently $P_{\a\b}=e_{\a\b}$. In that
case the matrix $P_{\a\b}$ has rank one, and  there are
$k-1$ real components of the $\kappa$-symmetry parameter that
participate with the $\kappa$-symmetry transformations. In the
special case of $D=4$, another special case of the action \eq{na}
was considered in \cite{BLS} which amounts to the following choices

\be
e_\a^1 = i\sqrt{1-a\over 2}\,(\l_A, -\bar\l_{\dot A})\ ,
\quad\quad
e_\a^2 = \sqrt{1+a\over 2}\,(\l_A, \bar\l_{\dot A})\ ,
\ee

where $a$ is an arbitrary real parameter. In real spinor notation,
this means $e_\a^1 \sim (\c_5 e^2)_\a$. As a result of this choice,
the components of $P_{\a\b}$ in $D=4$ become

\be
P_{A\dot A}= \l_A \l_{\dot A}\ , \quad\quad P_{AB} = a\,\l_A\l_B\ ,
\quad\quad P_{\dot A\dot B} = a\,{\bar \l}_{\dot A}{\bar \l}_{\dot B}\ .
\ee

For $a=0$ the model reduces to the standard superparticle model
without two-form coordinates, while for $a=1$, it reduces to the
special case where $P_{\a\b}=e_\a e_\b $. For $a\ne 0,1$, it is
a model with $2$ real component $\kappa$ symmetry parameters that
participate in the $\kappa$-symmetry transformations. The
quantization of these cases were carried out in detail in \cite{BLS}.

Model of massless superparticle with central charges for different dimensions
 was considered in  \cite{BLS}.
 For arbitrary dimension their model is given by:

\be
L=\l_\a \l_\b \Pi^{\a\b},
\la{BLS}
\ee

where $\Pi^{\a\b}$ is given by \eq{Pi}. The model \eq{na} in the spinorial
form \eq{L} after taking only one set of spinors $\l$ into consideration
coincides with the model \cite{BLS} which describes massless superparticle
with lowest rank of $P_{\a\b}$. This corresponds to maximal number of
$\kappa$ - symmetries.
 It is also possible to make a connection of the model \eq{L} with
particle in Lorentz-harmonic variables. It helps us to clarify the notion
that we
do not need to have additional constraint in the action. Where this
constraint describes the property
of
$\l$ such as $\l^{\a i} \l_{\a j} = 0$.

 The action of the superparticle with central charge coordinates in
Lorentz-harmonic formalism is \cite{BL}

\be
L= P_{AB}^{++} v_{A\a}^{-}v_{B\b}^{-} \Pi^{\a\b},
\la{harm}
\ee

where $\a$ is $SO(1,D)$ spinorial index, and A is index in spinor
representation of $SO(D-1)$. $(+,-)$ are $SO(1,1)$ weights and $v$ are
spinorial Lorentz harmonics. Rank of $P_{\a\b}$ is defined by the rank of
$P^{++}_{AB}$. Then later could be solved as 

\be
P^{++}_{AB} = \l^{+r}_A \l^{+r}_B,
\la{++-new}
\ee

where r runs from 1 to rank of the $P$. Then \eq{harm} is taking the form

\be
L = (\l^{+r}_A v_{A\a}^{-}) (\l^{+r}_B v_{B\b}^{-}) \Pi^{\a\b}.
\ee

The combination $\l^{+r}_A v_{A\a}^{-}$ gives nothing but set of the  
$SO(1,D-1)$ bosonic spinors $\l_\a^r$ i.e.

\be
  \l_\a^r = \l^{+r}_A v_{A\a}^{-}.
\la{lvv}
\ee

Finally, the Lagrangian \eq{harm} is equivalent to \eq{L}.

\bigskip


\subsection{\bf Superstring and Supermembrane with central charge coordinates}


 In this subsection we discuss generalization of superparticle with
central charge coordinates for the case of superstrings and supermembranes.
 The modifications of superstring action to include tensorial coordinates
have been studied quite extensively. See for example \cite{tc},
\cite{AmBa} (and references therein).
The motivation for construction of those models was the following. 
In $D=4$  one can introduce, in addition to target-space
string coordinates, the rank two tensor $z_{\m\n}$  that has six
independent 
degrees of freedom.
 Then, if it is possible to construct
an action with right number of symmetries, the conformal anomaly
could be canceled already in $D=4$. In this case
 string theory could be consistent,
at least on this level, in four dimensions.
Here we will follow \cite{RS-unpub}. To make a connection
with Green-Schwarz superstring one has to have kappa-symmetry of new
model. 
One of the problems appears already on this level. It is not easy to build
kappa-symmetric action of such a superstring. It means that even befor
quantizing one has to constract self-consistant classical model.
 By the other hand there are consistant
models
of superstring on group-manifold, where target space coordinates also can be
tensors, but problems appear in the attempt to generalize this
construction to cosets.
For the most recent review on strings with central charge
coordinates see \cite{LZnew} where null superstring with tensorial coordinates
was studied.
Here for the sake of completeness we just want to discuss
possible generalizations of the action \eq{na} for the case of strings and
membranes. As we saw before, the superparticle action with brane-charge
coordinates have nonusual feature: the first order formalism is simple and
well defined, but second-order one \eq{2-order-part} is troublesome and
very complicated. Here we argue that problem with models of strings with
brane-charge coordinates could have the same origin. The second order
formulations only were considered before, but here we would like to
examine the first order one using
Siegel's formulation of Green-Schwarz superstring \cite{Sstr}. This formulation
appeared useful for covariant quantization of superstring \cite{Berk1}.
It could be used to study first order formalism of superstring
where the generalized momenta for the string $P_i^\m$ is
introduced. Here $i =
0,1$ is
string
world-sheet coordinate index and $\m$ belongs to coordinates of target
space. 
 Because we want to realize superalgebra with brane charges on the level of
 superstring action, we introduce notion of extended superspace, exactly
as
before, by using generalized
 coordinates $X_{\a\b}, \t_\a$ and momenta $P_i^{\a\b}$.
Then generalization  of the action \eq{na} for the case of string is

\be
L = P^{\a\b i} \Pi_{\a\b i} + \frac{1}{2}e_{\a\b ij} P^{\a\c i}P_{\c}^{\b j}
 - i \e_{ij} \Pi_{i \a\b} \p_j \t^\a \t^\b ,
\la{Lag-str-cc}
\ee

where 

\be
\Pi_{\a\b i} = \p_i X_{\a\b} + i\t^{(\a} \p_i \t^{\b)}.
\la{Pistr-cc}
\ee

This action is invariant under global supersymmetry generated by the algebra
\eq{alg}. $e_{\a\b ij}$ is generalized vielbein that includes two
dimensional
metric $g_{ij}$ and following \cite{Sstr} $det(g) = 1$. In general case we have to
add this constraint in the Lagrangian but we just assumed that we already used some
symmetries of the action to fix $det(g) = 1$.
The model \eq{Lag-str-cc} could be reduced by separating
Lagrange multipliers $e_{\a\b}$ and  $g_{ij}$. It give origin to the
following
Lagrangian:

\be
L = P^{\a\b i} \Pi_{\a\b i} + \frac{1}{2}e_{\a\b} g_{ij} P^{\a\c i}P_{\c}^{\b j}
 - i \e_{ij} \Pi_{i \a\b} \p_j \t^\a \t^\b ,
\la{Lag-str-cc-1}
\ee

Important property of this model is that ordinary form of kappa-symmetry
should be modified (we denote new terms by ... )

\be
\d\t_\a = \k_\a + ... , \quad \d X^{\a\b} = i\d \t^\a \t^\b + ...,\quad
\d P_{\a\b}^i = -2i\e^{ij} (\k \p_i \t)_{\a\b} +... .
\la{kappa-s-cc}
\ee  

Unfortunately it is not completely clear how this kappa-symmetry transformation
should be modified to close on-shell.

The action \eq{Lag-str-cc-1} could be generalized for the case of p-branes.
In the simplest formulation one cane get rid of $e_{\a\b}$ by choosing
$e_{\a\b} = e_1 C_{\a\b}$. In this case Lagrangian for supermembrane is taking the
form:

\be
L = P^{\a\b i} \Pi_{\a\b i} + \frac{1}{e} e_{ij} P^{\a\b i}P_{\b}^{\a j}
  + e + L_{WZ} ,
\la{Lag-mem-cc-2}
\ee

where $L_{WZ}$ is Wess-Zumino term and for membrane $i= 0,1,2$.
It is also straightforward to rewrite action \eq{Lag-mem-cc-2} in the most general
form following \eq{Lag-str-cc}.
 Here we presented discussion on possible extension of the model given by \eq{na}
 to the cases of superstrings and superbranes. There are still a lot of questions
 about later models, because the complete symmetries of the actions
 \eq{Lag-str-cc}, \eq{Lag-str-cc-1} and  \eq{Lag-mem-cc-2} should be studied as well
 as quantization of the string models.

In this section we considered different interpretation of
p-brane charges not as integrals over world-volume of brane current but
 as coordinates of extended
superspace. Here we presented only particle realization of the
superalgebra with
brane charges. In the next chapter we will consider the quantization of 
the Universal model and give a field theory, that realizes superalgebra
with 
brane charges. 
In that case extra coordinates, coming from brane charges could be 
interpreted as pure twistorial ones and correspond to spin degrees of
freedom.

In this part of the chapter we discussed realization 
of the super Poincar\'e algebra
with brane charges in terms of super 0-brane models. We considered only
classical models and
discussed their symmetries and properties. In the next sections we will
consider 
different super 0-brane models realizing superalgebras beyond eleven
dimensional.

\bigskip




\section{\bf  Superparticles in $D>11$ }


The possibility of a super $p$-brane in $(10,2)$-dimensions was
conjectured long ago \cite{duff}, in the context of a generalized
brane-scan. More recently, there have been indications for the existence
of a (10,2) dimensional structure in $M$-theory \cite{vafa,km1,B1}.
Motivated by these considerations, super Yang-Mills equations of motion
in (10,2) dimensions were constructed in \cite{ns}. This result has been
recently generalized to describe the equations of motion of supergravity
in (10,2) dimensions \cite{n1}. Previously, possible existence of hidden
symmetries descending from (11,2) dimensions was pointed out \cite{b2}.
Recently \cite{B1}, it has been suggested that there may be a $(11,3)$
dimensional structure in the master theory, and even the possibility of
a $(12,4)$ dimensional structure has been speculated in \cite{km2}.
There is an
extension of the work presented in \cite{ns} to higher dimensions, and
it was found that the construction of \cite{ns} generalizes naturally to
$(11,3)$ dimensions. An extension beyond $(11,3)$ dimensions ran into an
obstacle \cite{es}, which has been removed in \cite{hn2}, where super
Yang- Mills equations have been constructed in $(8+n,n)$ dimensions, for
any $n\ge 1$. Here we follow \cite{RS}, \cite{RS1}.

The symmetry algebras realized in the field theoretic models just
mentioned are \cite{b2,ns,es,b5}: 
\bea
(10,2):\qquad \{ Q_\a, Q_\b\} &=& 
	(\c^{\m\n})_{\a\b}~P_\m~n_\n\ ,\la{an2}\\ 
(11,3):\qquad \{Q_\a, Q_\b\} &=& 
	(\c^{\m\n\r})_{\a\b}~P_\m~n_\n ~m_\r \ , \la{an3}
\eea 
where $n_\m$ and $m_\m$ are mutually orthogonal constant null vectors.
These break the $(10,2)$ or $(11,3)$ dimensional covariance. In order to
maintain this covariance, it is natural to replace the null vectors by
momentum generators \cite{b2} (see also \cite{B1}), thereby obtaining
\footnote
{
In $(8+n,n)$ dimensions, the full set of generators occurring on the
right hand side of $\{ Q_\a, Q_\b\}$ are $p$-form generators  with
$p=n_0, n_0+4,...,n+4$, where $n_0= n~{\rm mod}~4$. For example, in 
$(17,9)$ dimensions, there are $p$-form generators with $p=1,5,9,13$.
However, actions of the type considered here for $n$-particle systems 
naturally select the $n$\,th rank generator.
}
\bea
(10,2):\qquad \{ Q_\a, Q_\b\} &=& 
	(\c^{\m\n})_{\a\b}~P_{1\m}~P_{2\n}\ ,\la{alg2}\\ 
(11,3):\qquad \{Q_\a, Q_\b\} &=& 
	(\c^{\m\n\r})_{\a\b}~P_{1\m}~P_{2\n}~P_{3\r} \ . \la{alg3}
\eea 
The algebras \eq{an2} and \eq{alg2} first made their appearances in
\cite{b2}, and \eq{an3} and \eq{alg3} in \cite{es,b5}. In particular,
\eq{alg2} has been put to use in \cite{bars1,b2} in the context of higher
dimensional unification of duality symmetries; in \cite{b3} where four
dimensional bi-local field theoretic realizations are given and
interesting physical consequences such as family unification are
suggested; and in \cite{b6}, where a two-particle realization in
$(10,2)$ dimensions, in the purely bosonic context, was given. 

The purpose of this section is to present a supersymmetric extension of
the bosonic two-particle model of \cite{bars1}, and to extend further
these results to (11,3) dimensions, where a three-particle model arises.
We will construct multi-superparticle actions in which the algebras
\eq{alg2} and \eq{alg3} are realized. In doing so, we will find that the
multi-superparticle system has new bosonic local symmetries that generalize
the usual reparametrization \cite{B1}, and new fermionic local
symmetries that generalize the usual $\k$-symmetry of the single
superparticle. These symmetries will be shown to exist in
presence of super Yang-Mills background as well. 
 
These results can be viewed as preludes to the constructions of higher
superbranes in (10,2) and (11,3) dimensions. Since the latter should
admit superparticle limits, it is important to develop a better
understanding of the superparticle systems in this context.


\subsection{\bf Superparticles in (10,2) dimensions}



We consider two superparticles which propagate in their respective
superspaces with coordinates $X_i^\m (\tau_1,\tau_2)$ and
$\t_i^\a(\tau_1,\tau_2)$, with $i=1,2$, $\m=0,1,...,11$ and
$\a=1,...,32$. Working in first order formalism, we also introduce the
momentum variables $P^i_\m(\tau_1,\tau_2)$

The superalgebra \eq{alg2} can be realized in terms of supercharges
\be 
Q_\a=Q_{1\a}+Q_{2\a}\ , 
\ee
with $Q_i(\tau_1,\tau_2)$ defined as 
\footnote{
The spinors are Majorana-Weyl, their indices are chirally
projected, the charge conjugation matrix $C$ is suppressed in
$(\c^{\m_1\cdots \m_p}C)_{\a\b}$, which are symmetric for $p=2,3$ in
$(10,2)$ dimensions, and for $p=3,4$ in $(11,3)$ dimensions. This
symmetry property alternates for $p~{\rm mod}~2$. 
}

\be
Q_{i\a}=\p_{i\a}+\ft14~\c^{\m\n}_{\a\b}~\t_i^\b~P_{1\m}~P_{2\n} \ , 
\quad i=1,2 \ .
\la{q2} 
\ee 

The spinorial derivative is defined as $\p_{i\a}=\p/\p\t_i^\a$, acting
from the right. The transformations generated by the supercharges $Q_\a$
are

\bea
\d_\e X_i^\m = \ft14~\eb\c^{\m\n}(\t_1+\t_2)~\ve_{ij} P_{j\n}\ , 
\qquad  \d_\e \t_i =\e\ , \qquad \d_\e P_i^\m = 0\ , \la{s1} 
\eea

where $\eb\c^{\m\n}\t $ stands for $\e^\a\c_{\a\b}\t^\b$, and $\ve_{ij}$
is the constant Levi-Civita symbol with $\ve_{12}=1$.
Next, it is convenient to define the line element

\be
\Pi_i^\m=(\p_1 +\p_2) X_i^\m -\ft{1}{8} \tb_k \c^{\m\n}(\p_1+\p_2)~
\t_k~\ve_{ij} P_{j\n}\ .  \la{p1}
\ee

While this is not supersymmetric by itself, its product with $P^i_\m$ is
supersymmetric up to a total derivative term, and therefore it is a
convenient building block for an action. The fact that the sum of two
times occur in the line element is a consequence of maintaining
supersymmetry (in the sense just stated) {\it and} the fact that all
field depend on $\tau_1$ and $\tau_2$ (see the end of this section for a
discussion of a restricted time dependence, and its consequences). 

Introducing the symmetric Lagrange multipliers $A_{ij}(\tau_1,\tau_2)$
which are inert under supersymmetry, 

\be
\d_\e A_{ij}=0\ , \la{sa}
\ee
we consider the following action for a two-superparticle system in
$(10,2)$
dimensions
\be
I= \int d\tau_1 d\tau_2 \left(
	P_i^\m \Pi_{i\m} -\ft12 A_{ij} P_i^\m P_{j\m}\right)\ . \la{act1}
\ee  
The action \eq{act1} has a number of interesting symmetries. To begin
with, it is invariant under the target space rigid supersymmetry
transformations \eq{s1}, up to a total derivative term that has been
discarded. Furthermore, it has the local bosonic symmetry 
\be
\d_\L A_{ij} = (\p_1+\p_2)~\L_{ij}\ ,\qquad
\d_\L X_i^\m = \L_{ij} P_j^\m \ , \qquad
\d_\L P_i^\m = 0 \ ,\qquad
\d_\L \t = 0 \ , \la{b1}
\ee
where the transformation parameters have the time dependence
$\L_{ij}(\tau_1,\tau_2)$. Here too, a total derivative term, which has
the form $(\p_1+\p_2) \left(\ft12 \L_{ij} P^i_\m P^{j\m} \right)$, has
been dropped. The diagonal part of these transformations are the usual
reparametrizations, combined with a trivial symmetry of the action.
 The off-diagonal part of the symmetry are gauge symmetries
which are the first order form of those which are in the bosonic
two-particle model of \cite{b6}. Together with the reparametrization
symmetries, they allow us to eliminate the correct amount of degrees of
freedom to yield 8 bosonic physical degrees of freedom for each
particle.

The action \eq{act1} has also local fermionic symmetries which
generalize the usual $\k$-symmetries. Let us denote the $j$th symmetry
of the $i$th particle by $\k_{ij}(\tau_1,\tau_2)$. One finds that the
action \eq{act1} is invariant under the following transformations
\bea
\d_\k \t_i &=& \c^\m \k_{ij} P_{j\m}\ ,\nn\\
\d_\k X_i^\m &=& \ft{1}{4} \left(\tb_k\c^{\m\n}\d_\k \t_k\right)\ve_{ij} 
P_{j\n}\ ,  
\nn\\
\d_\k P_i^\m &=& 0\ ,  
\nn\\
\d_\k A_{ij} &=& \ft12 \kb_{ki}\c^\m (\p_1+\p_2) \t_k~
\ve_{\ell j} P^\ell_\m + (i\lra j)\ .
\la{k1}
\eea
In showing the $\k$-symmetry of the action, it is useful to note the
lemma
\be
P^i_\m (\d_\k \Pi_i^\m) = (\p_1+\p_2) \left( \ft18 \tb_k \c^{\m\n} \d_\k
\t_k \ve_{ij} P^i_\m P^j_\n \right) 
-\ft14 (\d_\k\tb) \c^{\m\n} (\p_1+\p_2) \t_k \ve_{ij} P^i_\m P^j_\n  \ . 
\la{lemma1}
\ee
The diagonal part of the $\k_{ij}$-transformations are the $\k$-symmetry
transformations that resemble the ones for the usual superparticle. The
off-diagonal $\k$-transformations are their generalizations for a
two-superparticle system. Just as the off-diagonal $\L_{ij}$
transformations are needed to obtain 8 bosonic degrees of freedom for
each particle, the off-diagonal $\kappa_{ij}$ symmetries are needed to
obtain 8 fermionic degrees of freedom for each particle. The commutator
of two $\k$-transformations closes on-shell onto the
$\L$-transformations
\be
[\d_{\k_{(1)}},\d_{\k_{(2)}}]= \d_{\L_{(12)}} \ , \la{close}
\ee
where the composite gauge transformation parameter is
\be
\L_{(12){ij}} = \ft12 \kb_{(2)ki}\c^{\m\n} \k_{(1)kj} P_{1\m}P_{2\n}
	 + (i \leftrightarrow j) \ . \la{c1}
\ee
It is clear that the remaining part of the algebra
is $~[\d_\k,\d_\L]=0~$ and $~[\d_{\L_1},\d_{\L_2}]=0~$.

The field equations that follow from the action \eq{act1} take the form 
\bea
&&
P_i^\m P_{j\m}= 0\ ,  \la{e1-new}\\
&&
\c^{\m\n}  P_{1\m} P_{2\n} (\p_1+\p_2) \t_i= 0\ , \la{e2}\\
&&
(\p_1+\p_2) X_i^\m = \left( A_{ij}\eta^{\m\n} 
+\ft14 \ve_{ij} \tb_k \c^{\m\n} (\p_1+\p_2) \t_k \right) P^j_\n\ ,
\la{e3}\\
&&
(\p_1+\p_2) P_{i\m} = 0\ . \la{e4-new}
\eea 
While the derivatives occur only in the combination $(\p_1+\p_2)$, the
fields can depend both on $\tau_+$ and $\tau_-$ defined by
$\tau_\pm \equiv \tau_1\pm \tau_2$. It is possible, for example, to restrict
the proper time dependences as follows
\be
X_i^\m (\tau_i)\ , \qq P_i^\m (\tau_i)\ ,\qq \t_i (\tau_i)\ . \la{rt}
\ee
The action still is given by \eq{act1}, with the line element now 
taking the form
\be
\Pi_i^\m=\p_i X_i^\m -\ft{1}{4} \tb_i \c^{\m\n} \p_i~
\t_i~\ve_{ij} P_{j\n}\ .  \la{po}
\ee
It is understood that the free indices of the left hand side are not not
be summed over on the right hand side. The bosonic and fermionic
symmetries discussed earlier retain their forms. In particular, the
global supersymmetry transformations \eq{s1} remain the same, and one
can express the bosonic symmetries \eq{b1} as
\be
\d_\L A_{ij} = \p_i\L_{ij} + \p_j \L_{ji}-\d_{ij}\p_i\L_{ij}\ ,
\qquad
\d_\L X_i^\m = \L_{ij} P_j^\m\ ,\qquad \d_\L P_i^\m =0\ ,
\qquad
\d_\L \t = 0\ , \la{gto}
\ee
where the parameters $\L_{ij}$ depend on $\tau_1$ and $\tau_2$.
Restricting the proper time dependence of the $\k$-symmetry parameter as
$\k_{ij}(\tau_i)$, the fermionic symmetries \eq{k1} can be simplified to
take the form
\bea
\d \t_i &=& \c^\m \k_{ij} P_{j\m}\ ,
\nn\\
\d X_i^\m &=& \ft14 \left(\tb_i\c^{\m\n}\d\t_i\right)\ve_{ij} P_{j\n}\ , 
\nn\\
\d P_i^\m &=& 0\ , 
\nn\\
\d A_{ij} &=& \ft12\left( \kb_{kj}\c^\m \p_k\t_k
+\kb_{ij}\c^\m \p_i\t_i\right)\ve_{ki} P_{k\m} + (i\lra j) \ , \la{kappa1}
\eea
which close on-shell as in \eq{close}, and the composite parameter now
takes the form 
\be
\L_{(12){ij}} = \ft12 \kb_{(2)ii}\c^{\m\n} \k_{(1)ij}P_{1\m}P_{2\n}
	 -(1 \lra 2) \ . \la{close1}
\ee
Similarly, the field equations become
\bea	
&& P_i^\m P_{j\m} = 0 \ , 
\qq 
\p_i P_{i\m} = 0 \ , 
\qq
\c^{\m\n} P_{1\m} P_{2\n} \p_i \t_i= 0 \ , \\
&&
\p_i X_i^\m =\left( A_{ij}\eta^{\m\n} 
+\ft12 \ve_{ij} \tb_i \c^{\m\n} \p_i\t_i \right) P_{j\n} \ .
\eea
Note that the $\k$-symmetry transformation of, say $X_1^\m $, maps a
function of $\tau_1$ to a function of $\tau_1$ {\it and} $\tau_2$. While
this may seem somewhat unusual, it does not present any inconsistency,
and in particular, there is no need to take the momenta to be constants.
The important point to bear in mind is that the symmetry transformations
close and that they are consistently embedded in a larger set of
transformations that map functions of $(\tau_1,\tau_2)$ to each other. 


\subsection{\bf  Putting the second particle on-shell}


In order to obtain an action for the first particle propagating in the
background of the second particle, we will follow the following
procedure. Recall that $\tau_\pm \equiv \tau_1\pm \tau_2$. Let us also
define $\t_\pm \equiv \t_1\pm \t_2$. We use the $ X_2, A_{22}, \t_- $
equations of motion in the action, thereby putting the second particle
on-shell. However, the remaining fields still have $\tau_-$ dependence.
In analogy with Kaluza-Klein reduction, we then set $\p_-=0$, and
integrating the action over $\tau_-$ to obtain:
\be
I=\int d\tau \left[ P_\m \left(\Pi^\m -A n^\m \right)
-\ft12 e P^\m P_\m \right]\ . \la{action2}
\ee
where the label $1$ has been suppressed throughout, the $\tau_-$
interval is normalized to $1$ and 
\bea
&& 
P_2^\m \equiv n^\m \ , \nn\\
&&
A_{11}\equiv e\ , \quad A_{12}\equiv A \ , \nn\\
&&
\Pi^\m = \p_\tau X^\m-\ft14\tb \c^{\m\n}\p_\tau \t~n_\n\ , \la{np}
\eea
where we have defined $\tau_+\equiv \tau$ and $\t_+ \equiv \t/\sqrt{2}$.
The vector $n^\m$ is constant and null as a consequence of $X_2$ and
$A_{22}$ equation of motion, and the fact that we have set $\p_- =0$. 

The action \eq{action2} is invariant under the local bosonic
transformations
\be
\d e = \p_\tau \xi\ ,\qquad \d A = \p_\tau \L\ , \qquad 
\d X^\m = \xi P^\m +\L n^\m \ ,\qquad \d P^\m=0\ ,
\qquad \d \t =0\ ,
\la{bos-trans-12}
\ee
obtained from \eq{b1} by setting $\p_-=0$ and using the notation $\xi
\equiv \L_{11}$ and $\L \equiv \L_{12}$. The action \eq{action2} is also
invariant under the global supersymmetry transformations \eq{s1}
\be
\d_\e X^\m = \ft14~\eb\c^{\m\n}\t~n_\n \ , 
\qquad  \d_\e\t =\e\ , \qquad \d_\e P^\m = 0\ , \qquad \d_\e e=0\ ,
\qquad \d_\e A=0\ ,
\la{s2} 
\ee
(we have rescaled $\e\ra \e/\sqrt{2}$ for convenience) and invariant
under the local fermionic $\k$ and $\eta$ transformations 
\bea
\d\t &=& \c^\m P_{\m}\kappa + \c^\m n_\m \eta\ ,
\nn\\
\d X^\m &=& \ft14 \tb\c^{\m\n} n_\n \left(\d_\k\t+\d_\eta \t\right)\ ,
\nn\\
\d P^\m &=& 0\ , 
\nn\\
\d e &=& -\kb\c^\m\p_\tau\t~n_\m\ ,
\nn \\
\d A &=& \ft12 \kb \c^\m P_\m \p_\tau \t 
	-\ft12 \etab \c^\m n_\m \p_\tau \t\ , 
\la{k2}
\eea
obtained from \eq{k1} by setting $\p_-=0$, using the field equations of
the second particle, setting $\k_{-i}=0$ , and using the notation
$\k_{+1}\equiv \sqrt{2} \k$ and $\k_{+2}\equiv \sqrt{2}\eta $. The
parameter $\k_{-i}$ has been set equal to zero, as it is associated with
the transformations of $\t_-$ that has been put on-shell, and
which has consequently dropped out in the action.

An alternative way to arrive at the same results is to start from the
restricted model described above by putting the
second particle on-shell, and this time integrating over $\tau_2$.


\subsection{\bf Introducing super Yang-Mills background}


First we start from description of super Yang-Mills in $(10,2)$ dimensions.
Yang-Mills also could be thought as some particular realization of superalgebra.
Moreover, using methods of first chapter it is possible to investigate
representations of superalgebras in $D>11$ and thus the models of SYM in
dimensions beyond eleven could be useful for understanding of the structure of
theory of everything.

Here we will start from the following superalgebra

\be
\{ Q_\a, Q_\b \} = (\C^{mn})_{\a\b} P_n n_m  \ ,
\la{alg12}
\ee

where $n_m$ is a constant null vector. We see that it is the same algebra with 
constant vector that was discussed in the beginning of this section.

The Yang-Mills equations of motion are given by \cite{ns}

\bea
&&\c^\m D_\m \l=0\ , \la{f1}\\
&&D^\m F_{\m[\r} n_{\s]} - \ft12 \bar\l\c_{\r\s} \l = 0 \ ,
\la{f12}
\eea

where the fields are Lie algebra valued and in the adjoint
representation of the Yang-Mills gauge group, and $D_\m
\l=\p_\m\l+[A_\m,\l]$. Due to the symmetry of $\c^{\m\n}C^{-1}$, the
last term in \eq{f12} involves a commutator of the Lie algebra generators. 
In addition to the manifest Yang-Mills gauge symmetry, these equations are 
invariant under the supersymmetry transformations \cite{ns}

\bea
\d_Q A_\m &=& \bar\e\c_\m\l \ , \la{t1} \\
\d_Q  \l &=& -\ft14 \c^{\m\n\r} \e F_{\m\n} n_\r \ , \la{t12}
\eea

and the extra bosonic local gauge transformation \cite{ns}

\be
\d_\Omega A_\m = \Omega~n_\m \ , \quad\quad \d_\Omega \l = 0\ , \la{to12}\la{ot12}
\ee

provided that the following conditions hold \cite{ns}

\bea
 n^\m D_\m \l &=& 0  \ ,   \la{c112}\\
 n^\m \c_\m \l &=&0 \ ,    \la{c212}\\
 n^\m F_{\m\n} &=& 0 \ ,   \la{c312}\\
 n^\m n_\m &=& 0 \ ,       \la{c412}\\
 n^\m D_\m \Omega &=& 0 \ ,\la{c512}
\eea

One can check that the field equations as well as the constraints are
invariant under supersymmetry as well as extra gauge transformations.

Finally, we use tha fact that that the commutator of two supersymmetry
transformations closes on shell, and yields a generalized translation,
the usual Yang-Mills gauge transformation and an extra gauge
transformation with parameters $\xi^\mu$, $\Lambda$, $\Omega$,
respectively, as follows:

\be
[ \d_Q(\e_1), \d_Q(\e_2) ] = \d_\xi +\d_\Lambda + \d_\Omega\ ,
\la{closure12}
\ee

where the composite parameters are given by \cite{ns}

\bea
\xi^\m &=&  \bar\e_2 \c^{\m\n}\e_1\ n_\n\ , \la{xi12}\\
\Lambda &=& -\xi^\m\ A_\m\ , \la{l12}\\
\Omega &=& \ft12 \bar\e_2 \c^{\m\n}\e_1\ F_{\m\n}\ .  \la{o12}
\eea

Note that the global part of the algebra \eq{t1} and \eq{t12}
 is given by \eq{alg12}.

The closure of the supersymmetry algebra on the fermion requires the
constraints \eq{c112} and \eq{c212}, while the supersymmetry and $\Omega$-
symmetry of the field equations and constraints require the remaining
constraints as well \cite{ns}. A superspace formulation of this model,
as well as its null reductions to $(9,1)$ and (2,2) can be found in
\cite{ns}.


The coupling of Yang-Mills background is best described in the second
order formalism. First of all we see that particle model we discussed before
realized superalgebra without constant vector, on the other hand, in the case 
when one puts second particle on-shell, i.e. uses equations of motion of
second particle in the Lagrangian \eq{act1}.

Elimination of $P^\m$ in \eq{act1} gives  
\be
I_0 =  \ft12 \int d\tau~e^{-1} \Pi^\m \left( \Pi_\m -A n_\m\right)\ .
\la{ac1}
\ee  
    
The bosonic and fermionic symmetries of this action can be read off from
\eq{bos-trans-12}, \eq{s2} and \eq{k2} by making the substitution $P^\m \rightarrow e^{-1}
(\Pi^\m-A n^\m)$. To couple super Yang-Mills background to this
system, we introduce the fermionic variables $\psi^r,\ r=1,...,32$,
assuming that the gauge group is $SO(32)$. The Yang-Mills coupling can
then be introduced as 
\be
I_1= \int d\tau~\psi^r \p_\tau \psi^s \p_\tau Z^M A_M^{rs} ,\la{ac12}
\ee
where $Z^M$ are the coordinates of the $(10,2|32)$ superspace, and $A_M^{rs}$
is a vector superfield in that superspace. 

The torsion super two-form $T^A=dE^A$ can be read from the superalgebra \eq{alg12}:

\be
T^c = e^\a \wedge e^\b\, (\c^{cd})_{\a\b}\, n_d\ , \quad\quad  T^\a =0\ ,
\la{tc12}
\ee

where the basis one-forms defined as $e^A=d Z^M E_M{}^A$ satisfy

\be
d e^c= e^\a \wedge e^\b\ (\c^{cd})_{\a\b}\ n_d\ ,\quad\quad d
e^\a = 0\ , \la{se12}
\ee

and $a,b,c,... $ are the (10,2) dimensional tangent space indices.

Using these equations, a fairly standard calculation \cite{w,bst} shows that the
total action $I=I_0+I_1$ is invariant under the fermionic gauge
transformations provided that the Yang-Mills super two-form is given by
\cite{ns}

\be
F=e^\a \wedge e^b \left[\,  n_b \chi_\a - 2(\c_b\l)_\a\, \right]
+ \ft12 e^a \wedge e^b\
F_{b a}\ , \la{fc12}
\ee

where we have introduced the chiral spinor superfield $\chi_\a$ and the
anti-chiral spinor superfield $\l$, and that the transformation rules
for $e$ and $A$ pick up the extra contributions 

\bea
\d_{\rm extra} e &=&  -4e \psi^r\p_\tau \psi^s (\kb \l_{rs})\ , \nn\\
\d_{\rm extra} A &=& \psi^r\p_\tau \psi^s \kb\left( 2\l_{rs}+(\Pi^a-An^a)
	\right)\c_a\chi_{rs}
	+ e \psi^r\p_\tau \psi^s \etab\left( 4\l_{rs}-
	\c_an_a\chi_{rs}\right)\ ,\nn\\
\d_{\rm extra} \psi^r &=& - \d \t^\a~A_\a^{rs} \psi^s\ .
\eea

In addition, the spinor $\l$ must satisfy the condition $ n^a\c_a \l
=0$. One can show that $F$ given in \eq{fc12} satisfies the
Bianchi identity $DF=0$, and the constraints on $F$ implied by \eq{fc12}
lead to the field equations of the super Yang-Mills system in (10,2)
dimensions \cite{ns}.


\subsection{\bf Superparticles in (11,3) dimensions}




In this subsection, we consider three superparticles propagating in
$(11,3)$ dimensional spacetime \cite{RS}, \cite{RSS},
and take their superspace coordinates to
be $X_i^\m (\vt), \t_i^\a(\vt)$ and momenta $P_i^\m (\vt)\ (i=1,2,3)$,
where $\vt=(\tau_1,\tau_2,\tau_3)$. Following the same reasoning
as for the case of tvelve dimensions we consider the following action

\be
I= \int d\tau_1 d\tau_2 d\tau_3~\left(
	P_i^\m \Pi_{i\m} -\ft12 A_{ij} P_i^\m P_{j\m}\right)\ , 
\la{act13}
\ee 
 
where

\be
\Pi_i^\m= \pth X_i^\m -\ft{1}{36} \tb_k \c^{\m\n\r}\pth \t_k~\ve_{ijk} 
P_{j\n}P_{k\r}\ .
\la{pi13}
\ee

The action is invariant under the local bosonic transformations

\be
\d_\L A_{ij} = (\p_1+\p_2+\p_3)~\L_{ij}\ ,\qquad
\d_\L X_i^\m = \L_{ij} P_j^\m \ , \qquad
\d_\L P_i^\m = 0 \ ,\qquad 
\d_\L \t = 0 \ . \la{b13}
\ee

The action is also invariant (modulo discarded total derivative terms)
under the global supersymmetry transformations

\bea
\d_\e X_i^\m &=& \ft1{12}~\eb\c^{\m\n\r} (\t_1+\t_2+\t_3)~\ve_{ijk}
P_{j\n}P_{k\r}\ , \nn\\ 
\quad  \d_\e\t_i &=&\e\ , \quad \d_\e P_i^\m \quad = \quad 0\ , 
\quad  \d_\e A_{ij} \quad = \quad 0\ ,
\la{susy13} 
\eea

and local fermionic transformations

\bea
\d_\k \t_i &=& \c^\m \k_{ij} P_{j\m}\ ,\nn\\
\d_\k X_i^\m &=& \ft1{12} \left(\tb_k \c^{\m\n\r}\d\t_k \right)\ve_{imn} 
	P^m_\n P^n_\r \ ,\nn\\
\d_\k P_i^\m &=& 0\ , \nn\\
\d_\k A_{ij} &=& \ft16 \kb_{ki}\c^{\m\n} \pth \t_k
	      \ve_{jmn} P^m_\m P^n_\n+ (i\lra j)\ .
\la{k13} 
\eea

The algebra closes on-shell as in \eq{c1}, with the composite gauge
parameter now given by 

\be
\L_{(12){ij}} = \ft13 \kb_{(2)ki}\c^{\m\n\r} \k_{(1)kj}
	        P_{1\m}P_{2\n}P_{3\r} +(i \lra j)\ . 
\la{c12}
\ee

The remaining part of the algebra is $~[\d_\k ,\d_\L]=0~$ and
$~[\d_{\L_1},\d_{\L_2}]=0~$. The equations of motion are similar to
\eq{e1-new}-\eq{e4-new}. In fact, all the formulae of this section are very
similar to those for two-superparticles and their
$n$ superparticle extension is straightforward.




To obtain an action which describes the propagation of the first
particle in the background of the other two particles, we follow the
steps described for the case of twelve dimensions. 
Let $\tau\equiv \tau_1+\tau_2+\tau_3$,
and $\t\equiv \t_1+\t_2+\t_3$, and denote the orthogonal
combinations by $\tau_\pm$ and $\t_\pm$. Using the equations of motion
for $\t_\pm, X_i, A_{ii}\ (i=2,3)$  and restricting the proper time
dependence of fields by setting $\p_\pm=0$, we obtain the action 

\be
I=\int d\tau \left[ -\ft12 e P^\m P_\m 
+ P_\m \left(\Pi^\m -A n^\m - B m^\m \right)\right]\ , \la{in13}
\ee

where $A\equiv A_{12}$ and $B\equiv A_{13}$ and

\be
\Pi^\m = \p_\tau X^\m-\ft16\tb \c^{\m\n\r}\p_\tau \t~n_\n m_\r\ .
\la{pin13}
\ee

The vectors $n^\m$ and $m_\m$ are mutually orthogonal constant null
vectors, as a consequence of $A_{ii}$ and $X_i^\m$ equations of motion
for $i=2,3$ resulting from the original action \eq{act13}, and having set
$\p_\pm =0$. The action \eq{in13} is invariant under the bosonic
transformations

\be
\d e = \p_\tau \xi\ ,\qquad \d A = \p_\tau \L\ , \qquad 
	\d B = \p_\tau \Sigma\ , \qquad
\d X^\m = \xi P^\m +\L n^\m + \Sigma m^\m \ , \la{gt13}
\ee

where $\xi\equiv \L_{11}$, $\L \equiv \L_{12}$ and $\Sigma\equiv \L_{13}$.
The action is also invariant under the global supersymmetry
transformations

\be
\d_\e X^\m = \ft1{12}~\eb\c^{\m\n\r} \t~n_\n m_\r \ , 
\qquad  \d_\e \t =\e\ , \qquad \d_\e P^\m = 0\ , 
\qquad \d_e A=0\ , \qquad \d_e B=0\ , \la{isusy13} 
\ee

and local fermionic $\k$, $\eta$ and $\omega$ transformations 

\bea
\d\t &=& \c^\m P_{\m}\kappa + \c^\m n_\m \eta
	+ \c^\m m_\m \omega\ , \nn\\
\d X^\m &=& \ft16 \tb\c^{\m\n\r} n_\n m_\r \left(\d_\kappa \t+\d_\eta\t 
	+\d_\omega \t\right)\ , \nn\\
\d P^\m &=& 0\ , \nn\\
\d e &=& -\ft23 \kb \c^{\m\n} \p_\tau \t n_\m m_\n\ , \nn \\
\d A &=&  -\ft13 \kb \c^{\m\n} \p_\tau \t m_\m P_\n 
	-\ft13 \etab \c^{\m\n} \p_\tau \t n_\m m_\n\ , \nn \\
\d B &=&  -\ft13 \kb \c^{\m\n} \p_\tau \t P_\m n_\n 
	-\ft13 \kb \c^{\m\n} \p_\tau \t n_\m m_\n
	-\ft13 {\bar \omega}\c^{\m\n} \p_\tau \t n_\m m_\n \ , \la{tr13} 
\eea

where $\k, \eta,\w$ are equivalent to $\k_{+i} (i=1,2,3)$ upto a
constant rescaling, and the irrelevant parameters $\k_{\pm i}$ have been
set equal to zero.

To describe an interaction between super Yang-Mills and 0-brane in
$(11,3)$ Dimensions we will start from discussion on realization of
$D=(11,3)$ superalgebra in terms of super Yang-Mills \cite{es}.
As for the case of twelve dimensions the super Yang-Mills is the realization of
the superalgebra not with constant vector but rather with constant tensor.

We begin by introducing the momentum generator to the superalgebra ,
by making use of a constant tensor $v_{\m\n}$ as follows:
\be 
\{Q_\a, Q_\b\} = (\c^{\m\n\r})_{\a\b}~P_\m~v_{\n\r} \ . 
\la{alg13}  
\ee

The next step is to postulate the supersymmetry transformation rules
which make use of $v_{\m\n}$. The strategy is then to obtain the field
equations, and any additional constraints by demanding the closure of
these transformation rules. At the end the (extra) gauge and
supersymmetry of all the resulting equations must be established. In
what follows, we will first present the results that emerge out of this
procedure. 

The super Yang-Mills equations take the form

\bea
&&\c^\m D_\m \l=0\ , \la{nf113}\\
&&D^\s F_{\s[\m} v_{\n\r]} + \ft1{12}\bar\l \c_{\m\n\r} \l = 0 \ .
\la{nf213}
\eea

In addition to the manifest Yang-Mills gauge symmetry, these equations
are invariant under the supersymmetry transformations

\bea
\d_Q A_\m &=& \bar\e\c_\m\l \ , \la{nt1} \\
\d_Q  \l &=& -\ft14 \c^{\m\n\r\s} \e F_{\m\n} v_{\r\s} \ , 
\la{nt113}
\eea

and the extra bosonic local gauge transformation

\be
\d_\Omega A_\m = -v_{\m\n}~\Omega^\n \ ,\quad\quad 
\d_\Omega \l = 0\ , \la{not13}
\ee

provided that the following conditions hold:

\bea
 v_\m{}^\n D_\n \l &=& 0  \ , \la{nc113}\\
 v_{\m\n} \c^\n \l &=&0 \ , \la{nc213}\\
 v_\m{}^\n F_{\n\r} &=& 0 \ , \la{nc313}\\
 v_\m{}^\r v_{\r\n} &=& 0 \ , \la{nc13}\\
 v_{[\m\n} v_{\r\s]} &=& 0 \ , \la{nc413}\\
 v_\m{}^\r v_\n{}^\s D_\r \Omega_\s &=& 0  \ , \la{nc513}\\
 v^{\m\n} D_\m \Omega_\n &=& 0\ . \la{nc613}
\eea

The commutator of two supersymmetry transformations closes on shell, and
yields a generalized translation, the usual Yang-Mills gauge
transformation and an extra gauge transformation with parameters
$\xi^\mu$, $\Lambda$, $\Omega^\m$, respectively, as follows:

\be
[ \d_Q(\e_1), \d_Q(\e_2) ] = \d_\xi +\d_\Lambda + \d_\Omega\ ,
\la{nclosure13}
\ee

where the composite parameters are given by

\bea
\xi^\m &=&  \bar\e_2 \c^{\m\n\r}\e_1\ v_{\n\r}\ , \la{nxi13}\\
\Lambda &=& -\xi^\m\ A_\m\ , \la{nl13}\\
\Omega^\m &=& \ft12 \bar\e_2 \c^{\m\n\r}\e_1\ F_{\n\r}\ .  \la{no13}
\eea

The global part of the algebra \eq{nt113} indeed agrees with
\eq{alg13}. Note the symmetry between the parameters $\xi^\m$ and
$\Omega^\m$. The former involves a contraction with $v_{\m\n}$, and the
latter one with $F_{\m\n}$.

The derivation of these results proceeds as follows. First, it is
easy to check that the closure on the gauge field requires an additional
local gauge transformation \eq{not13} with the composite parameter
\eq{no13}. Next, one checks the closure on the gauge fermion. In doing so, the following 
Fierz-rearrangement formula is useful:

\be
\e_{[1}\bar\e_{2]} = \ft1{64}
	\left( \ft1{3!}~\bar\e_2 \c^{\m\n\r}\e_1~\c_{\m\n\r} 
	+\ft1{7!2}~\bar\e_2\c^{\m_1\cdots \m_7}\e_1~\c_{\m_1\cdots\m_7}\right)\ ,
\la{f13}
\ee

where $\e_1$ and $\e_2$ are Majorana-Weyl spinors of the same chirality.
Using this formula, and after a little bit of algebra, one finds that:

 1.  The closure on the gauge fermion holds provided that the fermionic field
equation \eq{nf113}, along with the constraints \eq{nc113}
and \eq{nc213} are
satisfied.

 2. The supersymmetry of the constraint \eq{nc113} requires the constraint
\eq{nc313}, and a further variation of this constraint does not yield
new information.

 3. The supersymmetry of the constraint \eq{nc213} requires the further 
constraints \eq{nc413} and \eq{nc513}. 

 4. The equations of motion \eq{nf113} and \eq{nf213} transform into each other under 
supersymmetry. This can be shown with the use the constraints \eq{nc213} and
\eq{nc313}.

 5. Finally the invariance of the full system, i.e. equations of motion
and constraints, under the extra gauge transformation \eq{not13} has to be
verified. The invariance of the fermionic field equation \eq{nf113}, as
well as the constraints \eq{nc113} and \eq{nc213} do not impose new
conditions. However, the invariance of the constraint \eq{nc313} imposes
the condition \eq{nc513}, and the invariance of the bosonic field equation
\eq{nf213} imposes the condition \eq{nc613} on the parameter $\Omega^\m$.
Both of these conditions are gauge invariant.

In summary, equations \eq{alg13}-\eq{no13} form a consistent and closed
system of supersymmetric, Yang-Mills gauge and $\Omega$-gauge invariant
equations. The similarity of these equations to the corresponding ones in
$(10,2)$ dimensions is evident. One expects, therefore, a natural
reduction of these equations to those in $(10,2)$ dimensions. This will
indeed turn out to be the case.

The important next step is to establish that the constant tensor
$v_{\m\n}$ satisfying the conditions \eq{nc413} and \eq{nc13} actually
exists. The solution is

\be 
v_{\m\n}=m_{[\m} n_{\n]}\ , \la{vmn13} 
\ee 

where $m_\m$ and $n_\n$ are mutually orthogonal null vectors, i.e. they
satisfy 

\be
m_\m m^\m =0\ ,\quad\quad n_\m n^\m =0\ , \quad\quad m^\m n_\m = 0\ . \la{mn13} 
\ee 

Given the signature of the 14-dimensional spacetime, finding two
mutually orthogonal null vectors, of course, does not present a problem.
Indeed, this solution suggests a that an {\it ordinary} dimensional
reduction to $(9,1)$ dimensions should yield the usual super
Yang-Mills system. 

%
%

The simplest brane action in which the symmetry algebra 
\eq{vmn13},\eq{alg13} may be realized is that of a $0$-brane, i.e. a
superparticle \cite{RS,b7}
%
\footnote {We shall not the treat the realizations of the covariant
algebra \eq{alg3} here, but we refer the reader to \cite{RS,b8,RS1} for
their multi-superparticle realizations.}.
%
In \cite{RS}, an action for superparticle in the background of a second
and third superparticle was obtained essentially by putting the
background superparticles on-shell. The null vectors $m_\m$ and $n_\m$
satisfying \eq{mn13} are the constant momenta of the second and third
superparticles. We start from the action  \eq{in13}.

The action  as was mentioned before is invariant under the 
bosonic $\xi, \L$ and
$\Sigma$-transformations \eq{gt13}
and the global supersymmetry transformations \eq{isusy13}
The action is also invariant under the {\it local} fermionic $\k$,
$\eta$ and $\omega$ transformations \eq{tr13}.

Superparticle actions have also been constructed in \cite{b7}. Our
results essentially agree with each other. Some apparent differences in
fermionic symmetry transformations are presumably due to field
redefinitions and symmetry transformations proportional to the equations
of motion in \cite{RS1}.

We next describe the coupling of Yang-Mills background. To this end, it
is convenient to work in the second-order formalism. Elimination of
$P^\m$ in \eq{in13} gives as we saw before

\be
I_0=  \ft12 \int d\tau~e^{-1} \Pi^\m \left( \Pi_\m -A n_\m 
						-B m_\m\right)\ .
\la{ac113}
\ee      

The bosonic and fermionic symmetries of this action can be read off from
\eq{gt13} and \eq{tr13} by making the substitution $P^\m \rightarrow e^{-1}
(\Pi^\m-A n^\m-B m^\m)$. To couple super Yang-Mills background to this
system, we introduce the fermionic variables $\psi^r,\ r=1,...,32$,
assuming that the gauge group is $SO(32)$. The Yang-Mills coupling can
then be introduced as 

\be
I_1= \int d\tau~\psi^r \p_\tau \psi^s \p_\tau Z^M A_M^{rs} ,\la{ac213}
\ee

where $Z^M$ are the coordinates of the $(11,3|32)$ superspace, and $A_M^{rs}$
is a vector superfield in that superspace. 

The torsion super two-form $T^A=dE^A$ can be read from the superalgebra
\eq{alg13} :

\be
T^c = e^\a \wedge e^\b\, (\c^{cde})_{\a\b}\, v_{de}\ , \quad\quad  T^\a =0\ ,
\la{tc113}
\ee

where the basis one-forms defined as $e^A=d Z^M E_M{}^A$ satisfy

\be
d e^c= e^\a \wedge e^\b\ (\c^{cde})_{\a\b}\ v_{de}\ ,\quad\quad d
e^\a = 0\ , \la{se113}
\ee

and $a,b,c,... $ are the (11,3) dimensional tangent space indices.

Using these equations, a fairly standard calculation shows that the
total action $I=I_0+I_1$ is invariant under the fermionic gauge
transformations provided that the Yang-Mills super two-form is given by

\be
F=e^\a \wedge e^b \left[\,  (\c^c\chi)_\a v_{cb} - 2(\c_b\l)_\a\, \right]
+ \ft12 e^a \wedge e^b\
F_{b a}\ , \la{fc113}
\ee

and that the transformation rules for $e,A$ and $B$ pick up the extra
contributions. These contributions are determined by the requirement of
the cancellation of the terms proportional to $\Pi^2, \Pi\cdot n$ and
$\Pi\cdot m$, respectively, in the fermionic variation of the total
action. They are easy determine, but as their form is not particularly
illuminating we shall not give them here (see \cite{RS}, for the case of
superparticle in (10,2) dimensions). The fermionic field $\psi^r$ must
be assigned the fermionic transformation
rule

\be
\d  \psi^r =  -\d\t^\a A_\a^{rs} \psi^s\ .
\ee

In \eq{fc113}, we have introduced the chiral spinor superfield $\chi_\a$
and the anti-chiral spinor superfield $\l$. These fields and $F$ must
satisfy certain constraints so that the Bianchi identity $DF=0$ is
satisfied. These constraints are \cite{hn2}

\bea
&& 
n^a F_{a b} = 0 \ , \ m^a F_{a b} = 0 \ ,\la{s1113}\\
&& 
n^a \c_a \l = 0 \ , \  m^a \c_a \l = 0 \ , \la{s2113}\\
&&
n^a D_a \l = 0 \ , \  m^a D_a \l = 0 \ , \la{s3113}\\
&& 
D_\a\chi^\b= ( \c^{a b} )_\a{}^\b F_{a b} \ ,\la{s4113}\\
&&
D_\a \l^\b = \ft14(\c^{abcd})_\a{}^\b F_{ab}\ v_{cd}\ ,   \la{s5113}\\
&& 
D_\a F_{a b}  = 2\c^c (D_{[a}\chi)_\a v_{b]c}+4 \c_{[a} (D_{b]}\l)_\a 
\ . \la{s6113}
\eea

The above constraints are sufficient to solve the super Bianchi identity
$DF=0$, which can be shown \cite{hn2} to yield the the super Yang-Mills
system in (11,3) dimensions \cite{es}. Special $\c$-matrix identities
similar to those required for the existence of the usual super
$p$-branes are not needed here. In showing the vanishing of the term
proportional to $e^\a\wedge e^\b\wedge e^\c$, for example, it is
sufficient to do a Fierz rearrangement, and use the constraints \eq{s2113}.
One also find that the spinor superfield $\chi$ is unphysical, as it
drops out the equations of motion. 

The component form of the super Yang-Mills equations are \cite{es} and were 
descussed before.

In \cite{es}, an obstacle was encountered in extending the above
construction of super Yang-Mills system to higher than $14$ dimensions.
For example, in $(12,4)$ dimensions, while everything goes through in
much the same fashion as in $(11,3)$ dimensions, the supersymmetric
variation of the Dirac equation gave rise to a term proportional to
${\bar\l}\l$, which appeared to be non-vanishing, and hence problematic
in obtaining the correct Yang-Mills equation. However, as has been observed 
in \cite{n2}, this term actually vanishes due to the constraints
\eq{s2113}. In fact, super Yang-Mills systems in $(8+n,n)$ dimensions, for
any $n\ge 1$ have been constructed in \cite{n2}. More recently, an
action for $(10,2)$ dimensional super Yang-Mills, which can presumably
be generalized to higher dimensions, has also been found \cite{n3}.


We have presented simple action formulae for two- and
three-superparticle systems in $(10,2)$ and $(11,3)$ dimensions,
respectively. The symmetries of the action exhibit interesting
generalizations of reparametrization and $\k$-symmetries. An action
similar to \eq{ac1}-\eq{ac12} can easily be constructed for the $(11,3)$
dimensional superparticle and it implies the $(11,3)$ dimensional super
Yang-Mills equations \cite{RSS}. We also expect that the action
\eq{ac1}-\eq{ac12} can be generalized to obtain a heterotic string action
and possibly other superbrane actions. 

The $n$ particle models constructed here ($n=2,3)$ make use of $n$
fermionic coordinates $\t_i\ (i=1,...,n)$. The fact that they all
transform by the same constant parameter $\ve$ suggests that we can
identify them: $\t_i=\t$. It is also natural from the group theoretical
point of view to associate the coordinates $X_i^\m, \t_\a$ with the
generators $P_i^\m, Q_\a$. However, it is not necessary to do so, since
there are sufficiently many local fermionic symmetries to give $8$
physical fermionic degrees of freedom for each $\t_i$ .
Thus, it does not seem to be crucial to have one or many fermionic
variables. 

Another, and possibly more significant, feature of the models
constructed here is that they involve multi-times, in the sense that
fields depend on $\tau_i\ (i=1,...,n)$ over which the action is
integrated over. The derivatives occurring in the action come out to be
the sum of these times, which indicates that any (pseudo) rotational
symmetry among them is lost. One might therefore be tempted to declare
the fields to depend on the single time $\tau=\tau_1+\cdots \tau_n$.
Perhaps this is the sensible thing to do, however, we have kept the
multi-time dependence here, partially motivated by the fact that our
results may give a clue for the construction of an action that involves
an $(n,n)$ dimensional worldvolume. If an action can indeed be
constructed for $(n,n)$ brane, i.e. brane with an $(n,n)$ dimensional
worldvolume, one may envisage a `particle limit' in which the spatial
dependence is set equal to zero, yielding an $(n,0)$ dimensional
worldvolume. If the worldvolume diffeomorphisms are to be maintained,
then we may need to consider a term of the form $P^{ia}_\m \p_a X_i^\m$
in the action, where $a=1,...,n$ labels the $(n,0)$ worldvolume
coordinates. It is not clear, however, how supersymmetry and
$\k$-symmetry can be achieved in this setting.

Notwithstanding these open problems, one may proceed to view the
coordinates $(\tau_1,\sigma_1)$ and $(\tau_2,\sigma_2)$ as forming a
$(2,2)$ dimensional worldvolume embedded in $(10,2)$ dimensions, in the
case of a two-superstring system. Similarly, a three-superstring system
would form a $(3,3)$ dimensional worldvolume embedded in $(11,3)$
dimensions. 

It is of considerable interest to construct a string theory in $(3,3)$
dimensions which would provide a worldvolume for a $(3,3)$ superbrane
propagating in $(11,3)$ dimensions \cite{es}, thereby extending the
construction of \cite{vafa,km1} a step further \cite{RSS}. Indeed, a string
theory in $(n,n)$ (target space) dimensions has been recently constructed 
\cite{RSS}. This theory is based on an $N=2$ superconformal algebra for the
right-movers in $(n,n)$ dimensions, and an $N=1$ superconformal algebra
for the left-movers in (8+n,n) dimensions. The realizations of these
algebras contain suitable number of constant null vectors, which arise
in the expected manner in the algebra of supercharge vertex operators
\cite{km2}. Furthermore, the massless states are expected to assemble
themselves into super Yang-Mills multiplet in $(8+n,n)$ dimensions. Much
remains to be done towards a better understanding of this theory, and
its implications for the target space field equations that may exhibit
interesting geometrical structures that generalize the self-dual
Yang-Mills and gravity equations in $(2,2)$ dimensions \cite{hull3}. 

Further studies of supersymmetry in $D>11$ may also be motivated by the
fact that they contain both the type IIA and IIB supersymmetries of ten
dimensional strings \cite{b5,RSS}. Therefore, it would be interesting to
find a brane-theoretic realization of the $D>11$ symmetries that would
provide a unified framework for the description of all superstrings in
$(9,1)$ dimensions. 

\bigskip




\subsection{\bf  The model for $n$ superparticles in $(d-n,n)$ dimensions}


 Consider supercharges $Q_\a$ in $(d-n,n)$ dimensions where $n$ is the
number of time-like directions \cite{RS1}. Let the index $\a$ label the minimum
dimensional spinor of $SO(d-n,n)$. In the case of extended
supersymmetry, the index labelling the fundamental representation of the
automorphysm group is to be included. We will suppress that label for
simplicity in notation. Quite generally, we can contemplate the
superalgebra

\bea
\{ Q_\a, Q_\b \} &=& Z_{\a\b}\ , \nn\\
 \left[Z_{\a\b}, Z_{\c\d}\right] &=& 0 \ , 
\la{algnn}
\eea

where $Z_{\a\b}$ is a symmetric matrix which can be expanded in terms
of suitable $p$-form generators in any dimension by consulting Table 3
provided in Appendix A. The $Z$ generators have the obvious commutator
with the Lorentz generators, which are understood to be a part of the
superalgebra.

Next, consider dimensions in which the $n$\,th rank $\c$-matrix is
symmetric
\footnote
{For example, we can consider
$(n,n)$ and $(8+n,n)$ dimensions where one can have
(pseudo) Majorana-Weyl spinors, or $(4+n,n)$ dimensions where (pseudo)
symplectic Majorana-Weyl spinors are possible. In the latter case, the
tensor product of the antisymmetric $n$'th rank $\c$-matrices with the
antisymmetric invariant tensor of the symplectic automorphysm group is
symmetric. The spinors need not be Weyl, and thus we can consider other
dimensions as well (see Appendix  A).
}.
In $(d-n,n)$ dimensions, setting the $n$\,th rank generator equal to a
product of $n$ momentum generators \cite{b2} and all the other $Z$
generators equal to zero, one has

\be 
\{ Q_\a, Q_\b \} = 
	(\c^{\m_1\cdots \m_n})_{\a\b}~P_{\m_1}^1 \cdots P_{\m_n}^n 
\la{algnnn}\ .
\ee

Motivations for considering these algebras have been discussed elsewhere
(see \cite{B1,RS, RS1,RSS}, for example). The fact they can be realized
in terms of multi-particle systems was pointed out in \cite{b6}.
Superparticle models were constructed in \cite{RS1} and \cite{deli1},
with emphasis on the cases of $n=2,3$. In \cite{RS1} a multi-time model
was considered in which a particle of type $i=1,2,3$ depended on time
$\tau_i$ only. As a special case, an action for a superparticle in the
background of one or two other superparticles with constant momenta was
obtained. In the model of \cite{deli1}, where single time dependence was
introduced, a similar system was described. The results agree, after one
takes into account the trivial symmetry transformations that depend on
the equations of motion.

These models were improved significantly in \cite{deli2}, where all
particles are taken to have arbitrary (single) time dependence. In
\cite{deli2}, it was observed that the theory had $n$ first class
fermionic constraints, but one fermionic symmetry was exhibited (or $m$
of them for extended supersymmetry with $N=1,...,m$). Furthermore, the
form of the transformation rules given in \cite{deli2} appear to be
rather different than those found earlier in \cite{RS1} (albeit in the
context of a restricted version of the model).

Consider $n$ superparticles which propagate in $(d-n,n)$ dimensional
spacetime. Let the superspace coordinates of the particles be denoted by
$X_i^\mu (\tau)$ and $\theta_i^\a(\tau)$\, with $i=1,...,n$,
$\m=0,1,...,d-1$ and $\a = 1,..., dim\, Q$, where $dim\ Q $ is the
minimum real dimension of an $SO(d-n,n)$ or $SO(d-n,n)\times G$ spinor,
with $G$ being the automorphysm group. Working in first order formalism,
one also introduces the momentum variables $P^i_\m(\tau)$ and the Lagrange
multipliers $e_{ij}(\tau)$.

The superalgebra \eq{algnnn} can be realized in terms of the supercharge 

\be
Q_\a= \p_\a +\C_{\a\b}~\t^\b\ ,   \la{qnnn} 
\ee 

where, using the notation of \cite{deli2}, we have defined 

\be
\C  \equiv \frac{1}{n!}~\ve_{i_1\cdots i_n}~\c^{\m_1\cdots \m_n}~
	P_{\m_1}^{i_1} \cdots P_{\m_n}^{i_n}\ . \la{defcnn}
\ee 

The spinorial derivative is defined as $\p_\a =\p/\p\t^\a $, acting
from the right. The transformations generated by the supercharges
$Q_\a$ are \cite{deli2}

\be
\d_\ce X_i^\m = - \eb~V_i^\m \t \ , \qq
\d_\ce \t = \ce\ ,\qq
\d_\ce  P_i^\m =  0\ , \qq
\d_\ce e_{ij} = 0\ ,\la{snn} 
\ee

where, in the notation of \cite{deli2}, we have the definition 

\be 
 V_i^\m \equiv \frac{1}{(n-1)!}~\ve_{i i_2\cdots i_n}~
	\c^{\m \m_2\cdots \m_n}~ P_{\m_2}^{i_2}\cdots P_{\m_n}^{i_n}\ .
\la{defvnn}
\ee 

We use a convention in which all fermionic bilinears involve $\c
C$-matrices with the charge conjugation matrix $C$ suppressed, e.g. $\tb
\c^{\m\n} d\t \equiv  \t^\a (\c^{\m\n}C)_{\a\b}\,d\t^\b$. Otherwise
(i.e. when there are free fermionic indices), it is understood that
the matrix multiplications involve northeast-southwest contractions,
e.g. $(\C\,\t)_\a \equiv (\C)_\a{}^\b \t_\b$. Note that

\be
P^i_\m V_i^\m = n \C\ ,\quad\quad V_i^\m~dP^i_\m =d\C\ ,\quad\quad 
P^i_\m~dV_i^\m =(n-1) d\C\ ,
 \la{lemma1nn}
\ee

where $d \equiv \p /\p\tau$.

The action constructed in \cite{deli2} is given by

\be
I = \int d \tau~\left( P^i_\m \Pi_i^\m 
	-\ft12 e_{ij} P^i_\m P^{j\m} \right)\ , \la{ractnn}
\ee

where

\be
\Pi_i^\m = d X_i^\m +\ft{1}{n}\tb V_i^\m d \t \ . \la{lenn}
\ee

It should be noted that the line element \eq{lenn} is not invariant under
supersymmetry, but it is defined such that $P^i_\m \Pi_i^\m$ transforms
into a total derivative as $P^i_\m \left(\d_\ce \Pi_i^\m\right)=(1-n)
d(\eb~\C \t)$. Consequently, the action \eq{ractnn} is invariant under the
global supersymmetry transformations \eq{snn}. It also has the local
bosonic symmetry

\be
\d_\L e_{ij} =  d \L_{ij}\ , \qq 
\d_\L X_i^\m = \L_{ij} P^{j\m} \ , \qq
\d_\L P_i^\m = 0 \ , \qq
\d_\L \t     = 0 \ , \la{bnnn}
\ee

where the transformation parameters have the time dependence
$\L_{ij}(\tau)$. These transformations are equivalent to those given in
\cite{deli2} by allowing gauge transformations that depend on the
equations of motion.

The action \eq{ractnn} is also invariant under the following 
$\k$-symmetry transformations

\bea
\d_\k \t &=& \c^\m \k_i P^i_\m \ ,\nn\\
\d_\k X_i^\m &=& -\tb V_i^\m (\d_\k \t) \ , \nn\\
\d_\k P_i^\m &=& 0\ , \nn\\
\d_\k e_{ij} &=& -\ft{4}{(n+9)}~\kb_{(i}~V_{j)}^\m \c_\m~d\t \ .
\la{rknn}
\eea

In showing the $\k$-symmetry, the following lemmas are
useful:

\bea
P^i_\m \left(\d_\k \Pi_i^\m\right) &=&
\ft{(1-n)}{n}~d \left( \tb_k~\C \d_\k \t_k \right)
+ \ft{2}{n} (\d_\k \tb_k)~\C~d \t_k \ ,
\la{lemma2}\\[+0.25cm] 
\C \c^\m P^i_\m &=& \ft{1}{(n+9)}~M^{ij} V_j^\m \c_\m\ ,
\la{lemma3nn}
\eea

where

\be
M^{ij} \equiv P^i_\m P^{j\m}\ . \la{mijnn}
\ee

A special combination of the transformations \eq{rknn} was found in
\cite{deli2}, where it was also observed that there is a total of $n$
first class constraints in the model. The $\kappa$-symmetry
transformations given above realize the symmetries generated by these
constraints. 

The commutator of two $\k$-transformations closes on-shell onto the
$\L$-transformations 

\be
[\d_{\k_{(1)}},\d_{\k_{(2)}}]= \d_{\L_{(12)}} \ , \la{close-new}
\ee

with the composite parameter given by

\be
\L^{(12)}_{ij}= -4\kb^{(2)}_{(i}~\C~\k^{(1)}_{j)}\ . \la{rcnn}
\ee

It is clear that the remaining part of the algebra is
$~[\d_\k,\d_\L]=0~$ and $~[\d_{\L_1},\d_{\L_2}]=0~$.

Finally, we note that the field equations following from the action
\eq{ractnn} take the form

\be
P^i_\m P^{j\m} = 0\ ,\qq
dP^i_\m = 0\ , \qq
\C d\t  = 0\ , \qq
dX_i^\m + \tb V_i^\m d\t -e_{ij} P^{j\m}=0 \ . \la{reomnn}
\ee

The precise relation between the bosonic symmetry transformations
\eq{bnnn}, the $\kappa$-symmetry transformations \eq{rknn}, and those
presented in \cite{deli2} was found in \cite{RS1}.

In this section we have used a notation suitable to simple (i.e. $N=1$)
supersymmetry in the target superspace. One can easily account the
extended supersymmetry case by introducing an extra index $A=1,...,m$
for the fermionic variables, thereby letting $\e \ra \e^A$, $\theta \ra
\theta^A$ and $\k^i\ra \k^{Ai}$, etc. All the formulae of this section
still hold, since no need arises for any Fierz rearrangements that might
potentially put restrictions on the dimensionality of the target
superspace.

So far we considered only realizations of superalgebras in terms of super
0-branes. In the next sections of this chapter
we will discuss superstring models realized on
superalgebras in dimensions beyond eleven.

\bigskip




\section{\bf $N=(2,1)$ superstring in $(n,n)$ dimensions}


It is also possible to describe super Yang-Mills theory in
higher than $(10,2)$ dimensions. The $(10,2)$ dimensional super
Yang-Mills theory can be derived from a critical heterotic string theory
based on the $N=(2,1)$ superconformal algebra \cite{km1,ov}. In this
section we shall describe the underlying critical string theories of the
super Yang-Mills theories in higher than $(10,2)$ dimensions
\cite{RSS} using a
generalization of the heterotic $N=(2,1)$ string of
\cite{km1,km2,ov,km3}. The model is based on an $N=1$ superconformal
algebra for left-movers in $(8+n,n)$ dimensions and an $N=2$
superconformal algebra for right-movers in $(n,n)$ dimensions. Both
these algebras are extended with null currents
%
\footnote{For a construction which uses null-extended $N=(1,1)$
superconformal algebras realized in $(10,2)$ dimensions, see \cite{ff}.
}.
%
The null-extended $N=1$ algebra is realized in terms of free scalars
$X^{\hm}$ and fermions $\psi^{\hm}$ and makes use of $n-1$
mutually orthogonal null vectors 

\be
v_{\hm}^i~v^{\hm\,j}=0\ , \quad i,j=1,...,n-1, \ \ \hm =1,...,8+2n \ .
\ee

The left-moving $N=1$ algebra is realized as

\bea
T&=& -{1\over 2}~\eta_{\hm\hn}~\p X^{\hm} \p X^{\hn} + {1\over 2}~
\eta_{\hm\hn}~\psi^{\hm} \p \psi^{\hn}\ , \\
G &=& \sqrt {2}~\psi^{\hm}\p X^{\hn} \eta_{\hm\hn}\ ,\\
J^i &=& v_{\hm}^i~\p X^{\hm}\ ,\\
\C^i &=& v_{\hm}^i~\psi^{\hm}\ .
\eea 

The basic OPE's are $X^{\hm}(z) X^{\hn}(0)=-\eta^{\hm\hn}\log z$ and
$\psi^{\hm}(z) \psi^{\hn}(0)=-\eta^{\hm\hn}z^{-1}$. The OPE's of the
energy momentum tensor $T$ has central charge $c=12+3n$, and its OPE's
with the currents $G, J^i, \C^i$ imply that they have conformal spin
${3\over 2},1,{1\over 2}$, respectively. Thus, the ghost anomaly is
$c_g=-26+11-(n-1)\times( 2+1)=-(12+3n)$. 

The null-extended $N=2$ algebra is realized in terms of scalars $X^\m$,
and fermions $\psi^\m$ and makes use of a real
structure $I_\m{}^\n$ in $(n,n)$ dimensions obeying

\be
I_{\m\n}=-I_{\n\m}\ ,\qquad I_\m{}^\rho I_\rho{}^\n=\delta_\m^\n\ ,\quad
\m=1,...,2n , \la{rsss} 
\ee

where $I_{\m\n}=I_\m{}^\rho \eta_{\rho\n}$. The real structure has $n$
eigenvectors of eigenvalue $+1$, and $n$ eigenvectors of eigenvalue
$-1$. The crucial property of these eigenvectors that allows us to write 
down a critical algebra in $(n,n)$ dimensions is that the inner
products of two eigenvectors of the same eigenvalue vanishes. Hence, in
particular all the eigenvectors are null. Pick $(n-2)$ of these null
vectors, $\tv_{\m}^r$, say of eigenvalue $+1$. They satisfy 
\bea
I_{\m}{}^\n \tv_\n^r &=&\tv_{\m}^{r}\ , \la{pvss}\\
\tv_\m^r~\tv^\m_s &=& 0\ ,   \qquad r,s=1,...,n-2\ . \la{vvss}
\eea
The right-moving $N=2$ algebra is then realized as
\bea
\bar{T}&=& -{1\over 2}~\eta_{\m\n}~\bar{\p} 
X^\m \bar{\p} X^\n + {1\over 2}~ \eta_{\m\n}~\psi^\m \bar{\p} \psi^\n\ , 
\\
\bar{G}_\pm &=& {1\over \sqrt{2}}~(\eta_{\m\n}\pm 
I_{\m\n})~\psi^\m\bar{\p} X^\n \ ,
\\
\bar{J} &=& -{1\over 2}~I_{\m\n}~\psi^\m\psi^\n\ ,
\\
\bar{J}^r &=& \tv_\m^r~\bar{\p} X^\m\ ,
\\
\bar{\C}^r &=& \tv_\m^r~\psi^\m\ .
\eea
The energy momentum tensor $\bar{T}$ has central charge $\bar{c}=3n$,
and its OPE's with the currents $\bar{G}_\pm, \bar{J}, \bar{J}^r, \C$
imply that they have conformal spin ${3\over 2},1,1,{1\over 2}$,
respectively. Note that the closure of the algebra requires the eigen
property of the vectors $\tv^r$ as well as their nullness. Hence, in
this case the ghost anomaly is assumes the critical value
$\bar{c}_g=-26+2\times 11-2-(n-2)(2+1)=-3n$. 

Following the usual BRST quantization scheme one constructs the
right-moving supercharges \cite{km1,km2}:
\be
Q_{\ha}=\oint dz~\S_{gh}~S_{\ha}\ ,
\la{qss}
\ee
where $S_{\ha}$ are the right-moving spin fields of $\psi^{\hm}$, and
$\S_{gh}=\exp (-\phi/2-\phi_{1}/2-\cdots-\phi_{n-1}/2)$ is the spin
field of the commuting ghosts. The index $\ha$ labels the Majorana
spinor of $O(8+n,n)$. The single-valuedness of OPE algebra of fermionic
vertex operators in the Ramond sector require that $S_{\ha}$, and
therefore $Q_{\ha}$, are Majorana-Weyl.

BRST invariance requires the null-conditions
\be
\sh{v_{i}}Q =0\ ,  \quad i=1,...,n-1. \la{nullss}
\ee
The standard form of the target space superalgebra is obtained by
considering the anti-commutator of \eq{qss} with its picture-change $Q'$: 
\be
Q'=Z Z_{1}\cdots Z_{n-1} Q\ ,
\la{picss}
\ee
where $Z X=\{Q_{BRST},\xi X\}$ is the picture changing operation built
from the BRST charge and the zero-modes $\xi,\xi_1,...,\xi_{n-1}$ of the
$(\xi,\eta)$ systems used for bosonizing the commuting ghosts. The
supercharges \eq{qss} and \eq{picss} obey the algebra 
$\{Q_{\ha},Q'_{\hb}\}=(\sh{v_{1}}\cdots\sh{v_{n-1}}\sh{p})_{\ha\hb}$,
which reduces to
\be
\{Q_{\ha},Q'_{\hb}\}=(\c^{\hm_1...\hm_n})_{\ha\hb}~ 
v^{1}_{\hm_1}\cdots v^{n-1}_{\hm_{n-1}}P_{\hm_n}
\ee
in the BRST-invariant sector. The case of $n=2$ has also been discussed in
\cite{km3}.

The spectrum of states depend on the choices for the null vectors $v_i$
and $\tv^r$. The choices for $\tv^r$ break $SO(n,n)$ down to $SO(2,2)$,
and the choices for $v_i$ break $SO(2,2)$ down to $SO(2,1)$ or less.
Generically, one obtains massless states which assemble into a super
Yang-Mills multiplet in $(8+n,n)$ dimensions, which effectively has the
$8+8$ degrees of freedom of the usual $(9,1)$ dimensional super
Yang-Mills, after all the physical states conditions are imposed (see
\cite{km1,km2} for the $n=2$ case). There is a subtlety in the present
case, however, having to do with the spectral flows induced by the
null-currents in the left-moving sector. They shift the (nonchiral)
$(n,n)$ dimensional momentum with multiples of $\tv_r$. The physical
state conditions then force $\tv_r$ to be orthogonal to $v_i$. 

We do not yet know the exact feature of the target space field theory.
We expect, however, that it will be of the kind studied in \cite{hull2},
where the Yang-Mills field strength satisfies a generalized self-duality
condition.

In this chapter we discussed particle and string realizations of
different superalgebras. Using alternative
interpretation of the brane charges
we considered superparticle with brane-charge  coordinates realizing super
Poincar\'e
algebra with brane charges. In the next chapter we will  study properties
of
supersymmetric 
field theories that have supersymmetry generated by the  superalgebra with
brane charges.
We will  see that the quantization of 0-brane models gives us a lot of
information
about properties of the unknown field theory.
In this chapter we also discussed realization of nonlinear deformed
finite-dimensional 
superalgebras in  dimensions beyond eleven. Those superalgebras and
0-brane models 
could be useful  for understanding the unification of type IIA/B ,
heterotic
theories  as well as for describing properties of F-theory. Here we have
to notice that
in the later  case full analysis of  representation as well as field
theory
 of nonlinear-deformed algebras of this kind 
is not known  yet and consistent superparticle models could shed some
light on this subject.

\bigskip

\newpage




\chapter{\bf Quantization }

\section{\bf Quantization of particle in twistor-like coordinates}

In this chapter we will study quantization of models discussed in the
first section of  previous
chapter \cite{RSS-new}. 
The main goal is to study spectrum and equations of
motion coming from quantization of particle mechanics and then identify
them with
spectrum and linearized equations of motion of underlined field theory.
As we mentioned before, if the field theory, that realizes some particular
given superalgebra, is not known, it is possible to find its spectrum and
equations of motion using correspondence between quantized particle mechanics
and linearized field theory. As we will see this method is useful for
identifying spectrum and properties of linearized M-theory. Starting from
superalgebra with brane charges we use realization in terms of superparticle
with central charge coordinates and then quantization of this model gives
us
a filed-theory realization of M-theory superalgebra.
Here we will argue that this underlined field theory could be associated with
linearized perturbative version of M-theory. It is important to notice that we
do not have to consider low energy limit, as for the case of eleven
dimensional
supergravity, that is low energy limit of M-theory. The same methods we
will apply for  four dimensions where we will study
spectrum and linearized equations of motion of unknown field theory that
realized superalgebra with brane charges. 

 But before we consider particle with brane charges, we have to develop
a methods
 of quantization of complicated systems of 0-branes with central charge
 coordinates. To start with, we will discuss quantization of similar but
 bosonic models with twistor variables \cite{BMRS}.  Why do we have to use
twistors? 
 When one wants to apply covariant BRST quantization and use
Hamiltonian analysis,
 one may have mixture of second and first class constraints without
possibility
 to solve or decompose them covariantly. In this case twistors and conversion
 procedure, discussed in this and next chapter appear extremely useful.

In the last decades there has been an intensive activity in studying
(super)particles and (super) strings by use of
different approaches aimed at finding a formulation, which would be the most
appropriate for performing the covariant quantization of the models. Almost
all of the approaches use twistor or spinorial variables in one form or another
\cite{penrose} -- \cite{bpstv} and \cite{Berk1}.
This allowed one to better understand the geometrical and group--theoretical
structure of the theory and to carry out a covariant
Hamiltonian analysis (and in some cases even the covariant quantization)
of (super)particle and (super)string dynamics in space--time
dimensions $D=3,4,6$ and 10, where conventional twistor relations take place.

It has been shown that twistor--like variables appear in a
natural way as superpartners of Grassmann spinor coordinates in a doubly
supersymmetric formulation \cite{spinsup} of Casalbuoni--Brink--Schwarz
superparticles and Green--Schwarz superstrings \cite{gsw}, the notorious
fermionic $\kappa$--symmetry \cite{ks}
of these models being replaced by
more fundamental local supersymmetry on the worldsheet  supersurface  swept
by the superparticles and superstrings in target superspace \cite{stvz}
and for recent review see \cite{sor-rev}.
This has solved the problem of infinite reducibility of the fermionic
constraints associated with $\kappa$--symmetry \footnote{A
comprehensive
list of references on the subject the reader may find in \cite{bpstv}}.
As a result new formulation and methods of quantization of $D=4$
compactifications of
superstrings with manifest target--space supersymmetry have been
developed (see \cite{ber} for a review).
However, the complete
and simple solution of the problem of $SO(1,D-1)$ covariant quantization
of twistor--like superparticles and superstrings in $D~>~4$ is still lacking.

To advance in solving this problem one has to learn more on how to deal
with twistor--like variables
when performing the Hamiltonian analysis and the quantization of the
models. In this respect a bosonic relativistic particle in a twistor--like
formulation may serve as the simplest but rather nontrivial toy model.

The covariant quantization of the bosonic particle has been under intensive
study with both the operator and path--integral method
\cite{ferber,sg,bh,bfortschr,teit,polyakov,govaerts,sf}.
In the twistor--like approach the bosonic particle has been mainly quantized
by use of the operator formalism. For that different but classically
equivalent twistor--like particle actions have been considered
\cite{ferber,sg,bh,bfortschr,EiSo}.

The aim of the next section is to study some features of bosonic particle
path--integral quantization in the twistor--like approach
by use of the BRST--BFV quantization prescription \cite{bf} --
\cite{bff}.
In the course of the Hamiltonian analysis we shall observe links between
various
formulations of the twistor--like particle \cite{ferber,stvz,EiSo}
by performing a
conversion of the Hamiltonian constraints of one formulation to another.
A particular feature of the conversion procedure \cite{fs}
applied to turn the
second--class constraints into the first--class constraints is that the
simplest Lorentz--covariant way to do this is to convert a
full mixed set of the initial first-- and second--class constraints rather
than explicitly extracting and converting only the second--class constraints.
Another novel feature of the conversion procedure applied below
(in comparison with the conventional
one \cite{bff,fs}) is that in the case of the $D=4$ and $D=6$ twistor--like
particle the number of new auxiliary Lorentz--covariant  coordinates,
which one introduces to get a system of first--class constraints in an
extended phase space, exceeds the number
of independent second--class constraints of the original dynamical system, 
(but because of an appropriate amount of the first--class constraints
we finally get, the number of physical degrees of freedom remains the 
same). Here we will follow \cite{BMRS}.

{\it Notation.} We use the following signature for the space-time metrics:
$(+,-,...,-)$.


\subsection{\bf Classical Hamiltonian dynamics and the BRST-charge}


    The  dynamics of a massless bosonic particle in D=3,4,6 and 10 space--time
can be described by the action \cite{ferber}

\be 
S={1\over 2}\int d\tau \dot x^m ({\bar \l}\gamma _m \l ), \qquad 
\label{201}
\ee

where $x^m(\tau )$ is a particle space--time coordinate,
$\l ^{\a}(\tau )$ is an
auxiliary bosonic spinor variable, the dot stands for the time derivative
$\partial\over{\partial\tau}$ and  $\gamma ^m $ are the Dirac matrices.

The derivation of the canonical momenta \footnote{In what 
follows $P^{(..)}$ denotes the momentum conjugate to the
variable in  the brackets}

$P^{(x)}_m={{\partial L}
	\over{\partial{\dot x}^m}},~~
P^{(\l )}_{\a}={{\partial L}\over
{\partial}{{\dot \l}^{\a}}}$
results in a set of primary constraints 
$$ \Psi_m = P^{(x)}_m-{1\over 2}({\bar \l}\gamma _m \l )\approx 0, $$

\be 
P^{(\l )}_{\a}\approx 0.
\qquad 
\label{202}
\ee

They form the following
algebra with respect to the  Poisson brackets\footnote{The canonical

Poisson brackets are
$$
[P^{(x)}_m, x^n]_P=\delta ^n_m;
\qquad
[P^{(\l )}_{\a}, \l^{\b}]_P=\delta ^{\b}_{\a} $$}

\be 
[\Psi _m, \Psi _n]_P=0,~~ [P^{(\l )}_{\a},
P^{(\l )}_{\b}]_P=0,~~ [\Psi _m, P^{(\l )}_{\a}]_P=({\gamma _m}\l )_{\a}.
\qquad 
\label{203}
\ee

One can check that new independent secondary constraints do not appear
in the model.
In general,
Eqs. \eq{202} are a mixture of first-- and  second--class constraints.
The  operator quantization of this
dynamical system in $D=4$ (considered previously in \cite{sg,bh}) was
based on the Lorentz--covariant
splitting of the first-- and second--class constraints and on the subsequent
reduction of the phase space (either by explicit solution of the
second--class constraints \cite{bh} or,
implicitly, by use of the Dirac brackets
\cite{sg}), while
in \cite{bz0,bfortschr} a conversion prescription \cite{bff,fs} was used.
The latter consists in the extension of the phase space of the particle
coordinates and momenta with auxiliary variables in such a way, that new
first--class constraints replace the original second--class ones. Then the
initial
system with the second--class constraints is treated as a gauge fixing of a
 ``virtual" \cite{bff} gauge symmetry generated by the additional first--class
constraints of the extended system \cite{bff,fs}.
This is achieved by taking the auxiliary conversion degrees of freedom to be
zero or expressed in terms of initial variables of the model.

     The direct application of this procedure can encounter some technical
problems for systems, where the first-- and second--class constraints form a
complicated algebra
(see, for example, constraints of the $D=10$ superstring in a
Lorentz--harmonic formulation \cite{bzstr}).
Moreover, in order to perform the covariant
separation of the first-- and second--class constraints in the system
under consideration it is necessary either to introduce one more independent
auxiliary
bosonic spinor $\mu _{\a}$ (the second component of a twistor
$Z^A=(\l^{\a},\mu _{\a})$ \cite{penrose}) or to construct the second twistor
component from the variables at hand by use of a Penrose relation
\cite{penrose} ${\bar \mu}^{\dot \a}=ix^{\a {\dot \a}}\l _{\a} ~(D=4),~
\mu^{\a}=x^{\a \b}\l _{\b} ~(D=3)$. In the latter case the structure of the
algebra of the
first-- and second--class constraints separated this way
\cite{sfortschr,sg} makes the conversion procedure
rather cumbersome. To elude this one can try to simplify the procedure by
converting into the first class the whole set \eq{202} of the mixed
constraints. The analogous trick was used to convert fermionic constraints
in superparticle models \cite{EiSo,moshe}.

Upon carrying out the conversion procedure  we get a system characterized by
the set of first--class constraints $T_i$ that form (at least on the mass
shell) a closed algebra  with respect to the Poisson brackets
defined for all the variables of
the modified phase space. In order to perform the BRST--BFV quantization
procedure we associate with each constraint of Grassmann parity $\e $
the pair of canonical conjugate auxiliary variables (ghosts) $\eta _i,~
P^{(\eta )}_i$ with  Grassmann parity $\epsilon +1$ \footnote{If the
extended BRST--BFV method is used, with each constraint  associated are also
a Lagrange multiplier, its conjugate momentum of Grassmann parity $\epsilon
$ and an antighost and its momentum of Grassmann parity $\epsilon +1$ (see
\cite{bf,mhenn} for details).}. The resulting system  is required to be
invariant under gauge transformations generated by a nilpotent fermionic
BRST charge $\Omega $. This invariance substitutes the gauge symmetry,
generated by the first class constraints in the initial phase space.
The generator $\Omega $ is found as a series in
powers of ghosts
\be
\Omega ={\eta _i}T_i + higher~ order~ terms, 
\la{4Om}
\ee

where the structure of higher--order terms reflects the noncommutative
algebraic structure of the constraint algebra \cite{mhenn}.
Being the generator of the BRST symmetry $\Omega$ must be a dynamical invariant:

\be
{\dot \Omega}=[\Omega ,H]_P=0,
\la{4dotOm}
\ee

where $H$ is a total Hamiltonian of the system, which has the form

\begin{equation}
H=H_0+[\chi ,\Omega]_P.
\label{h}
\end{equation}

In \eq{h} $H_0$ is the initial Hamiltonian of the model and $\chi$ is a
gauge fixing fermionic function whose form is determined by admissible gauge
choices \cite{bfortschr,teit,govaerts,sf,west}.

Upon quantization $\Omega$ and $H$ become operators acting on quantum state
vectors. The physical sector of the model is singled out by the requirement
that the physical states are BRST invariant and vanish under the action of
$\Omega$. Another words, we
deal with a {\it quantum gauge theory}.

When the gauge is fixed, we remain only with physically nonequivalent
states, and the Hamiltonian $H$ is argued to reproduce the correct physical
spectrum of the quantum theory.

        When the model is quantized by the path--integral method, we also
deal with a quantum gauge theory.
The
Hamiltonian \eq{h} is used to construct an effective action and a corresponding
BRST-invariant generating functional
which allows one to get transition amplitudes between physical states of the
theory.

        Below we consider the conversion procedure and construct the BRST
charge for the twistor--like particle model in dimensions $D=3,4$ and $6$.


\subsubsection*{\bf D=3}

In $D=3$ the action \eq{201} is rewritten as 

\begin{equation}
  S={1\over 2}\int d\tau \l ^{\a}{\dot x}_{\a \b} \l ^{\b}, \qquad
\label{211}
\end{equation}

where $\l^\a$ is a real two-component commuting spinor (spinor indices are
risen and
lowered by the unit antisymmetric tensor $\epsilon _{\a \b}$) and $x_{\a
\b}=x_m\gamma ^m_{\a \b}$.

The system of
primary constraints \eq{202}

\be
\Psi_{\a \b} = P^{(x)}_{\a \b}-\l_{\a}\l_{\b} , \qq
  P^{(\l)}_{\a}\approx 0, 
\label{212}
\ee

is a mixture of a first--class constraint generating the
$\tau$--reparametrization transformations of $x$ 

\be
\phi = \l^{\a}P^{(x)}_{\a \b}\l^{\b},
\la{4lpl}
\ee

and four second--class constraints

\be 
(\l P^{(\l )}),~~ (\mu P^{(x)}\mu )-(\l \mu )^{2}, ~~(\mu P^{(\l )}),~~(\l
P^{(x)} \mu ), \qquad 
\label{2125}
\ee 

where $\mu ^{\a}=x^{\a \b}\l _{\b}$
(see \cite{sg} for details).

In order to perform a conversion of \eq{212} into a system of first--class
constraints we introduce a pair of  canonical conjugate bosonic spinors
$(\z^{\a},~P^{(\z )}_{\b})$, $[P^{(\z )}_{\b}, \z^{\a}]_P=\delta ^{\a}
_{\b}$, and take the modified system of constraints, which is of the first
class, in the following form:

\be
\Psi '_{\a \b} = P^{(x)}_{\a \b}-(\l_{\a}-\z_{\a})(\l_{\b}-\z_{\b}), \qq
\Phi '_{\a} = P^{(\l )}_{\a}+P^{(\z )}_{\a} .
\label{213}
\ee

Eqs. \eq{213} reduce to \eq{212}
by putting the auxiliary variables $\z^{\a}$  and $P^{(\z )}_{\a}$ equal to
zero. This reflects the appearance in the model of a new gauge symmetry with
respect to which $\z^{\a}$  and $P^{(\z )}_{\a}$  are pure gauge degrees of
freedom.

        It is convenient to choose the following phase--space variables as
independent ones:

\be
v^{\a}=\l^{\a}-\z^{\a}, \qquad
P^{(v)}_{\a}={1\over 2}(P^{(\l )}_{\a}-P^{(\z )} _{\a}), \qq
w^{\a}=\l^{\a}+\z^{\a},        
\qquad
P^{(w)}_{\a}={1\over 2}(P^{(\l )}_{\a}+P^{(\z )}_ {\a}), \qquad
\la{4vpwpl}
\ee

Then Eqs. \eq{213} take the following form 

\be
\Psi '_{\a \b}=P^{(x)}_{\a \b}-v_{\a}v_{\b}, \qq  P^{(w)}_{\a}\approx
0. 
\label{214} 
\ee

These constraints form an Abelian algebra.

One can see that $w^{\a}$ variables do not enter the constraint relations,
and their conjugate momenta are zero. Hence, the quantum physical states of
the model will not depend on $w^{\a}$ .

Enlarging
the modified phase space with ghosts, antighosts and Lagrange multipliers in
accordance with the following table

\begin{tabular}{cccc}
&&&\\
Constraint &  Ghost &     Antighost   &   Lagrange~multiplier   \\
${\Psi}'_{\a \b}$ & $c^{\a \b}$ & $\tilde c^{\a \b}$ & $e^{\a \b}$ \\
$P^{(w)}_{\a}$   &  $b^{\a}$  &   $\tilde b^{\a}$   &  $f^{\a}$ \\
&&&
\end{tabular}

\noindent

we write the classical BRST charges \cite{bf,mhenn} of the model in the
minimal and extended BRST--BFV version as follows 

\begin{equation} 
\Omega_{min}=c^{\a \b}\Psi '_{\b \a}+b^{\a}P^{(w)}_{\a}, \qquad \end{equation}
\begin{equation} \label{217}
\Omega =P^{(\tilde c)}_{\a \b}P^{(e)\b \a}+P^{(\tilde b) \a}P^{(f)}_{\a}+
\Omega _{min}. \qquad
\label{218}
\end{equation}


\subsubsection*{\bf D=4}

In this dimension we use two--component $SL(2,C)$ spinors
$(\l^{\a}=\epsilon^{\a \b}\l_{\b}; ~{\bar \l}^{\dot
\a}=\epsilon^{{\dot \a} {\dot \b}}{\bar \l}_{\dot \b}; ~\a ,{\dot
\a}=1,2; ~ \epsilon^{12}=-\epsilon_{21}=1)$. Other notation
coincides with that of the $D=3$ case. Then in $D=4$  the action
\eq{201} can be written as following

\begin{equation}
S= {1\over 2}\int d \tau \l^\a {\dot x}_{\a {\dot
\a}}\bar{\l^{\dot \a}}, \qquad
\label{221} 
\end{equation}

where $ x_{\a\dot \a}= x_{m}{\sigma}^{m}_{\a {\dot \a}} $, and
${\sigma}^m_{\a {\dot
\a}}$ are the relativistic Pauli matrices.  The set of the primary 
constraints \eq{202} in this dimension
 
\be
\Psi_{\a {\dot {\a}}}= P^{(x)}_{\a 
{\dot {\a}}}- \bar{ \l}_{\dot \a} \l_{\a} \approx 0,
\la{4psipll}
\ee

\be    
{P^{(\l )}}_{\a}\approx 0, \qquad 
\label{222} 
\ee 

\be
{\bar P}^{(\bar {\l} )}_{\dot 
{\a}}\approx 0 
\la{4barpl}
\ee

 contains two first--class constraints and three pairs of 
conjugate second--class constraints \cite{sg,sfortschr}.  One of the first 
class constraints  generates the $\tau$-reparametrization transformations 
of $x^{{\dot\a} \a}$ 

\be
 \phi= \l^{\a}P^{(x)}_{\a {\dot {\a}}}{\bar 
{\l}}^{\dot {\a}} ,
\la{philPla}
\ee
 
and another one generates  $U(1)$ rotations of the 
complex spinor variables

 \be 
   U=i(\l^{\a} P^{(\l )}_\a- \bar 
{\l}^{\dot{\a}}\bar P^{(\l )}_{\dot {\a}}). 
\label{2221}
 \ee

The form of the second--class constraints is analogous to that in the D=3
case (see Eq. \eq{2125} and \cite{sg}), and we do not present it explicitly
since it is not used below.

To convert the mixed system of the constraints \eq{222} into  first--class
constraints
one should introduce at least three pairs of canonical conjugate auxiliary
bosonic
variables, their number is to be equal to the  number of the second--class
constraints in \eq{222}. However, since we do not want to violate the
manifest Lorentz invariance, and the $D=4$ Lorentz group does not have
three--dimensional representations, we are to find a way round.
We introduce two
pairs of canonical conjugate conversion spinors $
(\zeta^{\a},P^{(\zeta )}_{\a}),~~
[\zeta^{\a},P^{(\zeta)}_{\b}]_P=-\delta^{\a}_{\b}, ~~[\bar {\zeta}^{\dot
{\a}},{\bar P}^{(\bar \zeta )}_{\dot {\b}}]_P=-\delta^ {\dot \a}_{\dot \b},
$
(i.e. four pairs of real auxiliary variables) and modify the constraints
\eq{222} and the $U(1)$ generator, which becomes an independent first--class
constraint in the enlarged phase space.
Thus we get the following system of the first--class constraints:

\be
\Psi^{'}_{\a \dot {\a}}=P^{(x)}_{\a \dot {\a}} -(\bar {\l}-\bar
{\zeta})_{\dot {\a}} (\l-\zeta)_{\a}\approx 0,
\la{psips-z}
\ee

\be 
{  \Phi_\a}=P^{(\l )}_\a+P^{(\zeta )}_{\a}\approx 0 ,
\label{223}
\ee

\be
\bar {\Phi}_{\dot {\a}}= \bar P^{(\bar \l )}_{\dot {\a}}+\bar P^{(\bar \zeta
)}_{\dot {\a}}\approx 0 ,
\la{PhiPP}
\ee

\be
U=i(\l^{\a} P^{(\l )}_\a+\zeta^{\a}
P^{(\zeta )}_{\a}- \bar {\l}^{\dot {\a}}\bar P^{(\l )}_{\dot {\a}}-\bar
{\zeta}^{\dot {\a}}\bar P^{(\zeta )}_{\dot {\a}}) .
\la{Ulplplp}
\ee

One can see
(by direct counting), that the number of independent physical
degrees of freedom of the particle  in the enlarged phase space is the
same as in the initial one. The latter is recovered by imposing gauge
fixing conditions on the new auxiliary variables

\be 
\z^{\a}=0,\qquad{\bar \z}^{\dot \a}=0,\qquad P^{(\z )}_{\a}=0, \qquad
P^{({\bar \z})}_{\dot \a}=0.
\label{2231}
\ee

By introducing a new set of the independent spinor variables analogous to
that in \eq{214} one rewrites Eqs. \eq{223} as follows

\be
\Psi'_{\a {\dot{\a}}}=P^{(x)}_{\a {\dot{\a}}}- v_{\a}{\bar
v}_{\dot{\a}}\approx 0, \quad  U=i(P^{(v)}_{\a}v^{\a}- P^{(\bar v)}_{\dot
{\a}}{\bar v}^{\dot {\a}}),
\la{Psipvv}
\ee

\be 
{P^{(w)}}_{\a}\approx 0, \qquad  P^{(\bar{w})}_{\dot{\a}}\approx 0.
\label{224}
\ee

Again, as in the $D=3$ case, $w_\a,~{\bar w}_{\dot\a}$ and their momenta
decouple from the first pair of the constraints \eq{224}, and can be
completely excluded from the number of the dynamical degrees of freedom  by
putting
\begin{equation}
w_\a=\l_\a+\z_\a=0, \qquad
{P^{(w)}}_{\a}={1\over 2}({ P^{(\l )}}_{\a}+{P^{(\z )}}_{\a})=0 
\label{es}
\end{equation}

in the strong sense. This gauge choice, which differs from \eq{2231},
reduces
the phase space of the model to that of a version of the twistor--like
particle dynamics, subject to the first pair of the first--class constraints
in \eq{224}, considered by Eisenberg and Solomon \cite{EiSo}.
The constraints \eq{224} form an abelian algebra, as in the $D=3$ case.
In compliance with the BRST--BFV prescription we introduce ghosts,
antighosts and Lagrange
multipliers associated with the
constraints \eq{224} as follows

\begin{tabular}{cccc}
Constraint  &  Ghost &   Antighost &  Lagrange~ multiplier  \\ ${\Psi}
'_{\a {\dot{\a}}}$ &  $c^{\dot{\a} \a}$ & ${\tilde c}^{{\dot \a} \a}$ &
$e^{\dot{\a} \a} $ \\ $U$ & $a$ & ${\tilde a}$ & $g$ \\ $P^{(w)}_{\a}$ &
$b^{\a}$ & ${\tilde b}^{\a}$ & $f^{\a}$ \\ $P^{(\bar{w})}_{\dot{\a}}$ &
${\bar b}^{\dot{\a}}$ & ${\tilde{\bar b}}^{\dot{\a}}$ & ${\bar f}^{\dot
{\a}}$ \\ \qquad 
\end{tabular}

  Then the BRST--charges of the $D=4$ model have the form

\begin{equation}
\Omega_{min}=c^{\dot {\a} \a}\Psi_{\a {\dot {\a}}}+b^{\a}P^{(w)}_{\a}+
{\bar b}^{\dot{\a}}P^{({\bar w})}_{\dot{\a}}+aU,
\qquad \end{equation}
\be \label{226}
\Omega=P^{({\tilde c})}_{\a {\dot{\a}}}P^{(e)
{\dot{\a}}\a}+P^{({\tilde b})} _{\a}P^{(f)\a}+P^{({\tilde{\bar
b}})}_{\dot{\a}}P^{({\bar f})\dot{\a}}+ P^{({\tilde
a})}P^{(g)}+\Omega_{min}.
\label{227} 
\ee


\subsubsection*{\bf D=6}

In $D=6$ a light--like vector $V^m$ can be represented in terms of commuting
spinors as follows

\be
V^m=\l^{\a}_i\gamma^m_{\a \b}\l^{\b i}, 
\la{Vlgl}
\ee

where $\l^{\a}_i$ is an $SU(2)$--Majorana--Weyl
spinor which has the $SU^*(4)$ index
$\a =1,2,3,4$ and the $SU(2)$ index $i=1,2$.
$\gamma^m_{\a \b}$ are $D=6$ analogs of the Pauli matrices (see
\cite{6spin,benght}). $SU(2)$ indices are risen and lowered by the unit
antisymmetric tensors $\epsilon_{ij},~~\epsilon^{ij}$. As to the $SU^*(4)$
indices, they can be risen and lowered only in pairs by the totally
antisymmetric tensors $\e_{\a \b \c \d}$, $\e^{\a \b \c \d}$
($\e_{1234}=1)$.

Rewriting the action \eq{201} in terms of $SU(2)$--Majorana--Weyl spinors,
one gets
\be 
S={1\over 2}\int d \tau {\dot x}^m \l^{\a}_i(\gamma_m)_{\a \b}\l^{\b i}.
\label{231} 
\ee

   The system of the primary constraints \eq{202} takes the form

\be
\Psi_{\a \b}=P^{(x)}_{\a \b}-\e_{\a \b \c \d}\l^{\c}_i \l^{\d i}, \quad 
{  ~P^{(\l )i}}_{\a}\approx 0,
\la{232}
\ee

where $P^{(x)}_{\a \b}=P^{(x)}_m \gamma^m_{\a \b}$. $\Psi_{\a \b}$ is
antisymmetric in $\a$ and $\b$ and contains six independent components.
(To get \eq{232} we used the relation
$(\c_m)_{\a\b}\c^m_{\c\d}\sim \e_{\a \b \c \d}$).

From Eqs. \eq{232}
one can separate four first--class constraints  by projecting  \eq{232}
onto $\l^{\a}_{i}$ \cite{sfortschr,benght}.  One of the first--class
constraints generates the $\tau$--reparametrizations of 
$x^{\a\b}$ 

\be
 \phi =\l^{\a}_{i}P^{(x)}_{\a \b}\l^{\b i}, 
\la{4repar-phi}
\ee

and another three ones form an
$SU(2)$ algebra 

\be
 T_{ij}=\l^{\a}_{(i}P^{(\l )}_{\a j)}. 
\la{4reapTlP}
\ee

 Braces denote
the symmetrization of $i$ and $j$.  All other constraints in \eq{232} are
of the second class.

   The conversion of \eq{232} into first--class constraints is carried out by
analogy  with the $D=4$
case. According to the conventional conversion prescription we had to
introduce five pairs
of canonical conjugate bosonic variables. Instead, in order to preserve
Lorentz invariance, we introduce the canonical conjugate pair of bosonic
spinors $\z^{\b}_j$, $P^{(\z )i}_{\a}$
($[P^{(\z )i}_{\a},\z^{\b}_j]_P=\d^{\b}_{\a}\d^i_j,$) modify the constraints
\eq{232} and the $SU(2)$ generators.
This results in the set of independent first--class constraints
$$
\Psi '_{\a \b}=P^{(x)}_{\a \b}-\e_{\a \b \c \d}
(\l^{\c}_i-\z^{\c}_i)(\l^{\d i}-\z^{\d i})\approx 0, $$

\be 
{  \Phi_{\a}}^i=P^{(\l )i}_{\a}+P^{(\z )i}_{\a}\approx 0, \qquad 
\label{233}
\ee

$$
T_{ij}=\l^{\a}_{(i}P^{(\l )}_{\a j)}-\z^{\a}_{(i}P^{(\z )}_{\a j)}\approx 0.
$$
In terms of spinors $v^{\a}_{i}$ and $w^{\a}_{i}$, and their momenta,
defined as in the $D=3$ case \eq{214}, they take the following form
$$
\Psi '_{\a \b}=P^{(x)}_{\a \b}-\e_{\a \b \c \d}v^{\c}_iv^{\d i}\approx
0, $$

\be 
{  T_{ij}}=v^{\a}_{(i}P^{(v)}_{\a j)}\approx 0, \qquad 
\label{234}
\ee

$$ P^{(w)i}_{\a}\approx 0. $$

    These constraints form a closed algebra with respect to the Poisson
brackets.
The only nontrivial brackets in this algebra are 

\be 
[T_{ij},T_{kl}]_{p}=\epsilon_{jk}T_{il}+\epsilon_{il}T_{jk}+
\epsilon_{ik}T_{jl} +\epsilon_{jl}T_{ik}, \qquad 
\label{235}
\ee

which generate the $SU(2)$ algebra.

We introduce ghosts, antighosts and Lagrange multipliers related to the
constraints \eq{235}

\begin{tabular}{cccc}
& & &\\
Constraint  &  Ghost  &  Antighost &  Lagrange~ multiplier  \\ ${\Psi} '_{\a
\b}$ & $c^{\a \b}$ & ${\tilde c}_{\a \b}$ & $e^{\a \b}$  \\ $T_{ij}$ &
$a^{ij}$ & ${\tilde a}_{ij}$ & $g^{ij}$ \\ $\Phi ^{i}_{\a}$ & $b^{\a}_{i}$ 
& ${\tilde b}^{i}_{\a}$ & $f^{\a}_{i}$ \\ &&& 
\end{tabular}

\noindent

and construct the BRST charges corresponding respectively, to the minimal
and extended BRST--BFV version, as follows 

\begin{equation}
\Omega_{min}=c^{\a \b}\Psi '_{\b \a}+b^{\a}_{i}P^{(w)i}_{\a}+a^{ij}T_{ji}+
\qquad
\label{238}
\ee

$$
(\epsilon_{jk}P^{(a)}_{il}+ \epsilon_{il}P^{(a)}_{jk}+
\epsilon_{ik}P^{(a)}_{jl}+\epsilon_{jl}P^{(a)}_{ik})a^{ij}a^{kl}.
$$

\be 
\Omega=P^{({\tilde c})}_{\a \b}P^{(e)\b \a}+P^{({\tilde b})\a}
_{i}P^{(f)i}_{\a}+P^{({\tilde a})ij}P^{(g)}_{ji}+\Omega_{min}, \qquad 
\label{237}
\ee

Higher order terms in ghost powers appear in \eq{238} and \eq{237} owing to the
noncommutative $SU(2)$ algebra of the $T_{ij}$ constraints \eq{235}.


\subsection{\bf Path Integral Quantization}

\subsubsection*{\bf Admissible gauge choice }

   One of the important problems in the quantization of gauge
systems is a correct gauge choice. In the frame of the BRST--BFV
quantization scheme gauge fixing is made by an appropriate choice of the
gauge fermion that determines the structure of the quantum Hamiltonian.
The Batalin and Vilkovisky theorem \cite{bf,mhenn} reads that the result of
path integration does not depend on the choice of the gauge fermions if they
belong
to the same equivalence class with respect to the BRST--transformations.
An analogous theorem
takes place in the operator BRST--BFV quantization scheme \cite{sf}.
Further
analysis of this problem for systems possessing the reparametrization
invariance showed that the result of path integration does not depend on
the choice of the
gauge fermion if only appropriate
gauge conditions are compatible with the  boundary conditions for the
parameters of the corresponding gauge transformations
\cite{bfortschr,polyakov,govaerts,sf,west}. In particular, it was shown
that the so--called ``canonical gauge", when the worldline gauge field of the
reparametrization symmetry of the bosonic particle is fixed to be a
constant, is not admissible in this sense.
(see \cite{bfortschr,govaerts}
for details). Anyway one can use the canonical gauge as a consistent limit
of an admissible gauge \cite{sf}.

Making the analysis of the twistor--like model one can show that admissible
are the following gauge conditions on Lagrange multipliers from the
corresponding Tables of the previous section in the dimensions $D=3,$ 4 and
$6$ of space--time, respectively,
\be 
D=3:\qquad {\dot e}^{\a\b}=0;\qquad f^{\a}=0; 
\label{261}
\ee

\be 
D=4:\qquad {\dot e}^{\a\dot\b}=0;\qquad f^{\a}=0; \qquad f^{\dot
\a}=0;\qquad g=0;
\label{262}
\ee

\be 
D=6:\qquad {\dot e}^{\a\b}=0;\qquad f_i^{\a}=0;\qquad g^{ij}=0; 
\label{263}
\ee

The canonical gauge

\be 
e=constant,
\label{264}
\ee

can be considered as a limit of
more general admissible gauge $e-\varepsilon {\dot e}=constant$ (at
$\varepsilon \rightarrow 0$) \cite{sf}.
Then the use of the gauge condition \eq{264} does not lead to any problems
with the operator BRST--BFV quantization.

Below we shall use the ``relativistic" gauge conditions \eq{261}, \eq{262} and
\eq{263} for the path--integral quantization. The use of the canonical
gauge \eq{264} in this case would lead to a wrong form of the particle
propagator.


\subsubsection*{\bf Path--integral BRST quantization }

In this section we shall use the
extended version of the BRST--BFV quantization procedure \cite{mhenn,bff}
and fix the gauge by applying the conditions \eq{261}, \eq{262}, \eq{263}.
The gauge fermion, corresponding to this gauge choice, is

     \be 
     \chi_D={1\over 2}P^{(c)}_me^m, ~~~D=3,4,6, \qquad
     \label{417}
     \ee

The Hamiltonians constructed with \eq{417} are \cite{bf,mhenn}
$$ {\it H}_D=[\Omega_D,\chi_D],\qquad D=3,4,6 $$ 

\be 
{H}_3=e^{m}(P^{(x)}_{m}-{1\over 2}v^{\a}(\gamma_{m})_{\a \b}v^{\b})
-P^{(c)}_{m}P^{({\tilde c})m}, \qquad
\label{420}
\ee

\be 
{H}_4=e^{m}(P^{(x)}_{m}-{1\over 2}{\bar v}^{\dot {\a}}(\sigma_{m})
_{\dot{\a}\a}v^{\a})-P^{(c)}_{m}P^{({\tilde c})m}, \qquad 
\label{421}
\ee

\be 
{H}_6=e^{m}(P^{(x)}_{m}-{1\over 2}v^{\a}_i(\c_m)_ {\a \b}v^{\b i})
-P^{(c)}_{m}P^{({\tilde c})m}, \qquad 
\label{422}
\ee

We shall calculate the coordinate propagator
$Z=\langle x_{1}^{m}\mid U_0\mid x_{2}^{m}\rangle$ (where 
$U_0=expiH(T_1-T_2)$ is the evolution operator), 
therefore boundary conditions for the phase space 
variables are fixed as follows:

\be  
x^{m}(T_1)=x_{1}^{m}, 
\qquad x^{m}(T_2)=x_{2}^{m}, \qquad 
\label{423}
\ee

 the boundary values of the ghosts, 
antighosts and canonical momenta of the Lagrange multipliers are put equal 
to zero (which is required by the BRST invariance of the boundary 
conditions \cite{mhenn}), and we sum up over all possible values of the 
particle momentum and the twistor variables.

The standard expression for the matrix element of the evolution  operator is

\be 
 Z_D=\int[D\mu DP^{\mu}]_D
 exp(i\int_{T_1}^{T_2} d \tau ([P^{\mu}{\dot {\mu}}]_D -{\it H}_D )),
\qquad D=3,4,6.
\label{424}
\ee

$[D\mu DP^{\mu}]_D $ contains functional Liouville measures
of all the canonical variables of the BFV extended phase space \cite{bf}.
$[P^{\mu}{\dot{\mu}}]_D $ contains a sum of products
of the canonical momenta with the velocities.

For instance, an explicit expression for the path--integral measure
in the $D=3$ case is
$$
[D\mu DP^{\mu}]=DxDP^{(x)}DvDP^{(v)}DwDP^{(w)}DeDP^{(e)}DfDP^{(f)} $$
$$
DbDP^{(b)}DcDP^{(c)}D{\tilde b}DP^{({\tilde b})}D{\tilde c}DP^{({\tilde
c})}.
$$

We can perform straightforward integration over  the all variables
that are not  present in the Hamiltonians \eq{420}, \eq{421}, \eq{422}
\footnote{ All calculations are done up to a multiplication constant,
which can always be absorbed by the integration measure.}.
Then \eq{424} reduces to the product of two terms

\be 
 Z_D=I_DG_D, \qquad
\label{425}
\ee

where

\be 
G_D=\int DcDP^{(c)}D{\tilde c}DP^{({\tilde c})} exp(i\int_{T_1}^{T_2}
d \tau (P^{(c)}_m{\dot c}^m+P^{({\tilde c})}_m{\dot{\tilde c}}^m
-{1\over 2} P^{({\tilde c})}_mP^{(c)m})), \qquad
\label{426}
\ee

and $I_D$ includes the integrals over bosonic variables entering \eq{420},
\eq{421}, \eq{422} together with their conjugated momenta.
We use the method analogous to that in \cite{rivelles}
for computing these integrals.

The calculation of the ghost integral $G_D$ results in 

\be 
G_D=(\Delta T)^D, \qquad
\Delta T=T_2-T_1, \qquad D=3,4,6.
\label{427}
\ee

Let us demonstrate main steps of the $I_D$ calculation in the $D=3$
case

\be
I_3 = \int [ D\m DP^\m ]exp(i\int _{T_1}^{T_2}d\tau
(P_m^{(x)}{\dot x}^m+P_m^{(e)}{\dot e}^m+
P_{\a}^{(v)}{\dot v}^{\a} 
  -e^{m}(P^{(x)}_{m}-
{1\over 2}v^{\a}(\c _m)_{\a \b} v^{\b}))
\label{801}
\ee

Integration over $P_m^{(e)}$ and $P_m^{(v)}$
results in the functional $\delta$-functions $\delta ({\dot e}),~ \delta
({\dot v})$ which reduce functional integrals over $e^m$ and $v^{\a}$ to
ordinary ones:

\be 
I_3=\int DxDP^{(x)}d^{3}ed^{2}v~exp(ip_m\Delta x^m- \\
i\int _{T_1}^{T_2}d\tau (x^m{\dot P}^{(x)}_m+e^m(P^{(x)}_m-
{1\over 2}v^{\a}(\c _m)_{\a \b}v^{\b})),
\qquad 
\label{802}
\ee

where $\Delta x^m=x_2^m-x_1^m$ \eq{423}. Since the
integral over $v^{\a}$ is a usual Gauss integral after
integrating over $x^m$ and $v^{\a}$ one obtains

\be 
I_3=\int d^{3}pd^{3}e{1\over {\sqrt{e^me_m-i0}}}exp(i(p_m\Delta x^m-
e^mp_m\Delta T)).
\label{803}
\ee

In general case of $D=3$, 4 and 6 dimensions, one obtains

\be 
I_D=\int d^Dpd^De{{1\over{(e^me_m-i0)^{{D-2}\over 2}}}} exp(i(p_m\Delta
x^m- e^mp_m\Delta T)),
\label{428}
\ee

that can be rewritten as

\be 
I_D=\int
d^{D}pd^{D}e\int_0^{\infty}dc~exp(i(p_m\Delta x^m-e^mp_m\Delta T
+(e^me_m-i0)c^{2\over {D-2}})),
\qquad 
\label{804}
\ee

where $c$ is an auxiliary variable.

Integrating over  $p^m$ and $e^m$ one gets

\be 
Z_D=\int_{0}^{\infty}dc{1\over{c^{D/2}}}
exp(i{{\Delta x^m\Delta x_m}\over{2c}} -c0), \qquad D=3,4,6,
\label{436}
\ee

or
$$
Z_D={1\over {(\Delta x^m\Delta x_m-i0)^{{D-2}\over 2}}}, $$
which coincides with  the coordinate propagator
for the massless bosonic particle in the
standard formulation \cite{govaerts}.

On the other hand integrating \eq{428} only over $e^m$ we get
the massless bosonic particle causal propagator in the form $$
Z_D=\int d^Dp{1\over{p^mp_m+i0}}exp(ip_m\Delta x^m).
$$


\subsubsection*{\bf Comment on the $D=10$ case}

Above we have restricted our consideration to the space--time dimensions
3, 4 and 6. The case of a bosonic twistor--like particle in $D=10$ is
much more sophisticated. The Cartan--Penrose representation of a
$D=10$ light--like momentum vector is constructed out of a
Majorana--Weyl spinor $\lambda^\a$ which has 16 independent components

\be
P^m=\l\Gamma^m\l.
\label{pc}
\ee

Transformations of $\l^\a$ which leave \eq{pc} invariant take values on
an $S^7$-- sphere (see \cite{EiSo,nispach,bfortschr}
and references therein). In contrast to
the $D=4$ and $D=6$ case, where such transformations belong to the
group $U(1)\sim S^1$ \eq{224} and $SU(2)\sim S^3$ \eq{233},
respectively, $S^7$
is not a Lie group and its corresponding algebra contains structure
functions instead of structure constants. Moreover, among the 10
constraints \eq{pc}  and 16 constraints $P^{(\l)}_\a=0$ on the
momenta conjugate to $x^m$ and $\l^\a$ $18=10+16-1-7$ (where 7 comes from
$S^7$ and 1 corresponds to local $\tau$--reparametrization) are of the
second class. They do not form a representation of the Lorentz group
and cause the problem for covariant Hamiltonian analysis.

One can overcome these problems in the framework of the
Lorentz--harmonic formalism (see \cite{bzstr,bpstv} and references 
therein), where to construct a light--like vector one introduces eight 
Majorana--Weyl spinors instead of one $\l^\a$. Such a spinor matrix
takes values in a spinor representation of the  double covering group
$Spin(1,9)$ of $SO(1,9)$ and satisfies second--class harmonic
conditions. The algebra of the constraints in this ``multi--twistor"
case is easier to analyze than that with only one commuting spinor
involved. The path--integral
BRST quantization of the $D=10$ twistor--like particle is in progress.

In the present section the BRST--BFV quantization
of the dynamics of massless bosonic particle in $D=3,4,6$
was performed in the twistor--like formulation.
To this end the initially mixed system of the first-- and
second--class constraints was converted into the system of
first--class constraints by extending the initial phase space
of the model with auxiliary variables in a Lorentz--covariant way.
The conversion procedure (rather than having been a formal trick)
was shown to have a meaning of a symmetry transformation
which relates different twistor--like formulations of the bosonic particle,
corresponding to different gauge choices in the extended phase space.

We quantized the model by use of
the extended BRST--BFV scheme for the path--integral quantization.
As a result we have
presented one of the numerous proofs of the equivalence between
the twistor--like and conventional formulation of
the bosonic particle mechanics.

This example demonstrates peculiar features of treating the
twistor--like variables within the course of the covariant
Hamiltonian analysis and the BRST quantization,
which one should take into account when studying more
complicated twistor--like systems, such as superparticles and
superstrings.

In the next section we will use methods developed in this section
for the case of quantization of superparticle with central charge
coordinates.

\bigskip



\section{\bf Quantization of the massive/massless particle with central charge
coordiantes }

  In this section we will discuss quantization of the model described by
Lagrangian \eq{na} in its spinorial form, i.e. \eq{L} as in \cite{RSS-new}.
This quantization is
different from one considered before \cite{BLS} in the sense that it is possible
to have universal model which includes massive as well as massless cases.
Usually quantization of massive particle is quite different from massless one,
for example for ordinary Brink-Schwarz superparticle massless model possess $\k$
symmetry but massive does not.  That is why in the massive case one
does not have
mixture of first and second class constraints. Therefore, for massive
case,  it is much
easier to
quantize the system and the problem of infinite reducibility of $\k$
symmetry
does not appear. In the Universal model described by \eq{na} in most
general case massive superparticle can have $\k$-symmetry which is given by 
\eq{nk}. The only difference in massive and massless case is that the
number of
$\k$ symmetries is different and is given by the rank of
$P_{\a\b}$. First, we will
start from Hamiltonian analysis of constraints and then proceed to Dirac
quantization. It is also possible to consider covariant BRST quantization but in
this section we will omit it.
Introduce $\hat P^{\a\b}$ as momentum  conjugated to $e_{\a\b}$, and $P^\a$ as
momentum for $\theta_\a$. The constraints for the action 

 \be
I= \int d\tau \left( P_{\a\b}\, \Pi^{\a\b} 
+\ft12 e_{\a\b}\, (P^2)^{\a\b} \right) \ ,
\la{na11}
\ee

are

\be
\hat P^{\a\b} = 0, \qquad (P^2)^{\a\b} = 0,\qquad \Psi_\a = P^\a + P_{\a\b}\theta^\b = 0.
\la{constr}
\ee

The first two are bosonic first-class  constraints and the last one is
fermionic
mixture of first and second class  constraints. Using usual terminology we
call constraint of a first
class if Poisson brackets  of it with all other constraints give either
zero or another constraint
otherwise constraint is of the second class. Second constraint in \eq{constr}
does not mean that
particle is massless. It could be massive and solving
$P^2_{\a\b} = 0$
give equations connecting components of $P_\m$ and tensorial momenta.
To see that start from 

\be
\{ \Psi_\a , \Psi_\b \} = 2P_{\a\b}.
\ee

 Then rank of $P_{\a\b}$ is equal to number of second class constraints
and
$(N-rank(P))$ is number of first class constraints, which  generate $\k$
- symmetry.
The problem of quantization is not only existence of second class constraints, 
but also  covariant separation of first class from second one.
The same situation appears for ordinary superparticle. To solve this problem one
can make a change of variables and consider system classically  equivalent
to previous one.
To start with let's introduce s commuting spinors $\l^i_\a$ where  i runs
from 1 to s, $\a = 1,...,N$
and $s \leq N/2$.
Then $P_{\a\b}$ can be chosen directly in form satisfying second  
equations in \eq{constr}:

\be
P_{\a\b} = \l^i_\a \l^i_\b,
\la{Pl}
\ee

where $\l^i_\a \l^{j \a}=0$.
 In this case rank of P is equal to s. Equation \eq{Pl} is generalization
of Penrose twistor decomposition
of massive momentum in four dimensions. Also it is possible to connect
$\l$ to spinorial Lorentz harmonics as it was shown in \eq{lvv}.

Using \eq{Pl}, the Lagrangian of superparticle is taking the form \eq{L}:

\be
L=\l^i_\a \l^i_\b \Pi^{\a\b},
\la{L1}
\ee

where  $\Pi^{\a\b} = dX^{\a\b} - \t^{(\a} d\t^{\b)}$. This Lagrangian
describes  massive as well as 
massless  superparticle with central charge coordinates in any dimension
and it's much easier to 
quantize it.

The constraint system for Lagrangian \eq{L1} is

\be
P^i_\a = 0, \qquad \phi_{\a\b} = P_{\a\b}-\l^i_\a \l^i_\b =0,\qquad
\Psi_\a = P_\a - P_{\a\b}\theta^\b=0,
\la{consl}
\ee

where all constraints are mixture of first and second class. After
studying 
properties of constraints we can proceed to quantization of the model.

To quantize the model of superparticle with central charge coordinates
 \eq{na} which is described by the constraints \eq{consl} we will use the method
 of conversion ,  see \cite{conv1} for recent review and geometrical
description of
the conversion . The problem with quantization of the system \eq{consl} is the
 following. The constraint system  is a mixture of the
first and
second class
 constraints and it's impossible to separate them covariantly. That is why
we
 prefer to convert them to the first class constraints only. 
 
  Usually, conversion is applied for the second
class constraints only but  the conversion in mixture could be
quite useful, see \cite{BMRS} and references there in.
The meaning of conversion  is that one extends phase space of the theory
by introducing 
additional variables, number
of which should coincide  with number of new first class constraints. Then
starting from extended 
phase space, one can reduce this extended space 
to one  equivalent to initial phase space. Therefore, as a result, all
constraint are going 
to be of the 
first class, moreover, some variables will not participate in
the theory. 
 Or, geometrically, 
\cite{conv1} one  start from symplectic manifold $ M$ which can be
represented as some 
fibre bundle $U$ restricted by the second class constraints condition. 
Then, to convert second class constraints into the first one we extend
phase space 
to direct sum of $U$ and tangent bundle $T M$. Now all
  constraints are of the first class . The  conversion appeared extremely
effective not
 only in particle models but also in the case  of field theories in
different dimensions as well as for
  strings and membranes in B-field \cite{R} and references therein.

  To convert
the system of constraints into first class we introduce additional  
bosonic variables
$\eta_\a ^i$ and fermionic ones  $\z^i$ such that

\be 
P^{(\l)i}_\a - P^{(\eta)i}_\a = 0,
\la{Peta}
\ee

\be
\hat \phi_{\a\b} = P_{\a\b}-(\l^i_\a  + \eta^i_\a) (\l^i_\b + \eta^i_\a) =0,
\la{phi}
\ee

\be
\hat \Psi_\a = P_\a - P_{\a\b}\theta^\b + \z^i(\l^i_\a  + \eta^i_\a)=0,
\la{Psi}
\ee

where $\{\z^i,\z^j \} = 2\d^{ij}$.
We see that in set of all constraints \eq{Peta} - \eq{Psi} momenta,
corresponding to $(\l - \eta)$ are all vanishing
and all other functions depend only $\tilde \l = \l + \eta$. That is why
one can reduce extended phase space with remaining coordinates $\tilde\l$, 
 $X$, $\t$, $\z$ and their conjugate momenta.
 If one has odd number of grassmanian variable it is worth to consider
$\z$
as matrix acting on the raw of wave functions \cite{BLS}.
Here we will slightly change this procedure to make it work for set of
spinors.
Therefore the wave function depends only on

\be
\Psi = \Psi(X^{\a\b},\tilde\l^i_\a,\t^\a,\z^i).
\la{wf}
\ee

Now we consider wave function to be a two-component: $\Psi = (\phi_1,
\phi_2)$. Then constraints acting on two component spinorial wave function
are

\be
\hat \phi_{\a\b} = (P_{\a\b}-\tilde\l^i_\a \tilde\l^i_\b) \times I =  0,
\la{phi1}
\ee
 
\be
\hat \Psi_\a = (P_\a - P_{\a\b}\theta^\b) \s^3 + (\z^i \tilde\l^i_\a)\s^1 
=0,
\la{Psi1}
\ee

where $\s$ are Pauli matrices and $I$ is unit matrix.

Imposing the first class constraints on the wave function

\be
\Bigl( {{\partial}\over{\partial X^{\a\b}}} - i\tilde\l^i_\a \tilde\l^i_\b\Bigr)
\times I  \Psi  = 0,
\la{wepsi}
\ee

\be
\Bigl( ({{\partial}\over{\partial\theta^\a}} - \t^\b
{{\partial}\over{\partial X^{\a\b}}})\s^3 
 - i\z^i\tilde\l^i_\a \s^1 \Bigr) \Psi = 0.
 \la{wepsi2}
 \ee

 The first equation can be solved as

\be
\Psi = e^{i\tilde\l^i \tilde\l^i X}\Phi(\tilde\l^i, \t, \z^i).
\la{phi-new}
\ee

 Using notation $\n^i = \l^i_\a \t^\a$   we can use $\tilde \Phi$ instead
of $\Phi$ as
 
\be
\tilde\Phi = \tilde\Phi(\n^i,\l^i_\a, \z^i).
\la{tilde}
\ee

Then equations imposed on $\tilde\Phi$ part of the wave function is

\be
\Bigl( {1\over 2} ({{\partial}\over{\partial\n^i}} - \n^i) \s^3 +
(\z^i)\s^1  \Bigr)\tilde\Phi = 0,
\la{eqfi}
\ee

where $\Phi = (\Phi_1,\Phi_2)$.
Then $\Phi_2$ can be solved in terms of $\Phi_1$ as

\be
\Phi_2 = \frac{1}{s}\n_i D^i \Phi_1
\la{Phi-1-2}
\ee

where $D^i = {{\partial}\over{\partial\n^i}} - \n^i$.

Because of arbitrary dependence of wave function on $\l$ one can decompose

\be
\tilde\Phi(\l^i_\a, \n^i)=\sum\nolimits_{k=1}\l^{i_1
\a_1}...\l^{i_k \a_k} \Phi_{\a_1 i_1...\a_k i_k}(\n^i).
\la{decomp}
\ee

It was shown \cite{BLS} that if one considers  reduction to a system
without central-charge coordinates,
then there is additional constraint on $\Psi$  which includes derivatives
of $\l$ and leads to cutting of the
infinite spectrum to states having spin 0 or 1/2 corresponding to ordinary Brink-Schwartz superparticle.
 It is useful to go to Weyl spinors from the Majorana ones. One can do that in
 even dimensions. 
 
 Consider case of $D=4$. In this dimension Majorana
spinor $\l_{\a}$ can be solved as pair of Weyl spinors $(\l_A, \bar\l^{\dot
A})$. In this case all terms in \eq{decomp} with $\l\bar\l$ components can
be expressed as $p_\m$, using Penrose identity $P_{A\dot A} = \l^i_A
\bar\l^i_{\dot A}$. Then the general solution \eq{decomp} is taking a
form

\bea
\tilde \Phi ( {\l_{A}}^i, \bar\l_{\dot A}^i ) & = &
\Phi_0(P_\m)+\sum\nolimits_{k=1}\l^{i_1 A_1}...\l^{i_k A_k}
\Phi_{A_1 i_1...A_k i_k}(P_\m, \n^i) \nonumber \\
&+& \sum\nolimits_{k=1}\bar\l^{i_1
\dot A_1}...\bar\l^{i_k\dot A_k}\Phi_{\dot A_1 i_1...\dot A_k i_k} 
(P_\m,\n^i)
\la{decomp1}
\eea
 
Using generalized Penrose decomposition we
acquire additional $U(1)$ and $SO(s)$ symmetry, because combination 
$\l^i_A \bar\l^i_{\dot A}$ is invariant under
those symmetries.

 Let us start from the case $ N=1$ and $D=4$. In this case the spectrum of
the particle
includes the following components of wave function:

\be
\Phi_0, \qquad \ \Phi_{A_1}, \qquad \Phi_{A_1 A_2}, \qquad \Phi_{A_1 A_2 A_3},
\qquad \Phi_{A_1 A_2 A_3 A_4}, \qquad ... \qquad,
\la{spectr}
\ee

and their complex conjugate, where all indices are symmetrized. 

One of the most important
 issues here is the absence of the mixed components, i.e. $\Phi$ with
dotted and undotted spinorial
indices.  It happens because every time we see such a combination we can
combine the $\l$'s together
to produce momentum $p_{\m}$. 
We have not consider any interaction yet that is why we can discuss only
linearized field
theory and in our spectrum we have infinite number of fields.
 
 Equations of motion come from the imposing the constraint \eq{wepsi} on the
wave function.  
But first, let us decompose each Majorana index into pair of Weyl
indices, i.e. $\a = (A, \dot A)$. Applying this decomposition to an
equation
\eq{wepsi} gives as one of the components of this equation

\be
 \biggl  ( \frac {\partial}{\partial X^{\dot A B}} - i \bar\l_{\dot A}
\l_B \biggr ) \Phi = 0,
 \la{EqMot}
 \ee
 
 Here we do not want to solve this equation as we did in \eq{phi} we would
rather following
 \cite{EiSo}
consider  $\Phi$ as function of $X$ and in the same time multiply 
  \eq{EqMot} by $\l^B \l^C \l^D...$ or their complex conjugated.
The resulting equations of motion are
 
 \be
 \frac {\partial}{\partial X^{\dot A B}} W^{B C D...  } = 0,
 \la{lll}
 \ee
 
 where $\l^B \l^C... \Phi = W^{B C D... }$.

Why does this method works? Every time when we multiply $\Phi$ by $\l$ or
by
$\bar\l$ the new field which we call $W$  depends on $\l$ through
only
combination of $\l\bar\l$ that is  $P$. Then $W$, and not
$\Phi$, are playing role of physical fields. Those fields correspond to
the supersymmetric field theory and equations \eq{lll} are linearized
equations of motion of that field theory.

Let us compare our result with result of \cite{EiSo}.
In the limit of zero brane-charges there is additional constraint in the
form of 

\be
 ( \l P^{\l} - \bar\l \bar P^{\l} + c) \Phi  = 0,
\la{lplpc}
\ee

where $c$ is a constant that takes positive or zero integer values up to
4.
If $c=1$ then solution of this constraint could be written as

\be
\Phi = \l^A A_A (p,x,... ),
\la{PhilApx}
\ee

where function $A_A$ does not depend on $\l$ anymore.
Then imposing constraint \eq{lll} give the following equation of motion.

\be
\p_{\dot B A} A^A = 0.
\la{spin1/2Eso}
\ee

This equation corresponds to the linearized field equation for spin 1/2  
field. In our case we do not have constraint \eq{lplpc}. Therefore, 
as we discussed before, we can have arbitrary number of $\l$'s in  $\Phi$
decomposition. To obtain $W$ fields that do not depond on $\l$ one has to
multiply $\Phi$ by different number of spinors each time and then use
equation \eq{wepsi} for each term with different number of $\l$'s
independently.

  So far we considered the four-dimensional case of the minimal
 supersymmetry. It is interesting to investigate what happens in $D=11$.
 In eleven dimensions we do not have Weyl spinors and all spinorial
indices are
 Majorana. The equation \eq{decomp} is valid for all dimensions and 
 arbitrary number
 of supersymmetries. In four dimensions it was easy to separate four-momentum
 $P$ from spin degrees of freedom described by the $Z_{\m\n}$ by using language
 of the Weyl spinors and correspondence $P_{A \dot A} = \l_A \l_{\dot A}$.
 In eleven dimensions this correspondence is a little more complicated, first
 because to express $P$ in terms of $\l$'s we have to use 
 
 \be
 P_{\m} = \l^{\a} (\C_{\m})_{\a\b} \l^{\b}.
 \la{P11}
 \ee
 
 Moreover, in the superalgebra we have not only two form brane-charge  but also
 self-dual five-form charge. In this case spectrum of the theory could be obtained by
 decomposition of all pairs of $\l$ in terms of $\c$-matrices. Here we have to use
  Fierz identities and combine all terms with $P$ into the wavefunctions 
  to exclude them from
 spectrum.  We saw how it works for the case of mixed components in four dimensions.
The more detailed treatment of the eleven-dimensional case and connection with
linearized M-theory will be given in the next section.

\bigskip


\section{\bf On spectrum of linearized M-theory }

It is known that M-theory is given by strong coupling limit of the type IIA string
theory. Interesting way to investigate the properties of the M-theory is to start from
the M-superalgebra, i.e. with eleven dimensional superalgebra with two and five form 
charges \cite{M}. Those higher form charges can describe 
nonperturbative
objects such as p-branes in eleven dimensions. M-theory also could be thought as Membrane
theory using parallels with string field theory.
In this work we use alternative interpretation of brane charges as a coordinates
of extended supermanifold for a superparticle model.
Here we argue that brane charges charges could play
an important role in identifying the properties of the perturbative spectrum of 
linearized M-theory.
 We can use connection between  quantized particle mechanics and  underlined field theory
 almost by the same way as was explained in \cite{green}. In this work,
  starting from the action of
 the eleven dimensional superparticle, linearized equations of motion
 of the eleven dimensional supergravity were obtained.
  In that case target space of the particle
was eleven dimensional supermanifold parametrized by $x$ and $\t$.
Here we propose that the quantization of eleven dimensional superparticle 
  which live on the supermanifold
 parametrized not only by ordinary coordinates but also by higher rank p-forms (two and five
 forms in this case) helps us investigate properties of linearized M-theory.
The supersymmetry of our superparticle model is defined by the full M-theory 
 superalgebra. It gives upon quantization not
 only spectrum of the field theory but also equations of motion for these 
 fields \cite{RSS-new}.
 Because there is only one field theory in eleven dimensions which is described by the
 superalgebra with two and five form charges, i.e. M-theory, we argue that the particle
 spectrum could be compared with the perturbative spectrum of the linearized M-theory.

   First let us recall ordinary connection between quantized particle mechanics and field 
 theory.

\be
\int Dx D\l D\theta exp\{iS(x,\l,\t)\} = \int D\phi \bar\phi \phi exp\{iS(\phi)\},
\la{x-phi}
\ee

where the left hand side describes the propagator of the superparticle and the right hand 
side gives the propagator of the field theory. The other formulation of this correspondence
could be expressed through identification of the wave functions of the quantized particle
mechanics and fields in the equivalent field theory. Then constraints imposed on particle
wave function become equations of motion of the field theory exactly in the same fashion as
appearance of Klein-Gordon equation in ordinary relativistic quantum mechanics.

    Let us start with \eq{decomp}

\be
\tilde\Psi(x,\l^i_\a, \n^i)=\sum\nolimits_{k=1}\l^{i_1
\a_1}...\l^{i_k \a_k} \Psi_{\a_1 i_1...\a_k i_k}(x, \n^i),
\la{decomp1-1}
\ee

which is the solution of \eq{wepsi} and \eq{wepsi2}. This spectrum describes massless as
well as
massive particle in arbitrary dimension. To have massless eleven dimensional model we
fix $i=1$ and $\a = 1,...,32$ is Majorana index of spinor representation of $SO(10,1)$.
In this case the wave function is taking the form:

\be
\tilde\Psi(x,\l_\a, \n^i)=\sum\nolimits_{k=1}\l^{
\a_1}...\l^{ \a_k} \Psi_{\a_1 ...\a_k }(x, \n^i).
\la{decomp2-2}
\ee

The right hand side of this equation gives spectrum of the correspondent field theory
where all fields are symmetric in all $\a$'s . The equations of 
motion for those fields are coming from the constraint \eq{wepsi} imposed on $\Psi$
by the same way as we saw in  previous sections.
 We have to notice that by this method one can
obtain only the spectrum of linearized and perturbative field theory.
But one more interesting observation could be found. Here we interpret two and five 
form charges as coordinates of the superparticle. 
On the other hand  tensorial charges could be described
by the brane charges, i.e. by the integrals of the brane currents on the worldvolume
of the membrane and five-brane correspondently. Comparing two different interpretations,
it should be connection between membrane spectrum and spectrum of the massive
particle defined on the M-algebra target-space. But in this case in \eq{decomp1-1} 
$i_k$ should  ran from 2 to 16 (as for massive case), describing particles with different
masses. It is not quite clear if one has to identify model with fixed $i$ or some of the
models with different $i$'s.
  
 To describe how to include momentum $P$ in $\Psi$ let us start from term in \eq{decomp2-2}
with $2N$ $\l$'s for arbitrary $N$. Then using definition

\be
\l_\a \l_\b = P_{\a\b} + Z^{(2)}_{\a\b} + Z^{(5)}_{\a\b},
\la{lP25}
\ee

where 

\be
P_{\a\b} = (\c^\m)_{\a\b} P_\m, \qquad Z^{(2)}_{\a\b} = (\c^{\m \n})_{\a\b} Z^{(2)}_{\m \n},
\qquad Z^{(5)}_{\a\b} = (\c^{\m_1 ... \m_5})_{\a\b} Z^{(5)}_{\m_1 ...  \m_5},
\la{def-PZZ25}
\ee

it is possible to write decomposition of each pair of $\l$'s 
using symmetry in $\l$ interchange.

\bea
\l^{\a_1} ... \l^{\a_{2N}} \Psi_{\a_1,...,\a_{2N} } & = & 
(P_{\a_1 \a_2} + Z^{(2)}_{\a_1 \a_2} + Z^{(5)}_{\a_1 \a_2} ) ... \nonumber \\
&& (P_{\a_{2N-1} \a_{2N} } + Z^{(2)}_{\a_{2N-1} \a_{2N} } + Z^{(5)}_{\a_{2N-1}  \a_{2N} } )
\Psi_{\a_1,...,\a_{2N} }.
\la{lPZZ-6}
\eea

Now if one uses contraction with $P$ it is convenient to define

\be
P \Psi^{(2r)} = \Psi^{(2r-2)} (p),
\la{PsiPr}
\ee

where $r=1,..,N$ and $\Psi^{(2r)}$ denotes components of  $\Psi$ with $2r$ spinorial
indices. In this case one can absorb all momenta $P$ in definition of $\Psi (p)$.
After summation of all possible $N$ one has the general decomposition of $\tilde \Psi$

\bea
\tilde \Psi (x, \l_\a, \n) & = & \Psi_0 (x, p, \n) + (Z^{(2)} + Z^{(5)} ) \Psi^{(2)} (x,p,\n)
+ ... \nonumber \\
 &+& (Z^{(2)} + Z^{(5)} )^N  \Psi^{(2N)} (x,p,\n) + ... ,
\la{tPsifin}
\eea

where $\Psi$ in right hand side with different number of spinorial indices do not
depend on $\l$ rather all $\l$ dependence is encoded in 
$Z^{(2)}_{\a\b}$ and $Z^{(5)}_{\a\b}$.
 On the
other hand all $P$ dependence is included into the definition of $\Psi (x, p, \n)$.
We see that each component of $\Psi$ in right hand side of \eq{tPsifin} is symmetric
in their spinorial indices.
Now we have fields that depend on $(x, p, \n)$. To make an direct connection with a field
theory one has to integrate $\Psi (x, p, \n)$ over all momenta with appropriate measure.
The integration gives fields of linearized M-theory.
In this example we considered only one case of massless spectrum of linearized M-theory
that corresponded to $i=1$. We have to mention that this particular case covers only subset
of linearized M-theory spectrum. The other values of $i$ must be included.
In most general case we will have fields with $i=1,...,16$ in the form of 
\eq{decomp1-1}. Those fields for different $i$ could be massive as well as massless.
One has to apply the same methods to absorb $P$ into the wave function for the case
of arbitrary $i$. The set of  fields for all $i=1,...,16$ for massive as well as massless
case gives us spectrum of Linearized M-theory. 
Here we see, that the  linearized M-theory must include massless as well as massive fields.
Equations of motion for those fields could be also produces by 
multiplying wave function by some combinations of $\l$'s and then applying \eq{wepsi}.

 In this section, we used an analogy with quantization of ordinary superparticle. 
This quantization produces spectrum and linearized equations of motion 
of ordinary supergravities and
supersymmetric Yang-Mills in different dimensions. Here we can conclude that quantization
of superparticle with brane charges in eleven dimensions produces perturbative
 spectrum of linearized M-theory given by \eq{tPsifin}.

    To make a connection with the representations of M-superalgebra that were discussed in
 Chapter 2 we have to clarify the following apparent contradiction. In the second Chapter
 we claimed that in $D=11$ for some particular values of brane-charges one can have other multiplets,
with the highest spin lower then two, 
apart from supergravity multiplet. On the other hand 
equation \eq{tPsifin} gives infinite set of fields of arbitrary spin
in the spectrum of linearized
M-theory. How can we explain this conflict? First of all, as it was mentioned in the 
second chapter, 
 the states of the supermultiplet depend on the values of extra Casimir operators
made from two and five-form charges. Therefore, in field theory realization of this 
multiplet,
fields will also depend on two and five-form charges playing role of extra coordinates of
extended superspace. However, to make a connection with ordinary field theory, where
all fields depend only $x$ and $\theta$ one needs to decompose each field in powers of
$Z^{(2)}_{\a\b}$ and $Z^{(5)}_{\a\b}$ as in equation \eq{tPsifin} then one sees that
this procedure generates infinite number of symmetric tensors with spinorial
indices which fall into the linearized M-theory multiplet.

\bigskip
\newpage





\chapter{\bf Superbranes}














\section{\bf L-branes}

\subsection{\bf Introduction}
In previous chapters we mostly considered 0-branes and only briefly 1-brane,
i.e. string. In this chapter we investigate a connection between Superalgebras
realization and geometry of the super p-branes. In some sense the realization of
superalgebra on the 0-brane give us some information about geometry of objects
embedded into the target space, but for the case of particles this geometry  was
almost trivial because the particle world-sheet is one-dimensional.
For the case of higher p-branes when world-sheet is is $p+1$ dimensional it is
important to understand how to embed the super world-sheet of the brane in the
super target space. Because we want to make a connection with Superalgebra
realization, we have to consider embedding of one super-object, which is
realization of world-sheet superalgebra into the target space, that gives 
the target space realization of the superalgebra.
We will start from brief introduction to embedding formalism and then proceed
to the using an embedding to identify new objects such as L-branes
\cite{L-brane}.

In recent years there has been renewed interest in superstring theory as
a candidate theory that unifies all the fundamental forces in nature.
The interest was sparked by the realization that the five different
consistent ten-dimensional superstring theories are in fact related to
each other by duality transformations. Furthermore, it is believed that
they are related to a new theory in eleven dimension which has been
called M-theory. A crucial r\^{o}le in this development has been played
by the soliton solutions of superstrings and eleven-dimensional
supergravity which correspond geometrically to multi-dimensional objects
present in the spectra of these theories. These are generically referred
to as $p$-branes.

One way of studying the dynamics of $p$-branes is to use the theory of
superembeddings \cite{se1, se2, hs}. In this framework the worldvolume
swept out by a $p$-brane is considered to be a subsupermanifold of a
target superspace. A natural restriction on the embedding gives rise to
equations which determine the structure of the worldvolume
supermultiplet of the $p$-brane under consideration and which may also
determine the dynamics of the brane itself. There are several types of
worldvolume supermultiplets that can arise: scalar multiplets, vector
multiplets, tensor multiplets which have 2-form gauge fields with
self-dual field strengths, and multiplets with rank 2 or higher
antisymmetric tensor gauge fields whose field strengths are not
self-dual \cite{hs}. Although this last class of multiplets can be
obtained from scalar or vector multiplets by dualisation (at least at
the linearized level) it is often the case that the version of the
multiplet with a higher rank gauge field is more natural in a given
geometrical context. For example, D-branes in type II string theory have
worldvolume vector multiplets but, in the context of D$p$-branes ending
on D$(p+2)$-branes, the dual multiplet of the latter brane occurs
naturally \cite{chs1}.

In this chapter we discuss a class of $p$-branes whose members are
referred to as L-branes \cite{L-brane}. By definition, these branes are those which
have worldvolume supermultiplets with higher rank non-self-dual tensor
gauge fields which are usually referred to as linear multiplets, whence
the appellation. According to \cite{hs} there are two sequences of
L-branes: the first has as its members a 5-brane in $D=9$, a 4-brane in
$D=8$ and a 3-brane in $D=7$ which all have eight worldvolume
supersymmetries, while the second has only one member, the 3-brane in
$D=5$ which has four worldvolume supersymmetries. These sequences and
their worldvolume bosonic field contents are tabulated below, where $A_p$
denotes a $p$-form potential and $(S,T)$ are auxiliary fields.

\begin{center}
\begin{tabular}
{|c|c|c|l|}
\hline
8 world susy & \ \ L5-brane  & \ \  D=9 &\ \ $3\phi, \ A_4$  \\
         & \ \ L4-brane  & \ \  D=8 &\ \ $3\phi, \ A_3, S $  \\
         & \ \ L3-brane  & \ \  D=7 &\ \ $3\phi, \ A_2, S,T$  \\
\hline
4 world susy  & \ \ L3-brane  & \ \  D=5 &\ \ $\phi,  \ A_2$  \\
\hline
\end{tabular}
\end{center}

The 3 and 4-branes of the first sequence can be obtained by double
dimensional reduction from the first member of the sequence, namely the
5-brane in $D=9$, and the latter can be interpreted as arising as a
vertical reduction of the geometrical sector of the heterotic/type I
5-brane, followed by the dualisation of the scalar field in the
compactified direction. By the geometrical sector we mean the sector
containing the worldvolume fields corresponding to the breaking of
supertranslations. The relevant target space field theory in this
context is the dimensional reduction of the dual formulation of $N=1,
D=10$ supergravity followed by a truncation of a vector multiplet. The
L5-brane is expected to arise as a soliton in this theory. Another
possible interpretation of this brane is as the boundary of a D6-brane
ending on a D8-brane. In this case the target space geometry should be
the one induced from the embedding of the D8-brane in type IIA
superspace. However, in this section we shall take the target space to be
flat $N=1, D=9$ superspace for simplicity. The generalization to a
non-trivial supergravity background is straightforward while the
generalization to an induced D8-brane background geometry is more
complicated and would require further investigation.

The L3-brane in $D=5$ also has two possible interpretations. On the one
hand it may arise as the geometrical sector of a soliton in
heterotic/type I theory compactified on $K3$ to $D=6$, vertically reduce
to $D=5$ and then dualised in the compactified direction. Alternatively,
it could be related to the triple intersection of D-branes \cite{bdr}
over a 3-brane with one overall transverse direction. This brane will be
studied in \cite{cod1} as an example of a brane of codimension one.

A feature of L-branes is that their worldvolume multiplets are off-shell
multiplets in contrast to many of the branes that have been studied
previously such as M-branes and D-branes. In particular, this is true
for the worldvolume multiplets of the first sequence of L-branes, even
though their dual versions involve hypermultiplets which are unavoidably
on-shell. The standard embedding constraint does not lead to the
dynamics of L-branes and imposing the Bianchi identity for the
worldvolume tensor gauge field does not change the situation. As a
consequence the equations of motion of such branes have to be determined
by other means, either by directly imposing an additional constraint in
superspace or by using the recently proposed brane action principle
which has the advantage of generating the modified Born-Infeld term for
the tensor gauge fields in a systematic way \cite{hos}. We note that the
heterotic 5-brane in $N=1, D=10$, which is related to the L5-brane in $D=9$ as
we described above, would normally be described by a worldvolume hypermulitplet
which is on-shell. However, as noted in \cite{hs}, one can go off-shell using
harmonic superspace methods. This has been discussed in detail in a recent
paper \cite{bik}.

In the foregoing discussion we have assumed throughout that the target space
supersymmetry is minimal. It is possible to relax this. For example, one can
obtain an L2-brane  in an $N=2, D=4$ target space by double dimensional
reduction of the L3-brane in $D=5$. This brane and the non-linear dynamics of
the associated linear multiplet has been studied from the point of view of
partial breaking of supersymmetry in reference \cite{bg}.

The organization of the section is as follows: in the next subsection we give
a brief introduction to the theory of superembeddings; then we study
the L5-brane in $D=9$ at the linearized level; next the torsion
and Bianchi identities are solved in the non-linear theory; after that
we construct the action and in section 6 we derive the Green-Schwarz
equations of motion and determine how these can be expressed in
superspace. Finally the L3-brane in $D=7$ and L4-brane in $D=8$ are
studied.


\subsection{\bf Superembeddings}


In the superembedding approach to $p$-branes both the target
space and the worldvolume swept out by the brane are superspaces. This
is different to the Green-Schwarz formalism where only the target space
is taken to be a superspace while the worldvolume is purely bosonic. The
local $\k$-symmetry of the GS formalism can be understood as arising
from the local supersymmetry of the worldvolume in the superembedding
approach upon gauge-fixing.

The geometric principles underlying the superembedding approach were
given in \cite{hsw}. The embedding $f:M\rightarrow \unM$, which maps the
worldvolume $M$ into the target superspace $\unM$, is chosen to break
half of the target space supersymmetries so that the fermionic dimension
of $M$ must be chosen to be half the fermionic dimension of $\unM$. More
general embeddings which break more of the supersymmetries are possible
but will not be considered here. We adopt the general convention that
worldvolume quantities are distinguished from target space quantities by
underlining the latter. Coordinates on the worldvolume (target space)
are denoted by $z^M=(x^m,\th^{\m})$ and $z^{\unM} = ({x}^{\unm} ,
{\th}^{\umu})$ respectively; the tangent bundles are denoted by $T$
($\unT$) and local preferred bases are denoted $E_A=(E_a,E_{\a})$ and
$E_{\unA} = (E_{\una} , E_{\ua}$). The associated cotangent bundles
$T^*$ ($\unT^*$) are similarly spanned by the dual basis one-forms
$E^A=(E^a, E^{\a})$ and $E^{\unA} = (E^{\una}, E^{\ua}$). The target
space supervielbein $E_{\unM}{}^{\unA}$ and its inverse
$E_{\unA}{}^{\unM}$ are used to change from a preferred basis to a
coordinate basis. The worldvolume supervielbein and its inverse are
similarly denoted, but without underlining of the indices. In the
foregoing Latin indices are bosonic and Greek indices are fermionic.

The embedding matrix $E_A{}^{\unA}$ specifies the relationship between
the bases on $T$ and $\unT$

\be
E_A = E_A{}^{\unA}E_{\unA}\ .
\label{1}
\ee

Expressed in local coordinates the embedding matrix is given by

\be
E_A{}^{\unA} = E_A{}^{M}\p_{M}z^{\unM}E_{\unM}{}^{\unA}\ .
\label{2}
\ee

The basic embedding condition is that the purely fermionic part of $T$
is determined only by the pullback of the purely fermionic part of
$\unT$ and does not involve the pullback of the bosonic part of $\unT$.
This means that the embedding matrix should satisfy the constraint

\be
E_{\a}{}^{\una} = 0\ .
\label{3}
\ee

This basic embedding constraint determines the supermultiplet structure
of the brane and in many cases will also be enough to put the brane
on-shell. In some cases, however, an additional constraint involving
forms on the worldvolume and the target space is necessary. The
L-branes discussed in this paper are particularly interesting in this
regard because the embedding constraint \eq{3} is not sufficient to
determine the dynamics of these branes so that the equations of motion
must be derived from additional constraints or actions.

In order to make further progress it is convenient to introduce the
normal tangent bundle $T'$ which has a basis denoted by $E_{A'}=(E_{a'},
E_{\a'})$. This basis is related to the basis of $\unT$ by the normal
matrix $E_{A'}{}^{\unA}$. Note that normal indices are distinguished from
tangent indices by primes. There is considerable freedom in the choice
of the components of $E_{A}{}^{\unA}$ and $E_{A'}{}^{\unA}$. A simple and
convenient choice is

\be
\ba{rclrcl}
E_{a}{}^{\una} & = & u_{a}{}^{\una} &\qquad\qquad
E_{\a}{}^{\una}& = & 0 \\
E_{a}{}^{\ua} &= &\L_{a}{}^{\b'}u_{\b'}{}^{\ua}&
E_{\a}{}^{\ua} & = & u_{\a}{}^{\ua} + h_{\a}{}^{\b'}u_{\b'}{}^{\ua}\\
&&&&&\\
E_{a'}{}^{\una}& = & u_{a'}{}^{\una}&\qquad\qquad
E_{\a'}{}^{\una}&=&0 \\
E_{a'}{}^{\ua}&=&0&
E_{\a'}{}^{\ua} & = & u_{\a'}{}^{\ua}\ .
\ea
\label{4}
\ee

where $\L_{a}{}^{\b'}$ is an arbitrary vector-spinor. The matrices
$u_{a}{}^{\una}$ and $u_{a'}{}^{\una}$ together make up an element of
the Lorentz group of the target space and, similarly, the matrices
$u_{\a}{}^{\ua}$ and $u_{\a'}{}^{\ua}$ make up the corresponding element
of the spin group.
With this choice the components of the inverse matrices
$((E^{-1})_{\unA}{}^{A}, (E^{-1})_{\unA}{}^{A'})$ are given by

\be
\ba{rclrcl}
(E^{-1})_{\una}{}^{a} & = & (u^{-1})_{\una}{}^{a} &\qquad\qquad
(E^{-1})_{\una}{}^{\a}& = & 0 \\
(E^{-1})_{\ua}{}^{a} &= &0   &
(E^{-1})_{\ua}{}^{\a} & = & (u^{-1})_{\ua}{}^{\a}\\
&&&&&\\
(E^{-1})_{\una}{}^{a'} & = & (u^{-1})_{\una}{}^{a'}&\qquad\qquad
(E^{-1})_{\una}{}^{\a'} & = & -(u^{-1})_{\una}{}^{b}\L_{b}{}^{\a'}\\
(E^{-1})_{\ua}{}^{a'}&=&0&
(E^{-1})_{\ua}{}^{\a'} & = & (u^{-1})_{\ua}{}^{\a'} -
(u^{-1})_{\ua}{}^{\b}h_{\b}{}^{\a'}\ .
\ea
\label{6}
\ee

The consequences of the basic embedding condition can be conveniently
analyzed by means of the torsion identity which is simply the equation
defining the target space torsion tensor pulled back onto the
worldvolume by means of the embedding matrix. Explicitly, one has

\be
\nabla_A E_B{}^{\unC}-(-1)^{AB}\nabla_B E_A{}^{\unC}+T_{AB}{}^C E_C{}^{\unC}
= (-1)^{A(B+\unB)}E_B{}^{\unB}E_A{}^{\unA}T_{\unA \unB}{}^{\unC}.
\label{25}
\ee

where $\nabla$ denotes a covariant derivative which acts independently
on the target space and worldvolume indices. There are different
possibilities for this derivative and we shall specify our choice later.
With the embedding condition \eq{3} the dimension zero component of the
torsion identity does not involve any connection terms and reduces to
the simple form

\be
T_{\a \b}{}^{c}E_{c}{}^{\unc} = E_{\a}{}^{\ua}E_{\b}{}^{\ub}
T_{\underline{\a \b}}{}^{\unc}
\label{12}
\ee

In a sense one can consider the dimension zero component of the torsion
tensor (Frobenius tensor) as the basic tensor in superspace geometry and
the above equation specifies how the target space and worldvolume
Frobenius tensors are related.

The flat target superspace geometry for L$p$-branes $(p=3,4,5)$ also
includes the following differential forms

\bea
G_2 &=& dC_1\ , \nn\\
G_{p+1} & = & dC_p\ ,\nn\\
G_{p+2} &=& dC_{p+1}-C_1 G_{p+1} \ ,  \qquad p=3,4,5,
\eea

which obey the Bianchi identities

\bea
dG_2 &=& 0\ , \nn\\
dG_{p+1} &=& 0\ ,\nn\\
dG_{p+2} &=& G_2 G_{p+1}\ . \la{ggg}
\eea

In the flat target space under consideration here the non-vanishing
components of the forms $G_q$ are

\be
G_{\ua \ub \una_{1} \ldots \una_{(q-2)}} = -i(\C_{\una_{1}
\ldots \una_{(q-2)}})_{\ua \ub}\ ,
\label{10}
\ee

except for $G_{\ua \ub \una\unb\unc} $ which arises for the L$4$-brane in
$D=8$, in which case a factor of $\C_9$ is needed so that
$(\C_{\una\unb\unc}\C_9)_{\ua\ub}$ has the right symmetry. In $D=8,9$
the spinor indices label $16$ component pseudo-Majorana spinors while in
$D=7$ they represent a pair of indices, one of which is an $Sp(1)$
doublet index, which together label a $16$ component symplectic-Majorana
spinor.

In proving the Bianchi identities $dG_{p+1}=0$ for $Lp$-branes in
$(p+4)$ dimensions, the following $\C$-matrix identities are needed:

\be
(\C_{\una_1})_{(\underline{\a\b}}\,(\C^{\una_1 \cdots
\una_{(p-1)}}){}_{\underline{\c\d})}=0\ . \la{ci}
\ee

These identities are well known in the context of the usual
$p=2,3,4,5$-branes in $D=7,8,9,10$, respectively \cite{bst,act}. To
prove the Bianchi identity $dG_{p+2}=G_2G_{p+1}$ for an L$p$-brane in
$(p+4)$ dimensions, on the other hand, one needs to use a $\C$-matrix
identity resulting from the dimensional reduction of \eq{ci} from one
dimension higher. For example, to prove $dG_7=G_2G_6$ for the L$5$-brane
in $D=9$, one needs the following identity

\be
(\C_{\una_1})_{(\underline{\a\b}}\,(\C^{\una_1 \cdots
\una_5}){}_{\underline{\c\d})}
+ C_{(\underline{\a\b}}\,(\C^{\una_2\cdots \una_5})_{\underline{\c\d})}=0\
,\la{ci2}
\ee

which follows from the identity \eq{ci}, which holds in $D=10$,
by a dimensional reduction to $D=9$.

In addition to the geometrical quantities for each L$p$-brane there is a
$(p-1)$-form worldvolume gauge field ${\cA}_{p-1}$ with modified field
strength $p$-form ${\cF}_{p}$ defined by

\be
{\cF}_p = d{\cA}_{p-1}- \unC_p\ ,\qquad p=3,4,5\ ,
\ee

where $\unC_{p}$ is the pull-back of a target space $p$-form
$C_{p}$. This field strength obeys the Bianchi identity

\be
d{\cF}_p=-\unG_{p+1}\ ,
\ee

where $\unG_{p+1}$ is the pull-back of a target space $(p+1)$-form
$G_{p+1}$. In the first sequence of L$p$-branes this identity is a
consequence of the basic embedding condition \eq{3}, while for the
L3-brane in $D=5$ this is not the case. For this brane the ${\cF}$
Bianchi identity is required in order to completely specify the
worldvolume supermultiplet as a linear multiplet.


\subsubsection{\bf The linearized theory}


Let us consider the linearization of a flat brane in a flat target
space. The target space basis forms are

\bea
E^{\una} &= & dx^{\una}
- {i\over 2} d\th^{\ua}(\C^{\una})_{\ua\ub}\th^{\ub}\ , \nn\\
E^{\ua} &=& d\th^{\ua}\ .
\la{Ealfa}
\eea

In the physical gauge we have

\bea
x^{\ua} &= &(x^a,x^{a'}(x,\th){})\ ,\nn\\
\th^{\ua} &= &(\th^{\a},\Th^{\a'}(x,\th){})\ .
\la{theta}
\eea

where it is supposed that the fluctuations of the brane which are
described by the transverse (primed) coordinates as functions of the
(unprimed) brane coordinates are small. In addition we write the
worldvolume basis vector fields in the form $E_A{}^M\p_M = D_A-H_A{}^B
D_B$ where $D_A = (\p_a,D_{\a})$ is the flat covariant derivative on
the worldvolume. In the linearised limit, the elements of the embedding
matrix take the form

\be
E_a{}^{\unb} = (\d_a{}^b{},{}\p_a X^{b'}),
\la{E1}
\ee

\be
E_{\a}{}^{\ub} = (\d_{\a}{}^{\b}{},{}D_{\a}\Th^{\b'}),
\la{E2}
\ee

\be
E_a{}^{\ub} = ({}0{},{}\p_a \Th^{\b'}).
\la{E3}
\ee

Consequently, \eq{3} tells us that the deformation $H_A{}^B$ of the
supervielbein vanishes and that

\be
D_{\a} X^{a'} = i(\C^{a'})_{\a\b'}\Th^{\b'},
\la{lin basic}
\ee

where

\be
X^{a'} = x^{a'} \pl {i\over 2} \th^{\a}
(\C^{a'})_{\a\b'}\Th^{\b'}.
\la{X}
\ee

For the  L-branes of the first sequence \eq{lin basic} takes the form

\be
D_{\a i} X^{a'} = i(\c^{a'})_{ij} \Th^j_{\a}\ , \qquad a'=1,2,3;\ i=1,2\ .
\la{basic L}
\ee

Defining

\be
X_{ij} = (\c^{a'})_{ij}X_{a'}\ ,
\ee

substituting into \eq{basic L} and multiplying by $\c_{a'}$
we get

\be
D_{\a i} X_{jk} =-i(\e_{ij}\Th_{\a k} +
\e_{ik} \T_{\a j}).
\la{mult X}
\ee

This equation describes the linear multiplet with eight supersymmetries
in $d=6,5,4$ corresponding to the $L5$, $L4$ and $L3$-branes of the first
sequence. The field content of this multiplet consists of 3 scalars, an
8-component spinor and a divergence-free vector in all cases together
with an additional auxiliary scalar in $d=5$ and two auxiliary scalars
in $d=4$. The dual of the divergence-free vector can be solved for in
terms of a $4,3$ or 2-form potential in $d=6,5,4$ dimensions
respectively. For example, in $d=6$, spinorial differentiation of
\eq{mult X} gives

\be
D_{\a i}\Th_{\b j} = \e_{ij}(\c^a)_{\a\b}h_a
+\ft12  (\c^a)_{\a\b}\p_a X_{ij},
\la{DG}
\ee

where $h_a$ is the conserved vector in the multiplet, $\p^a h_a=0$.
This field, together with the 3 scalars $X_{ij}$ and the 8 spinors
$\Th_{\a i}$ (evaluated at $\th=0$) are the components of the
(off-shell) linear multiplet. At the linearized level the field
equations are obtained by imposing the free Dirac equation on the spinor
field $(\c^a )^{\a \b} \p_a \Th_{\b}{}^j = 0$. One then finds the
Klein-Gordon equation $\p_a \p^aX_{ij}=0$ for the scalars and the
field equation for the antisymmetric tensor gauge field $\p_{[a}
h_{b]}=0$.


\subsection{\bf The L5-brane in D=9}


We now turn to a detailed discussion of the L5-brane in $D=9$ in a flat
target superspace. We begin with a brief discussion of the target space
geometry. This can be derived most simply by dimensional reduction from
the flat $N=1, D=10$ supergeometry. In this way or by a direct
construction, one can establish, in addition to the usual supertorsion,
the existence of the $2,6,7$ forms $G_2,G_6,G_7$ defined in \eq{ggg}.

In $N=1,D=9$ flat superspace, the only non-vanishing component of the
torsion tensor is

\be
T_{\ua \ub}{}^{\unc}=-i(\C^{\unc})_{\ua\ub}\ .
\la{tt}
\ee

With this background we can now start to study the details of the
L5-brane using the torsion identity \eq{25}. From this starting point
there are now two equivalent ways to proceed. The first one is to fix
some of the components of the torsion tensor on the worldvolume
$T_{AB}{}^C$ in a convenient form. The second one is to fix the
connection used in \eq{25} by specifying some of the components of
the tensor $X_{A, B}{}^{C}$ defined by

\be
X_{A,B}{}^C \equiv (\nab_A u_B{}^{\unC})(u^{-1})_{\unC}{}^C\ .
\label{28}
\ee

Using this method all the components of the worldvolume torsion can be
found in terms of the vector-spinor $\Lambda_b{}^{\beta'}$ introduced in
\eq{4}. Although the two methods are equivalent, we have found that in
practice the second method is more efficient and will be used here. The
components of $X_{A, B}{}^{C}$ can be chosen to be

\begin{eqnarray}
X_{A,b}{}^c & = & X_{A,b'}{}^{c'} = 0 \nonumber \\
X_{A,\beta}{}^\gamma & = & X_{A,\beta'}{}^{\gamma'} = 0.
\label{29}
\end{eqnarray}

Note that, since $X_{AB}{}^C$ takes its values in the Lie algebra of the
target space Lorentz group, the components with mixed primed and
unprimed spinor indices are determined by the components with mixed
vector indices. Thus we have

\bea
X_{A,\b}{}^{\c'} &=& \ft14 (\C^{bc'})_{\b}{}^{\c'} X_{A,bc'}\ ,\nn\\
X_{A,\b'}{}^{\c} &=& \ft14 (\C^{bc'})_{\b'}{}^{\c} X_{A,bc'}\ .
\eea

We shall now analyze the torsion identity \eq{25} order by order in
dimension starting at dimension zero.


{\bf Dimension 0}

We recall that the dimension 0 component of the torsion identity is

\be
T_{\a \b}{}^{c}E_{c}{}^{\unc} = E_{\a}{}^{\ua}E_{\b}{}^{\ub}
T_{\underline{\a \b}}{}^{\unc}
\label{12'}
\ee

Projecting \eq{12'} with $(E^{-1})_{\unc}{}^{c'}$ and using the expressions for
the embedding matrix given in \eq{4} we find

\be
h_{\a i \b}{}^{k}(\c^{c'})_{kj} + h_{\b j \a}{}^{k}(\c^{c'})_{ik} = 0\ ,
\label{15}
\ee

which can be solved for $h_{\a}{}^{\b'}$ to give

\be
h_{\a}{}^{\b'} = h_{\a i \b}{}^{j} =
\d_{i}{}^{j}h_{a}(\c^{a})_{\a \b}\ ,
\label{16}
\ee

where $h^2 \equiv h^a h_a$. Projection with $(E^{-1})_{\unc}{}^{c}$ on
the other hand yields

\be
T_{\a \b}{}^{a} = -i \e_{ij}(\c^{b})_{\a \b}m_{b}{}^a\ ,
\label{17}
\ee

where $m_{a}{}^{b}$ is given by

\be
m_{a}{}^{b} = (1-h^2)\d_{a}{}^{b} + 2h_{a}h^{b}\ .
\label{18}
\ee

At the linearized level the field $h_a$ coincides with the
divergence-free vector field in the linear multiplet discussed in the
previous section. As we shall see later the divergence-free condition
receives non-linear corrections in the full theory.


{\bf Dimension 1/2}

At dimension $1 \over 2$ equation \eq{25} gives rise to two equations

\be
\nab_\a E_b{}^{\unc}+T_{\a b}{}^c E_c{}^{\unc} = -E_b{}^{\ub}
E_\a{}^{\ua}T_{\underline{\a \b}}{}^{\unc}
\label{33}
\ee

and

\be
\nab_\a E_{\b}{}^{\uc} + \nab_\b E_\a{}^{\uc} + T_{\a \b}{}^{\c}
E_\c{}^{\uc} = -T_{\a\b}{}^c E_c{}^{\uc}.
\label{34}
\ee

Projection of \eq{33} with $(E^{-1})_{\unc}{}^c$ gives

\be
T_{\a b}{}^c = i h_a (\c^a\c^c)_{\a}{}^{\b} \L_{b \b i}\ ,
\label{36}
\ee

while projection  with $(E^{-1})_{\unc}{}^{c'}$  gives

\be
X_{\a i,b}{}^{c'} = -i(\c^{c'})_{ij} \L_{b \a}{}^j\ .
\label{37}
\ee

The dimension  ${1 \over 2}$ component of $X_{A,B}{}^C$,
$X_{\a,\b}{}^{\c'}$,  is then determined due to the fact that

\be
X_{\a,\b}{}^{\c'}
= -{1 \over 2}(\c^{b})_{\b \c}(\c^{c'})_{j}{}^{k}X_{\a i, bc'}\ .
\label{38}
\ee

Similarly one finds that

\be
X_{\a,\b'}{}^{\c}= {1 \over 2}(\c^{b})^{\b \c}(\c^{c'})_{j}{}^{k}X_{\a i, bc'}\
{}.
\ee

Projecting \eq{34} with $(E^{-1})_{\uc}{}^{\c}$ we find

\be
T_{\a \b}{}^{\c} = -h_{\b}{}^{\d'} X_{\a,\d'}{}^{\c}-h_{\a}{}^{\d'}
X_{\b,\d'}{}^{\c}\ ,
\label{39}
\ee

while projecting onto the transverse space with $(E^{-1})_{\uc}{}^{\c'}$ we
find

\be
\nab_{\a i} h_{a} = {i \over 2}(\tilde{\L}_{\a i a}
- (\c_{a}\c^{b})_{\a}{}^{\b} \tilde{\L}_{\b i b})\ ,
\label{40}
\ee

where $\tilde{\L}_{\a i b}$ is defined by

\be
\tilde{\L}_{\a i a} \equiv m_{a}{}^{b} \L_{\a i b}\ .
\label{41}
\ee

All dimension ${1 \over 2}$ quantities on the worldvolume can therefore
be expressed in terms of the vector-spinor $\L_{b}{}^{\b'}$ and the
worldvolume vector $h_a$.


{\bf Dimension 1}

At dimension $1$ equation \eq{25}  again gives two equations but now
involving spacetime derivatives of the embedding matrix
$E_{A}{}^{\unA}$, namely

\be
\nab_a E_b{}^{\unc} - \nab_b E_a{}^{\unc} + T_{ab}{}^c E_c{}^{\unc}
= E_b{}^{\ub} E_a{}^{\ua} T_{\underline{\a \b}}{}^{\unc}
\label{42}
\ee

and

\be
\nab_a E_{\b}{}^{\uc} - \nab_{\b} E_a{}^{\uc} + T_{a \b}{}^c E_c{}^{\uc}
+ T_{a \b}{}^{\c} E_{\c}{}^{\uc} = 0\ .
\label{43}
\ee

The first of these equations, when projected onto the transverse space, gives

\be
X_{a,b}{}^{c'} = X_{b,a}{}^{c'}\ .
\label{44}
\ee

This also determines $X_{a,\b}{}^{\c'}$ because

\be
X_{a, \b}{}^{\c'}
= -{1 \over 2}(\c^{b})_{\b \c}(\c^{c'})_{j}{}^{k}X_{a,bc'}.
\label{45}
\ee

Furthermore,

\be
X_{a, \b'}{}^{\c} ={1 \over 2}(\c^{b})^{\b \c}(\c^{c'})_{j}{}^{k}X_{a,bc'}.
\ee

Projection onto the worldvolume on the other hand yields

\be
T_{ab}{}^c = -i\L_{b \b}{}^i (\c^c )^{\b \a} \L_{a \a i}.
\label{46}
\ee

Equation \eq{43}, when projected onto the worldvolume with
$(E^{-1})_{\uc}{}^{\c}$,
gives

\be
T_{a \b}{}^{\c} = \L_a{}^{\d'} X_{\b,\d'}{}^{\c} - h_{\b}{}^{\d'}
X_{a,\d'}{}^{\c}.
\label{47}
\ee

The analysis of the projection of \eq{43} onto the transverse space,
however, is more difficult. The resulting equation can be analyzed more
easily if we multiply by $m_{a}{}^{b}$; we then find that

\begin{eqnarray}
\nab_{\b j}\tilde{\L}_{b \c k} &=& -{1 \over 2}(\c_{c'})_{jk}
m_{b}{}^{a}m_{d}{}^{e}(\c^{d})_{\b \c}X_{a,e}{}^{c'}
+\e_{jk}(\c_{c})_{\b \c} m_{b}{}^{a}\nab_{a}h^c \nonumber \\
&& -ih_b \tilde{\L}_{c \a j}(\c^{ca})^{\a}{}_{\b}\L_{a \c k}
+ih^a \tilde{\L}^{c}{}_{\a j}(\c_{bc})^{\a}{}_{\b}\L_{a \c k}
-ih_d \tilde{\L}_{c \a j}(\c^{dc})^{\a}{}_{\b}\L_{b \c k} \nonumber \\
&& -{i \over 2}h_c \tilde{\L}_{b \a j}(\c^{dc})^{\a}{}_{\c}\L_{d \b k}
-{i \over 2} h^c \tilde{\L}_{b \c j} \L_{c \b k}
+ih_c \tilde{\L}_{b \a j}(\c^{dc})^{\a}{}_{\b}\L_{d \c k}
+ih^c \tilde{\L}_{b \b j} \L_{c \c k} \nonumber \\
&& +{i \over 2}\e_{jk}h_c \tilde{\L}_{b \a}{}^i (\c^{dc})^{\a}{}_{\c}
\L_{d \b i} +{i \over 2}\e_{jk}h^c \tilde{\L}_{b \c}{}^i \L_{c \b i}.
\label{48}
\end{eqnarray}

Using the fact that $[\nab_A , \nab_B ]h_c = -T_{AB}{}^{D}\nab_{D} h_c
-R_{ABc}{}^{d}h_d$ it can now be shown that

\be
T_{\a \b}{}^{d}\nab_{d}h_{c} = -T_{\a \b}{}^{\c}\nab_{\c}h_c
- \nab_{\a}\nab_{\b}h_c - \nab_{\b}\nab_{\a}h_c
+X_{\a,c}{}^{a'}X_{\b,a'}{}^{d}h_d
+X_{\b ,c}{}^{a'}X_{\a ,a'}{}^{d}h_{d}.
\label{49}
\ee

{}From \eq{49} and \eq{40} it can be seen that $m_b{}^a \nab_a h_c$ is in
fact a function of $h_a$, $\L_b{}^{\b'}$ and
$\nab_{\b j}\tilde{\L}_{b\c k}$. With this substitution \eq{48} can be
rewritten to give a rather
unwieldy expression for $X_{a,e}{}^{c'}$ in terms of
$\nab_{\b j}\tilde{\L}_{b\c k}$ and terms of order $\L^2$.


{\bf Dimension 3/2}

The dimension $3 \over 2$ component of \eq{25} is given by

\be
\nab_a E_{b}{}^{\uc} - \nab_b E_{a}{}^{\uc}
+ T_{ab}{}^{c}E_{c}{}^{\uc} + T_{ab}{}^{\c}E_{\c}{}^{\uc} = 0\ .
\label{50}
\ee

Its projection onto the worldvolume determines $T_{ab}{}^{\c}$ to be

\be
T_{ab}{}^{\c} = \L_{a}{}^{\b'}X_{b,\b'}{}^{\c} - \L_{b}{}^{\b'}X_{a,\b'}{}^{\c}
\label{51}
\ee

while the projection onto the normal space gives

\be
\nab_{[a}\L_{b]}{}^{\c'} = -\L_{[a}{}^{\b'}
X_{b],\b'}{}^{\d}h_{\d}{}^{\c'} - T_{ab}{}^{c}\L_{c}{}^{\c'}
\label{52}
\ee

so that all dimension $3 \over 2$ components are expressible as
functions of lower dimensional quantities. No further components exist
at higher dimensions. The torsion identity \eq{25} therefore
determines all the fields on the worldvolume of the brane off-shell.

\bigskip


{\bf The $\cF$ Bianchi Identity}

In addition to the torsion identities all L-branes should satisfy a
further condition which relates superforms on the target space and
worldvolume. In the case of the L5-brane there is a worldvolume 4-form gauge
potential ${\cA}_4$ with corresponding field strength 5-form ${\cF}_5$.
The explicit form for ${\cF}_5$ is

\be
{\cF}_5= d{\cA}_4 -\unC_5
\ee

where $\unC_5$ is the pull-back onto the worldvolume of the target space
5-form potential. The corresponding Bianchi identity is

\be
d{\cF}_5 = - \underline{G}_6
\label{53}
\ee

where $\underline{G_6}$ is the pull-back of $G_6=dC_5$. Equation \eq{53}
can then be solved for ${\cF}_5$. All the components of this tensor are
zero except at dimension zero where we find

\be
{\cF}_{abcde} = -2(m^{-1})_{e}{}^{g}h^{f}\e_{abcdfg}
\label{54}
\ee

where $(m^{-1})_{a}{}^{b}$ is the inverse of $m_{a}{}^{b}$ which is given
explicitly by

\be
(m^{-1})_{a}{}^{b} = {1
\over{1-h^4}}[(1+h^2)\d_{a}{}^{b}-2h_a h^b]\ ,
\ee

where $h^2\equiv h^ah_a$ and $h^4\equiv (h^2)^2$. Furthermore, it is
easy to see that the positive dimension components of the Bianchi
identity are also satisfied. To see this it is convenient to define a
6-form $I_6$ as $I_6 = d{\cF}_5 - \underline{G_{6}}$. It is then
straightforward to show that $dI = 0$ if we remember that the pullback
commutes with the exterior derivative. All components of $I_6$ itself
with more than two spinorial indices must vanish on dimensional grounds.
To show that the other components are also zero we then have to use the
fact that $dI_6=0$ at each dimension independently.
Doing this recursively proves $I_6 = 0$ and thereby establishes \eq{53}.

We conclude this section by noting the relation between the Hodge dual
of ${\cF}_{abcde}$ and $h^a$ which follows from \eq{54}:

\be
{\cF}^a= {2h^a\over 1-h^2}\ , \quad\quad {\cF}^a =
\ft1{5!}\, \e^{abcde} {\cF}_{bcde}\ , \quad\quad h^2 \equiv h^a h_a \ .
\la{fh}
\ee


\subsubsection{\bf The construction of the action}


Recently it was shown how GS-type actions can be systematically constructed for
most branes starting from the superembedding approach \cite{hos}. The only
brane actions that cannot be constructed are those of the $5$-branes in
D=7 and D=11 which both have self-dual anti-symmetric tensors as
components of their supermultiplets. For all other $p$-branes the
starting point for the construction of the action is a closed
$(p+2)$-form, $W_{p+2}$, the Wess-Zumino form, on the worldvolume. This
$(p+2)$-form is given explicitly as the exterior derivative of the (locally
defined) Wess-Zumino potential $(p+1)$-form,
$Z_{p+1}$, $W_{p+2} = dZ_{p+1}$. Since the de Rham cohomology of a
supermanifold is equal to the de Rham cohomolgy of its body, and since
the dimension of the body of the worldvolume superspace is $p+1$, it
follows that $W_{p+2}$ is exact, so that there is a globally defined
$(p+1)$-form $K_{p+1}$ on the worldvolume satisfying

\be
dK_{p+1} = W_{p+2}
\label{55}
\ee

The Green-Schwarz action of the $p$-brane can
then be defined as

\be
S = \int_{M_{o}}L_{p+1}^{0}
\label{56}
\ee

where $M_o$ denotes the (bosonic) body of $M$,

\be
L_{p+1}=K_{p+1}-Z_{p+1}\ ,
\la{lkz}
\ee

and $L_{p+1}^0$ is defined by

\be
L_{p+1}^{0} =dx^{m_{p+1}}\wedge dx^{m_p}\ldots dx^{m_1} L_{m_1\ldots
m_{p+1}}|\ ,
\label{57}
\ee

where the vertical bar indicates evaluation of a superfield at $\th=0$.

By construction $dL_{p+1} = 0$. The $\k$-symmetry of the GS-action is
ensured because, under a worldvolume diffeomorphism generated by a worldvolume
vector field $v$,

\begin{eqnarray}
\d_{v} L_{p+1} &=& {\cal{L}}_{v}L_{p+1} = i_v dL_{p+1} + di_v L_{p+1} \nonumber
\\
&=& d(i_v L_{p+1})\ .
\label{58}
\end{eqnarray}

As explained in \cite{hos}, reparametrizations and $\k$-symmetry
transformations on $M_o$ are essentially the leading components in a
$\th$ expansion of worldvolume diffeomorphisms so that the action given
above is invariant under these transformations by construction.

In the case of the $L5$-brane the Wess-Zumino $7$-form is given by

\be
W_7 = d Z_6=\underline{G_7} + \underline{G_2}{\cF}_5\ .
\label{59}
\ee

where $Z_6$ can be chosen to be

\be
Z_6
= \underline{C_6} + \underline{G_2}{\cA}_4 + \underline{C_1 C_5}\ .
\label{65}
\ee

The globally defined $6$-form $K_6$ needed for the construction of the
action can be solved from

\be
W_7 = dK_6\ .
\label{60}
\ee

The only non-zero component of $K_6$ is the purely bosonic one which is
found to be

\be
K_{abcdef} = -{{1+h^2}\over{1-h^2}}\,\e_{abcdef}.
\label{61}
\ee

Using \eq{fh} we may rewrite the function appearing in \eq{61} as

\be
{{1+h^2}\over{1-h^2}} = \sqrt{1+{\cF}^2}\ ,
\label{62}
\ee

where ${\cF}^2 \equiv {\cF}^a {\cF}_a$. This is the L-brane analogue of
the Dirac-Born-Infeld term in the D-brane action. Therefore $K\equiv
{1\over 6!}\e^{a_1\ldots a_6} K_{a_1\ldots a_6}|$ is given by

\bea
K &=& \sqrt{-g} \sqrt{1+{\cF}^2}\nn\\
  &=& \sqrt { -{\rm det}\ (g_{mn} + {\cF}_m{\cF}_n) }\ ,
\label{63}
\eea

where $g = \det(g_{mn})$ is the determinant of the metric on the
bosonic worldvolume induced by the embedding

\be
g_{mn}={\cE}_m{}^{\una}{\cE}_n{}^{\unb} \eta_{\una\unb}
	  =e_m{}^a e_n{}^b \eta_{ab}\ ,
\ee

where

\be
{\cE}_m{}^{\una}= E_m{}^A E_A{}^{\una}|\ ,
\ee

and ${\cF}_a$ is related to ${\cF}_m$ through the worldvolume vielbein
$e_m{}^a$ as ${\cF}m=e_m{}^a {\cF}_a$. The final form for the L5-brane action
is therefore given by

\be
S = \int_{M_{o}} d^6 x\,{\cL}
\ee

where the Green-Schwarz Lagrangian is

\be
{\cL}=\left(\sqrt {-{\rm det}\ (g_{mn} + {\cF}_m{\cF}_n})
- {1\over{6!}}\e^{m_1\ldots m_6} Z_{m_1\ldots m_6}\right)|\ ,
\label{64}
\ee

with  $Z_{m_1\cdots m_6}$ given in \eq{65}.

\subsubsection{\bf The equations of motion}

{}From the action given in the last section it is straightforward to
derive the equations of motion for the L5-brane. The dynamical variables
in the action are the worldvolume gauge potential $A_{mnpq}$ and the
coordinate $z^{\underline{M}}$ on the target space manifold. The
variation with respect to the worldvolume gauge field is straightforward
and yields

\be
\sqrt{-g} \nabla_m \left({1 \over \sqrt{1+{\cF}^2}}{\cF}^{mnpqr}\right)
= {1 \over 2} \e^{m_1 m_2 npqr}{\cal{E}}_{m_2}{}^{\unM_2}
{\cal{E}}_{m_1}{}^{\unM_1}
G_{\underline{M_1 M_2}}
\label{66}
\ee

where ${\cal{E}}_{m}{}^{\unM} \equiv \p_m Z^{\unM}$, $g$ is again the
induced GS metric on $M_o$ and the covariant derivative is formed using
the Levi-Civita connection of the metric $g$. Note that the Green-Schwarz
embedding matrix (often denoted by $\P$) ${\cE}_m{}^{\unA}$ is given by

\be {\cE}_m{}^{\unA}=\p_m z^{\unM} E_{\unM}{}^{\unA}={\cE}_m{}^{\unM}
E_{\unM}{}^{\unA}\ .
\ee

The
variation of the action with respect to $z^{\underline{M}}$, however, is
rather more involved. Defining $V^{\unM} \equiv \d z^{\unM}$ and
$V^{\unA}=V^{\unM}
E_{\unM}{}^{\unA}$ we find that the
variation of the metric $g_{mn}$ is given by

\be
\d g_{mn} = 2(\p_m V^{\una} + {\cal{E}}_{m}{}^{\ub}
V^{\ua}T_{\underline{\a \b}}{}^{\una}){\cal{E}}_{n \una}.
\label{68}
\ee

Similarly, the variation of ${\cF}_{m_1\cdots m_5}$ gives

\be
\d{\cF}_{m_1\cdots m_5} =
{\cE}_{m_5}{}^{\unA_5} \ldots {\cE}_{m_1}{}^{\unA_1} V^{\unA}
G_{\unA\unA_1\ldots\unA_5}
+ 5 \p_{m_5} \left( V^{\unA_5} {\cE}_{m_4}{}^{\unA_4}
\ldots {\cE}_{m_1}{}^{\unA_1} C_{\unA_1\ldots \unA_5} \right)\ .
\la{70}
\ee

The complete variation of the Green-Schwarz Lagrangian ${\cL}$ (up to
total derivatives) with respect to the $z^{\unM}$ is then given by

\bea
\d{\cL} &=& \sqrt{-g}\,t^{mn}\left(
\p_{m}V^{\una} -i {\cal{E}}_{m}{}^{\uc}
V^{\ub}(\C^{\una})_{\underline{\b \c}} \right)
{\cal{E}}_{n}{}^{\unb} \eta_{\underline{ab}} \nonumber \\
&+& {1 \over 5!}\sqrt{-g}{1 \over \sqrt{1+{\cF}^2}}{\cF}^{m_1
\ldots m_5}{\cal{E}}_{m_5}{}^{\underline{M}_5}
\ldots {\cal{E}}_{m_1}{}^{\underline{M}_1} V^{\underline{N}}
G_{\underline{N}\underline{M}_1 \ldots \underline{M}_5} \nonumber \\
&-& {1 \over 6!}\e^{m_1 \ldots m_6} {\cal{E}}_{m_6}{}^{\underline{M}_6}
\ldots {\cal{E}}_{m_1}{}^{\underline{M}_1} V^{\underline{N}}
G_{\underline{N}\underline{M}_1 \ldots \underline{M}_6} \nonumber \\
&-& {1 \over 5!} \e^{m_1 \ldots m_6}{\cal{E}}_{m_1}{}^{\underline{M}_1}
V^{\underline{N}} G_{\underline{N}\underline{M}_1}{\cF}_{m_2 \ldots m_6}\ ,
\label{71}
\eea

where we have used \eq{66} and the tensor $t^{mn}$ is given by

\be
t^{mn} = {1 \over \sqrt{1+{\cF}^2}}(g^{mn} + {\cF}^m {\cF}^n)\ .
\label{72}
\ee

It is straightforward to read off the equations of motion from \eq{71}.
For the case of a flat target space one finds, from the vanishing of the
coefficent of $V^{\una}$,

\bea
\nab_m\left( t^{mn} {\cal{E}}_{n}{}^{\unb}
\eta_{\underline{ab}}\right)&=& -{i \over 3!2!}{1 \over
\sqrt{1+{\cF}^2}}{\cF}^{m_1
\ldots m_5}{\cal{E}}_{m_5}{}^{\ub_5} {\cE}_{m_4}{}^{\ub_4}
{\cE}_{m_3}{}^{\unb_3}\ldots {\cal{E}}_{m_1}{}^{\unb_1}
(\C_{\una \unb_1 \ldots \unb_3})_{\ub_4\ub_5} \nonumber \\
&& + {i\over{4!2!}}{\e^{m_1 \ldots m_6}\over \sqrt{-g}}
{\cal{E}}_{m_6}{}^{\ub_6} {\cal{E}}_{m_5}{}^{\ub_5}
{\cE}_{m_4}{}^{\unb_4}\ldots {\cal{E}}_{m_1}{}^{\unb_1}
(\C_{\una \unb_1 \ldots \unb_4})_{\ub_5\ub_6}\ ,
\label{71a}
\eea

where the covariant derivative is again the Levi-Civita derivative with
respect to the GS metric, and, from the vanishing of the coeficient of
$V^{\ua}$,

\bea
t^{mn} {\cal{E}}_{m}{}^{\ub}
(\C^{\una})_{\underline{\b \a}} {\cal{E}}_{n}{}^{\unb}
\eta_{\underline{ab}}&=& -{1 \over 4!}{1 \over \sqrt{1+{\cF}^2}}{\cF}^{m_1
\ldots m_5}{\cal{E}}_{m_5}{}^{\unb_5}
\ldots {\cE}_{m_2}{}^{\unb_2} {\cal{E}}_{m_1}{}^{\ub}
(\C_{\unb_2\ldots\unb_5})_{\underline{\a \b}} \nonumber \\
&& + {1 \over 5!}{\e^{m_1 \ldots m_6}\over \sqrt{-g}}
{\cal{E}}_{m_6}{}^{\unb_6}
\ldots{\cE}_{m_2}{}^{\unb_2} {\cal{E}}_{m_1}{}^{\ub}
(\C_{\unb_2\ldots\unb_6})_{\underline{\a\b}} \nonumber \\
&& +{1 \over 5!} {\e^{m_1 \ldots m_6}\over \sqrt{-g}} {\cal{E}}_{m_1}{}^{\ub}
C_{\underline{\a \b}}{\cF}_{m_2 \ldots m_6}\ ,
\label{71c}
\eea

We shall now compare these equations of motion with the equations of motion one
derives for the L-brane in superspace, i.e. with both worldvolume and target
superspaces. The simplest case to consider is the fermion equation of motion in
a flat target space, equation \eq{71c}, and we shall retrict the discussion to
this example.

The most general Dirac-type equation we can write down in superspace is

\be
M^{ab}E_{b}{}^{\ub}u_a{}^{\una}(\C_{\una} )_{\underline{\b \a}} = 0
\label{73}
\ee

where $M^{ab} = A \eta^{ab} + B h^a h^b$, A and B being scalar
functions of $h_a$. It turns out that this equation (evaluated at $\th=0$) is
equivalent to \eq{71c} provided that we choose the tensor $M_{ab}$ to be equal
to $m_{ab}$. In fact, the equation then reduces to the Dirac equation for the
spinor $\tilde\L$, i.e. the superspace analogue of the linearized
Dirac equation introduced in \eq{41}.

To show that this is the case we need first to evaluate \eq{73} at $\th=0$.
We note that

\be
E_a{}^{\ua}|={\cE}_a{}^{\ub} Q_{\ub}{}^{\ua}
\la{73a}
\ee

where ${\cE}_a{}^{\ua}=e_a{}^m {\cE}_m{}^{\ua}$, $e_a{}^m$ being the inverse
vielbein for the GS metric, and where $Q$ is a projection operator given by

\be
Q_{\ua}{}^{\ub}=(E^{-1})_{\ua}{}^{\c'} E_{\c'}{}^{\ub}|\ .
\la{73b}
\ee

It is easy to evaluate $Q$ explicitly; one finds

\be
2Q=1 + {1\over 6!}\e^{a_1\ldots a_6} \hat\C_{a_1\ldots a_6} -h^a\hat\C_a
-{1\over5!} \e^{a b_1\ldots b_5} h_a \hat\C_{b_1\ldots b_5}\ ,
\la{73c}
\ee

where

\be
\hat\C_a :={\cE}_a{}^{\una}\C_{\una}
\la{73d}\ .
\ee

To show the equivalence of the two  fermionic equations one simply computes
\eq{73} and then right multiplies it by $2(1-h^4)^{-1}(1-h^a\hat\C_a)$. The
resulting equation then has the same form as equation \eq{71c} when this is
expressed in terms of $h$ rather than ${\cF}$. This equation takes the form

\bea
\left({1-h^2\over 1+h^2} \eta^{ab} + 4{h^a h^b\over
1-h^4}\right){\cE}_b{}^{\unb}(\hat\C_a)_{\ua\ub} &=& -{1\over4!}{2\over
1-h^4}\e^{ba_1\ldots a_5} h_{a_5}{\cE}_b{}^{\ub}(\hat\C)_{a_1\ldots
a_4})_{\ua\ub}\nonumber \\
&-&{1\over5!} \e^{b a_1\ldots a_5} {\cE}_b{}^{\ub}(\hat\C_{a_1\ldots
a_5})_{\a\b} \nonumber \\
&-& {2h^b\over 1-h^2}{\cE}_b{}^{\ub} C_{\ua\ub}\ .
\la{73e}
\eea

One might wonder whether other choices of the tensor $M_{ab}$ could lead to a
different consistent set of equations of motion. Although we have not checked
this we believe that it is unlikely. In other words, if one were to make a
different choice for $M_{ab}$ one would find non-linear inconsistencies at
higher dimension arising as a consequence.


\subsection{\bf L-branes in D=7 and D=8}


The L3-brane in $D=7$ and the L$4$-brane in $D=8$ can in principle be
derived by double dimensional reduction from the L5-brane in $D=9$.
However, it is simpler to construct them directly using the same
techniques that were used for the L5-brane. To derive the action we only
need to analyse the torsion and Bianchi identities at dimension zero as
this information is sufficient to compute the $(p+1)$-form $K_{p+1}$
which, together with the Wess-Zumino form $Z_{p+1}$ determines the
action from \eq{lkz} and \eq{56}.


\subsubsection*{\bf The L$4$-Brane in $D=8$}


The analysis of the dimension zero torsion identity is similar to the
L5-brane case; from \eq{12'} one again finds the constraint \eq{15} on
the field $h_{\a}{}^{\b'}\rightarrow h_{\a i}{}^{\b j}$ which is solved
by

\be
h_{\a i\b j}=\e_{ij}\left( C_{\a\b} S + (\c^a)_{\a\b}\, h_a \right)\ .
\la{p4}
\ee

The scalar field $S$ can be identified with the auxiliary field of the
linear multiplet in $d=5$. The dimension 0 components of the
worldvolume torsion are found to be

\be
T_{\a i,\b j}{}^a = -i \e_{ij} ( (\c^b)_{\a\b} m_b{}^a + C_{\a\b} m^a ) \ ,
\la{79}
\ee

where

\bea
m_b{}^a &=& (1 - h^2 + S^2)\d_b{}^a \pl 2h^a h_b\ ,\\
m_a &=& 2S\,h_a\ . \la{m's}
\eea

{}From the Bianchi identity

\be
d{\cF}_4=-{\unG}_5\ ,
\ee

it follows that

\be
{\cF}_a = -{2h_a\over 5(1-h^2+S^2)}\ , \la{fh4}
\ee

where ${\cF}_{abcd}= \e_{abcde} {\cF}^e$.

To construct an action, we need to consider the Wess-Zumino form

\be
W_6 = d Z_5=\underline{G_6} + \underline{G_2}{\cF}_4\ .
\label{59a}
\ee

with  $Z_5$ given by

\be
Z_5
= \underline{C_5} + \underline{G_2}{\cA}_3 + \underline{C_1 C_4}\ .
\label{65a}
\ee

The globaly defined $5$-form $K_5$ needed for the construction of the
action can be solved from $W_6= dK_5$. We find that the only non-zero

component of $K_5$ is the purely bosonic one given by

\be
K_{abcde} = -\left( {1+h^2-S^2\over 1-h^2+S^2}\right)\,\e_{abcde}.
\label{61a}
\ee

Using the action formula $ S = \int_{M_{o}} d^5x\,{\cL} $ where
${\cL}=K_5-Z_5 $ and recalling \eq{fh4} and \eq{61a} we find that the
Lagrangian can be written as

\be
{\cL}= \left( \sqrt{-{\rm det}\, \left(g_{mn} + 25(1-S^2){\cF}_m{\cF}_n
\right)}
- {1\over{5!}}\e^{m_1\ldots m_5} Z_{m_1\ldots m_5}\right) \ ,
\ee

with $Z_{m_1\cdots m_5}$ given in \eq{65a}.


\subsubsection*{\bf The L$3$-Brane in $D=7$ }


The construction of the L$3$-brane action in $D=7$ parallels exactly the
constructions presented above. We find that the analogues of the equations
\eq{p4}-\eq{65a} for this case are

\be
h_{\a i\b j}=\e_{ij}\left(C_{\a\b} S + (\c_5)_{\a\b} T +
(\c_5\c^a)_{\a\b}\, h_a\right)\ ,
\la{p3}
\ee

where $S,T$ are the auxilary fields, and

\be
T_{\a i, \b j}{}^a = -i\e_{ij}\left((\c^b)_{\a\b} m_b{}^a \pl
                      (\c^{ba})_{\a\b} m_b\right)\ ,
\la{80}
\ee

where

\bea
m_a{}^b &=& (1 -h^2+S^2+T^2)\d_b{}^a \pl 2h^a h_b,\\
m_a &=& 2T h_a\ .
\eea

Furthermore, starting from the Bianchi identity

\be
d{\cF}_3=-{\unG}_4\ ,
\ee

we find that

\be
{\cF}_a = -{h_a\over 3(1-h^2-S^2-T^2)}\ , \la{fh3}
\ee

where ${\cF}_{abc}= \e_{abcd} {\cF}^d$ and that

\be
W_5 = d Z_4=\underline{G_5} + \underline{G_2}{\cF}_3\ .
\label{59b}
\ee

\be
Z_4
= \underline{C_4} + \underline{G_2}{\cA}_2 + \underline{C_1 C_3}\ .
\label{65b}
\ee

\be
K_{abcd} = -\left( {1+h^2-S^2-T^2\over 1-h^2+S^2+T^2}\right) \,\e_{abcd}.
\label{61b}
\ee

Again, using the action formula $ S = \int_{M_{o}} d^4 x\,{\cL} $ where
${\cL}=K_4-Z_4 $ and recalling \eq{fh3} and \eq{61b} we find that the
Lagrangian for the L$3$-brane in $D=7$ can be written as

\be
{\cL}= \left(\sqrt{-{\rm det}\,\left(g_{mn} + 36(1-S^2-T^2){\cF}_m{\cF}_n
\right)}
- {1\over{4!}}\e^{m_1\ldots m_4} Z_{m_1\ldots m_4}\right) \ ,
\ee

with $Z_{m_1\cdots m_4}$ given in \eq{65b}.


\subsection{\bf Properties of L-branes.}


We have seen that L-branes are examples of a class of $p$-branes with an
unconventional worldvolume supermultiplet, namely the linear
multiplet. These branes arise naturally within the superembedding
approach but have so far been neglected in the literature. In marked difference
to most other branes we have seen
that the linear multiplets are off-shell. One of the consequences is that the
usual torsion
equations are not the equations of motion for the branes. In fact these
have to be derived from an action. We have illustrated the dynamics of
the L$5$-brane by solving the highly non-linear torsion equations in
terms of a divergence-free vector $h_a$ and a vector-spinor
$\L_{a}{}^{\a'}$ in a flat target space background. For the L$4$-brane in
$D=8$, an additional auxiliary scalar $S$ and for the L$3$-brane in $D=7$, the
additional auxiliary scalars $(S,T)$ were shown to arise. Using a general
action principle which is valid for the
construction of actions for most branes we have found the Green-Schwarz
action of the L-branes. For the L$5$-brane we derived the Green-Schwarz
equations of motion and we illustrated the relationship between the
equations of motion in superspace and those derived from the action in
the case of the spinor equation.

We have noted that the L$5$-brane can be viewed as the dimensional reduction of
a superfivebrane in $D=10$ dimensionally reduced to $D=9$, followed by
dualisation of the scalar corresponding to the extra dimension to a 4-form
potential. This relation between a fivebrane in $D=10$ and L$5$-brane in $D=9$
is similar to the relation between the M$2$-brane in $D=11$ and D$2$-brane in
$D=10$. The latter relation has been called M-duality \cite{pkt4} which relates
Type IIA string theory to M-theory in the strong coupling limit. Other
worldvolume duality transformations have also been studied. Indeed, the
worldvolume $U(1)$ gauge fields arising in D$p$-branes have been dualised to
$(p-2)$-form gauge fields for $p \le 4$. We refer the reader to \cite{jhs4} for
various aspects of these dualizations and for an extensive list of references
for earlier works on the subject. The point we wish to emphasize here is that,
while the methods employed in the literature so far for worldvolume
dualizations become forbiddingly complicated beyond $p=4$, the superembedding
approach provides an alternative and simpler method which seems to apply
universally to all possible branes. Regardless of the approach taken, results
obtained in the area of worldvolume dualizations are hoped to provide further
connections among a large class of branes that arise in the big picture of
M-theory.

As mentioned earlier, we have focused our attention on flat target
superspaces in this paper. The generalization of our work to curved target
superspace is straightforward. Consider the case of the L$5$-brane, for
example. The results presented in subsection $3.a$ and $3.b$  remain the same in
curved superspace. The superforms occurring in the Wess-Zumino term
\eq{65}, however, now live in a curved target superspace. The field
strength forms $G_2,G_6,G_7$ still obey the Bianchi identities \eq{ggg},
but they will have more non-vanishing components than those given in
\eq{10}. The expected solution to the full set of Bianchi identities is
$N=1,D=9$ supergravity coupled to a single vector multiplet, as
resulting from the dimensional reduction of the $N=1,D=10$ supergravity
theory in its dual formulation. Thus, the $D=9$ field content is

\bea
&& {\rm Supergravity:}  \quad\quad
(g_{mn}, C_{m_1\cdots m_5},\,C_{m_1\cdots m_6},\, \phi)\,
(\psi_m,\, \chi)  \nn\\
&& {\rm Maxwell}: \quad\quad\quad \ \ (C_m, \s)\, (\l)
\eea

where we have grouped the bosonic and fermionic fields separately, in a self
explanatory notation. It should be noted that the fields $C_5$ and $C_6$ come
from the dimensional reduction of a 6-form potential, and $(C_1,\s)$ come from
the Kaluza-Klein reduction of the metric in $D=10$.  The coupling of $n$ vector
multiplets to $N=1,D=9$ supergravity has been determined in \cite{gns} in a
formalism that contains $(B_2,B_1)$ which are the duals of $(C_5,C_6)$ and in
\cite{at-new} in a formalism which has the fields $(C_5,B_1)$.  It would be
interesting to study the brane solitons of these theories and to determine the
maximum possible symmetries they may exhibit. It is known that the 5-brane
solution of $N=1,D=10$ supergravity involves a non-constant dilaton and
consequently it does not give rise to an $AdS_7 \times S^3$ geometry in the
near horizon limit (see \cite{duff1} for a detailed discussion of this matter
and for earlier references). On the other hand, there exists a singleton field
theory which is a candidate for the description of a fivebrane in this
background \cite{nst}. Given the close relation between the L$5$-brane in $D=9$
and the 5-brane in $D=10$, it is natural to study the L$5$-brane solution of
the Einstein-Maxwell supergravity in $D=9$ and to determine if it permits a
constant dilaton, thereby possibly giving rise to an $AdS_7 \times S^2$ near
horizon geometry. It should be noted that the candidate singleton field theory
in this case would be characterized by a a superconformal field theory of a
free linear supermultiplet in six dimensions discussed in section 3, and whose
superconformal transformations can be found in \cite{bps-new}.

The field content of the target space Einstein-Maxwell supergravities
relevant to the L$4$-brane in $D=8$ and L$3$-brane in $D=7$ remains to be
worked out in detail as well. The expected results are various versions
of Einstein-Maxwell supergravities in which certain fields have been
dualised.

In the next section we will investigate properties of string and p-branes
in background fields. We will see that the low energy limit gives us
noncommutative field theories and that Seiberg-Witten map between
noncommutative and ordinary theory could be generalized for non-associative 
case.

\bigskip



\section{\bf Strings and branes as Hamiltonian systems, noncommutativity
and non-associativity}

 In this section we will use Hamiltonian methods developed in previous chapter
to investigate properties of strings and branes in background field \cite{R}.
Although, the main topic of the discussion will be bosonic part of the brane
action it is also possible to generalize it to fermionic case.

 String theory appeared extremely useful in studying the properties of
non-commutative field theories. Moreover, studying properties of
 strings and membranes
in  B-field  not only helps to investigate 
the low-energy field theories \cite{Wit1} but also gives new information
about  superstring theory and 
M-theory. One of examples of new models that attracted much attention recently
is OM-theory
\cite{OM1},\cite{SST},\cite{Berg1},\cite{Berg2}.
It is known that noncommutative Yang-Mills appears in decoupling limit 
of string theory in constant
B-field.
 It interesting to study correspondence 
between noncommutative and ordinary theories from
superstring theory point of view.

    In this section we apply Hamiltonian/BRST formalism
to study properties of string theory in constant B-field.
One of the ways to investigate the 
correspondence between non-commutative and ordinary Yang-Mills 
is to 
consider instantons in both theories \cite{Wit1}. 
But how can we see this correspondence  
from the point of view of the string theory?
Discussion on possible ways
to answer this question will be main topic of this section.

   If one uses the world-sheet approach, non-commutativity appears from 
interpreting 
time ordering as operator ordering 
in two-point functions as well as in product of vertex operators \cite{Sch1}. 
In the Hamiltonian treatment, non-commutativity arises from modification of the
boundary constraint.
This modification appears due to additional contribution coming from
modified boundary conditions in constant 
B-field \cite{Chu1}, \cite{Chu2}, \cite{Jab1} \cite{Jab2}, \cite{Oh1} and  \cite{Lee}.
 These new boundary constraints are of the second class and 
for a first sight spoil commutational relations between the  
coordinates and momenta on the D-brane. Apparently, one of the possible
ways to quantize theory with the second class constraints is to introduce the Dirac 
brackets \cite{Chu2}, \cite{Jab2}, \cite{Oh1} and \cite{Lee}. 
On the level of Dirac 
brackets inconsistency between commutators of 
coordinates and momenta disappear, instead, string endpoints 
on the D-brane do not commute anymore.

   There is another way to quantize a system 
with the second class constraints, 
the conversion method \cite{Fad}, \cite{Manv}, \cite{BF1}.
 Usually, conversion means the extension of the phase space 
by new auxiliary degrees of freedom together with modification of the second 
class constraints in such a way that they become of the first class. 
 These new first class constraints correspond
 to the gauge symmetries of the theory.
This framework could be understood from BRST-quantization point of view.
In BRST invariant theory, one can treat additional variables as ghosts 
and new modified action acquires new gauge symmetry in the same fashion as
appearance 
of local fermionic symmetry in ordinary BRST approach.
The main difference is that in the case of the conversion new symmetry could be bosonic.
The connection of BRST conversion procedure with Fedosov quantization was considered in
\cite{Lyakh}.

    In this work, we apply conversion to the system with second class constraints. 
 Finally, we have model  with all 
 constraints of the first class. In this case we do not have to use the 
 Dirac brackets
and all target space coordinates become commuting. 
 For the models of interest,
 it is not necessary 
to apply BRST quantization, it is enough to consider ordinary Dirac 
quantization of the system with first class constraints only.  
The only reason to have
BRST quantization procedure is
to cancel ambiguity in appearance of the new variables and to interpret 
these extra 
coordinates of the extended phase space as a ``ghosts''.
In this section we consider bosonic part of the string/membrane actions only.
It is possible to extend this approach to the supersymmetric
case following \cite{CZ}, \cite{LW}, \cite{LZ} and references therein.
As we mentioned before, in this chapter we present stringy version
of Seiberg-Witten map.
There are different field-theoretical 
approaches to study Seiberg-Witten map between
 noncommutative and ordinary Yang-Mills \cite{JS}.
In this section we  follow \cite{R}.

\subsection{\bf String in constant B-field}

\subsubsection{\bf Decoupling limit and constraints}
 
 The bosonic part of the worldsheet  action of the string ending on D-brane
  in constant B-field background is given by

\be
S = \frac {1}{4\pi \a'} \int_{ M^2}^{ab}\p X^\m \p X_\m 
   - \frac{1}{2 \pi \a'} \int_{\p M^2} B_{ab}X^a \p_t X^b,
\la{S1}
\ee

where  $B_{ab}$ has non-zero direction only along D-brane. In general case $B_{ab}$
includes B-field
together with $U(1)$ field on D-brane. The boundary conditions
in the presence of D-brane are

\be
 g_{ab} \p_n X^b + 2 \pi \a' B_{ab} \p_t X^b = 0,
\la{bound1}
\ee

where $\p_n$ is derivative normal to the boundary of the world-sheet.

The effective metric seen by the open string is

\be
G^{ab} = \Biggl ( \frac {1}{g+2\pi \a'B}  g  \frac {1}{g-2 \pi \a' B} \Biggr )
^{ab},
\la{G1}
\ee

\be
G_{ab} = g_{ab} - (2 \pi \a')^2 (B  g^{-1} B)_{ab},
\la{G2}
\ee

and noncommutativity parameter is given by

\be
 \th^{ab} =- (2 \pi \a')^2 \Biggl ( \frac {1}{g+2 \pi \a'} B \frac {1}{g-2 \pi 
\a' B} \Biggr )^{ab}.
\la{th}
\ee

In the decoupling, zero slope, limit \cite{Wit1} one can take $\a' \ra 0$ keeping fixed
open string parameters such as 

\be
G^{ab} = - \frac{1}{(2 \pi \a')^2} \biggl ( \frac {1}{B} g \frac {1}{B}
 \biggr )^{ab},
\nonumber
\la{G3}
\ee

\be
G_{ab} = - (2 \pi \a')^2 (B g^{-1} B)_{ab},
\la{G4}
\ee

\be
\th^{ab} =  \biggl( \frac {1}{B} \biggr )^{ab}.
\nonumber
\la{th2}
\ee

In this limit the kinetic term in \eq{S1} vanishes. The remaining part, that
governs the dynamics, is the second term in the \eq{S1}. Equation \eq{S1} in this case
describes the evolution of the string boundary, i.e. particle living
on the world-volume of D-brane.
This part of the action is given by

\be
S = \frac {1}{2} \int_{\p M^2} dt B_{ab} X^a \p_t X^b.
\la{Slim}
\ee

For simplicity and without loss of generality one can take $B_{ab}$ nonzero only 
in two space directions on 
the D-brane, i.e. $B_{ab} = 2b \e_{ab}$ where $a,b = 1,2$ and the rest of string coordinates
do not give any contribution to \eq{Slim}.

Let us discuss how noncommutativity appears on the level of Hamiltonian 
formalism. The string action in decoupling limit is

\be
S = b \int dt \e_{ab} X^a \dot X^b.
\la{Sb}
\ee

Canonical momentum is given by

\be
 P_a = -b \e_{ab}X^b.
\la{PP}
\ee

Therefore, the constraints that completely describe this model are

\be
\phi_a = P_a + b \e_{ab} X^b.
\la{constrP}
\ee

All these constraints are of the second class, i.e. they do not commute with
each other

\be
[\phi_a, \phi_b] = 2b \e_{ab}.
\la{phiP}
\ee

Using the second class constraints \eq{constrP} the Dirac brackets are 

\be
[X^a,X^b]_D = [X^a,X^b]  - [X^a, \phi^c] \frac {\e^{cd}}{2b} [\phi^d, X^b],
\la{XX1}
\ee

and 

\be
[X^a, X^b]_D = \frac {1}{2b}\e^{ab}.
\la{XX2}
\ee

We see that string coordinates  on the D-brane do not commute.
It means that
 in the decoupling limit one has noncommutative Yang-Mills on the world-volume of D-brane.
It is possible to calculate a two-point function of the fields 
propagating on the boundary of the
string worldsheet. The two-point function is 

\be
< X^a (t), X^b (0) > = \frac {1}{2b} \e^{ab}.
\la{prop}
\ee

Using correspondence between time 
and operator ordering it  leads to noncommutative coordinates on the D-brane.

Another way to investigate properties
of the model described by the constraints \eq{constrP} is to extend the phase space 
 and to introduce additional pair
of canonically conjugated variables $\x$ and $ P_{\x}$ with commutator
 $[\x, P_{\x}]  = 1$ in the framework of BRST quantization 
In this case it is possible 
to convert the second class constraints \eq{constrP} into the first one.

\be
 \hat \phi_1 = P_1 + b (X^2 - \sqrt{2}  \x) , \qquad \hat \phi_2 = (P_2 + \sqrt 2  P_{\x}) - b X^1,
\la{Chi-hat}
\ee

where new constraints are of the first class.
 Let us count number of degrees of freedom to be sure that we did not loose
any information. 
Each of the second class constraints eliminate one degree of freedom, but
each of the first class one eliminates two. Thus, instead of two second class
constraints we have two first one now. Therefore we 
need to add two independent degrees of freedom 
(i.e. $\x$ and $ P_{\x}$ ).

Now we can identify new variables. The variables $P_1, X_1$ are the same as before
and the new ones are

\be
P^{+} = -P_2 - \sqrt 2  P_{\x}, \qquad X^{ -} = X^2 - \sqrt 2 \x,
\la{P+}
\ee

and 

\be
P^{-} = -P_2 + \sqrt 2  P_{\x}, \qquad X^{ +} = X^2 + \sqrt 2 \x.
\la{P+1}
\ee

 We have to mention that new momenta are canonically conjugated to new
coordinates.
 The variables $ P^{-}, X^{+}$
are not dynamical because they do not participate in the constraints.
The equation \eq{Chi-hat} could be reformulated using new variables 
\eq{P+},\eq{P+1} in the following way

\be
\hat\phi_1 = P_1 + b X^{-}, \qquad  \hat \phi_2 = P^{+} + b X^1.
\la{constr-P}
\ee

These two constraint are of the first class. Moreover, 
the propagator \eq{prop} is not
antisymmetric anymore, but rather symmetric in interchange of $ a$ and 
$b$ that leads to commutative 
coordinates not only on the level of Poisson brackets but also on the level of 
two-point functions. This could be seen using correspondence 
between operator and time
ordering \cite{Sch1}. Therefore, the two-point functions between new fields 
are

\be
< X^1 (t) , X^{-} (0) >  = < X^{-}(t), X^1(0) >,
\la{prop1}
\ee

and could be calculated using path-integral BRST approach.
Thus, we  removed the noncommutativity of the  string end-point coordinates by introducing
new
 variables, and applying the conversion of the constraints. 
Applying conversion procedure we converted all second class constraints into the
first class ones.
Now we do not have to use Dirac brackets and all coordinates commute on the level of the Poisson
brackets.
After field redefinition, the new coordinates of the string boundary are described
 by  
$X^1  , X^{-}$, that commute, because propagator is symmetric. 
Therefore, using time ordering product gives
commutativity on the boundary of string world-sheet. 
This could also be shown using BRST approach.

 The Hamiltonian of the new system that is
equivalent to the model described  by  \eq{S1} is 

\be
H = \l \hat \phi_1 + \l_1 \hat \phi_2,
\la{Ham}
\ee

where $\l$ and $\l_1$ are Lagrange multipliers.

The BRST charge is given by 

\be
Q =  c (P_1 + b X^{-}) + c_1 (P^{+} + b X^1)
\la{QP}
\ee

where $c$ and $c_1$ are ghosts corresponding the first class constraints.
  It is also possible to include information that boundary action comes
 from string as was shown in \cite{Wit1}, but the main result
 of this section is  that conversion together
 with change of variables and field redefinition gives ordinary, commutative behavior
 for the string boundary.

     It is possible to conduct conversion for string in decoupling limit in more covariant way.
To do so, let us start from system of constraints \eq{constrP} and then
 introduce new variables $\x^a$ and $P^\x_a$, $[\x^a, P^\x_b] = \d^a_b$ where $a = 1,2$.
Then generalizing \eq{P+} it is convenient to define

\be
P^+_a = P_a + P^\x_a, \qquad X^{+a} = X^a + \x^a,
\la{++}
\ee

\be
P^-_a = P_a - P^\x_a, \qquad X^{-a} = X^a - \x^a,
\la{--}
\ee

 Then \eq{constrP} could be modified into

\be
\hat \phi_a = P^-_a + b \e_{ab} X^{+b}.
\la{-+}
\ee 

Now all the constraints $\hat\phi_a$ are of the first class, i.e. they commute. 
In this case the situation is different from
previous example of conversion. The counting of 
degrees of freedom  tells us that we need to add one more 
first class constraint, otherwise modified system is not going to be equivalent to initial one.
 This constraint could be chosen in the form

\be
\Psi = P^+ X^+ - P^- X^-.
\la{-+2}
\ee

This choice in some sense reminds constraint that could be obtained
from \eq{constrP} by projecting by $X^a$. If all new variables are equal to zero, then
  constraints
\eq{-+} transform to \eq{constrP}, and \eq{-+2} is  satisfied.

In this subsection we presented two examples of conversion that produced 
equivalent systems 
with commuting variables (i.e. commuting string end-points). Those examples 
could be of the strong suggestion that
noncommutativity
could be removed not only on the level of constraints/Dirac brackets, but 
also on the level of two-point
functions.

\bigskip

\subsubsection{\bf General case }

Here we step aside from decoupling limit of string  and analyze
complete set of constraints coming from the action \eq{S1}.
It is more convenient to start not from Lagrangian \eq{S1} but rather from one without B-dependent
term \cite{Wit1} and impose boundary condition \eq{bound1} as additional boundary constraint.
The starting action is

\be
S = \frac {1}{4 \pi \a'} \int d^2 \s \p X^\m \p X_\m,
\la{S2S}
\ee

and boundary constraint

\be
X'^a + \dot X^b B_b{}^a = 0, \qquad X'^{m} = 0,
\la{X'S}
\ee

where $X^\m$ is full set of target-space coordinates for string, $X^a$ are
coordinates of string endpoints on the D-brane and $X^m$ is the rest of coordinates. 

As usually, we have first class constraints which follows from \eq{S2S} 

\be
H = 2 \pi \a' P^2 - \frac {1}{2 \pi \a'}  X'^2, \qquad H_1 = P_\m X'^\m.
\la{HamS}
\ee

The Hamiltonian is given by 

\be
H_t = \int d\s \biggl ( NH + N_1 H_1 \biggr).
\la{Hamiltonian}
\ee

The momentum in this case is 

\be
\dot X^a = 2 \pi \a' P^a.
\la{P2S}
\ee

Then the boundary condition \eq{X'S} could be rewritten as constraint

\be
\Phi^a  =  X'^a + 2 \pi \a' P^c B_c{}^a.
\la{Phi2S}
\ee

If B-field is equal to zero, the boundary condition $X'=0$ leads to infinite number of constraints 
in the form 

\be
N^{(2k+1)} = 0, \qquad N_1^{(2k)} = 0, \qquad  X^{(2k+1)} = P^{(2k+1)} = 0,
\la{N}
\ee

where $(k)$ denotes k's derivative in respect to $\s$. All these constraints are
equivalent to extending $\s$ to $ [-\pi, \pi] $ and taking the orbifold projection
\cite{Lee},\cite{BH} 

\be
X(- \s) = X(\s), \quad P(-\s) = P(\s), \quad N(-\s) = N(\s), \quad N_1 (-\s) =- N_1(\s).
\la{orb}
\ee

In the presence of constant B-field the secondary constraints appear, as usually, from the fact that 
 commutator of $\Phi^a$  with Hamiltonian gives either constraint or condition for Lagrange
multipliers. First of all, in constant B-field, conditions for the Lagrange multipliers are
the same as in \eq{N} and using the linear combinations of $\Phi$'s leads to the same boundary
conditions for $X^a$ and $P_a$ as in \eq{N} except that for $X^a$ if $k=0$ we have constraint \eq{Phi2S}.

Because we start from \eq{S2S} but not \eq{S1} in difference from 
 \cite{Chu2}, \cite{Jab2}, \cite{Oh1} and \cite{Lee} we do not have modification of the
higher derivative constraints but rather, after using  Dirac brackets, contribution to the 
noncommutativity is given only due to $\Phi^a$  i.e.

\be
[X^a (\s), X^b(\s') ]_D = [X^a, X^b]  - \int d \s'' [X^a, \Phi^c (\s'')] C^{-1}_{cd} [P'^d(\s''),
X^b],
\la{D2S}
\ee

where $C^{-1}_{cd}$ is inverse matrix of commutator coefficients between $\Phi$ and $P'$. 
We also have to use regularization for the endpoints (see \cite{Chu2}, \cite{Jab2}) , i.e.
for $\s, \s' = 0$ or $  \pi $. The same contribution is given by considering higher odd derivatives
of $P_a$.
We see that coordinates of endpoints do not commute but momenta do.

    Let us investigate the nature of this noncommutativity. 
Here we will not use conversion of the system of
second class constraints into the first ones. It is more transparent to modify the system by 
the way that
second class constraints $\Phi, X^{(2k+3)} $ and $ P^{(2k+1)}$ are taking similar form  as
the constraint system with zero
B-field. The noncommutativity on the level of Dirac brackets appears because of the nonzero commutator
of $X^a$ and $\Phi^b$. If B-field is zero we see that they commute and Dirac brackets are the same as
Poisson
ones. Let us modify  constraints to obtain the system that is equivalent to initial one. 
For this example we do not have to
consider higher derivative constraints, but it is straightforward to incorporate them into the
 picture.
We will start from

\be
\Phi^a = X'^a + 2 \pi \a' P_c B^{ca}, \qquad P'^a = 0,
\la{Phi2SP}
\ee

where a runs from 1 to r. We have 2r second class constraints here.
To modify them introduce 2r new canonical variables $c^a$ and $P^{(c)}_a$, $[c^a, P^{(c)}_b] = \d^a{}_b$ 
and 2r additional second class
constraints. With modified initial ones we have

\be
\hat \Phi^a = X'^a + c^a, \qquad \phi_1^a = c^a - 2 \pi \a' P_c B^{ca},
\nonumber
\la{phi1S}
\ee

\be
P'_a = 0, \qquad P^{(c)}_a = 0.
\la{PPS}
\ee

Now, instead of 2r constraints we have 4r ones but we added 2r new variables, so
 number of degrees of freedom remains the same. The Dirac brackets \eq{D2S} are identically
equal to zero because $\hat \Phi$ commute with $X^a$, as in the case of string without B-field, and
the rest of constraints \eq{phi1S}, \eq{PPS} also do not give any contribution to Dirac brackets.
Procedure, which we used here, is different from conversion. Conversion 
assumes that
one can obtain initial constraints after fixing some particular values of additional variables.
In previous example we showed that after fixing all new variables equal to zero one has initial
constraint system. Here it is not the case. We can not fix c, otherwise it gives $P=0$.
One can ask why new model is equivalent to the previous one. 
It could be argued that what we did is just
redefining the variables. We do not have to fix any particular value
of $c^a$ or $P^{(c)}_a$ because additional constraints are not of the first class,
 but rather of the second class
and that is why they can be solved algebraically to produce initial system.

    In this section we considered simplified system comparing 
to one of \cite{Chu2}. The main difference in sets of 
constraints
is that in the \cite{Chu2} there were considered even higher derivatives of $\phi^a$ rather then $X'^a$.
Even for the model of \cite{Chu2} the 
application of the same technique is straightforward.

           Now we see that introduction of new variables even 
without changing the nature of constraints (from 
second to the first class) produces the
 Dirac brackets for the string endpoints which are equal to zero.
 Therefore we are able to show that endpoints of the string expressed in new variables
 commute between each other. Analysis of two-point functions is not so straightforward as
 for the case of the decoupling limit. After change of variables and 
  field redefinition $< \hat X^a, \hat X^b >$,  where $\hat X$ are new variables,
they are becoming effective string coordinates on the boundary. Therefore, two-point functions 
  becomes symmetric and using the
interpretation of
time ordering as operator ordering gives commutativity.

     Here we argue that noncommutativity of string endpoints is not fundamental but rather
removable and depends on the change of basis. It also possible to consider modified 
BRST charge and perform analysis of mode decomposition to show that it is possible
to introduce new algebraic variables which produce commutativity of the string endpoints on the
level of Dirac brackets. It is straightforward to show if one starts from constraints for the string
modes given in \cite{Lee} and modify them by introducing c and $P^{(c)}$ by the same way as in 
\eq{phi1S}.

\bigskip


\subsection{\bf  Membrane in constant C-field, decoupling limit }

In this section we discuss how to remove noncommutativity for the decoupling limit of 
membrane
ending on the M-5 branes. First of all, there are some crucial differences between taking 
decoupling 
limit of the string and membrane \cite{Berg1}, \cite{KS}. 
The only constant in eleven dimensions is the Plank
constant.
 In this case we can not take flat background metric generated by the  five-branes as was  
 explained in \cite{Berg1}. 
Moreover we have to consider stock of  five branes and probe membrane ending on one of the
five-branes. Then the decoupling limit could be found from the following :\\
  1.Bulk modes of the membrane must decouple and disappear.\\
2.String, that lives on the boundary, i.e. on the surface of the five-brane is 
completely described
by Wess-Zumino term only .\\
3.Then, after decoupling, all dynamics of the world-volume theory of five-brane is governed
by $(2,0)$ six-dimensional tensor multiplet.\\
  As was shown in \cite{Berg1} and \cite{Berg2} by rewriting membrane kinetic term in terms of
Lorentz harmonics it is possible to produce decoupling explicitly in the limit when $l_p \ra 0$ but 
open membrane metric is fixed. 
Following \cite{Berg1} one can find convenient form of Wess-Zumino membrane term
in the case of constant C-field.
The Wess-Zumino term for membrane has two contributions - from pullback of eleven-dimensional three
form $A^{(3)}$ to membrane surface and from pullback of five-brane two-form $B^{(2)}$ to membrane
boundary

\be
S_{WZ} = \int_{M^3} A^{(3)} + \int_{\p M^3} B^{(2)}.
\la{CBM1}
\ee

The constant C-field on the five-brane is given by $C^{(3)} = dB^{(2)} + A^{(3)}$, where 
$A^{(3)}$ is pullback to five-brane world-volume. The notion of nonlinear self-duality for C-field
on the five-brane is extremely important here \cite{ES1}.
At least locally we can write $A^{(3)} = da^{(2)}$. 
Excluding part of $A^{(3)}$ which gives zero pullback
to the five-brane world-volume and using the fact that C is constant one has

\be
C^{(3)}_{\m\n\l} X^\l = 3(B^{(2)}_{\m\n} + a^{(2)}_{\m\n}).
\la{CBM}
\ee

The components of constant C-field could be chosen in the form \cite{Wit1}

\be
C^{(3)}_{012} = - \frac {h}{\sqrt{1+l_p^6 h^2}}, \qquad C^{(3)}_{456} = h.
\la{ChM}
\ee

The action of the membrane ending on M-5-brane in decoupling limit is given by two terms \cite{Berg1}

\be
S = \int d^2 \s \frac {h}{3 \sqrt{1+l_p^6 h^2}} \e_{ijk} X^i \dot X^j X'^k + \int d^2 \s \frac {h}{3}
\e_{abc} X^a \dot X^b X'^c,
\la{a1-brane}
\ee

where 5-brane world-volume is decomposed into the two three-dimensional  pieces, where
$X^i$ lies on the membrane and $X^a$ are  five-brane coordinates normal to the membrane. 
 Prime denotes differentiation in respect to $\s$ and dot in respect to time.

Consider the second part of the \eq{a1-brane}

\be
S = \int d^2 \s \frac {h}{3} \e_{abc} X^a \dot X^b X'^c,
\la{a-brane}
\ee

The momentum is given by

\be
P_a = - \frac{h}{3} \e_{abc} X^b X'^c.
\la{mP}
\ee

With constraints

\be
\phi_a = P_a + \frac{h}{3} \e_{abc} X^b X'^c,
\la{mc}
\ee

and Poisson brackets

\be
[ \phi_a (\s), \phi_b (\s') ] = {2 \over 3}h \e_{abc} X'^c \d (\s - \s').
\la{mPhPh}
\ee

Among three constraints two are of the second class and one is the first class.
To see that project $\phi_a$ by $X, X', P$. Then first class constraint is

\be
\Psi = X'^a P_a,
\la{mc1}
\ee

and the second ones

\be
X^a P_a = 0, \qquad P^2 + {h \over 3} \e_{abc} P^a X^b X'^c = 0.
\la{mc2}
\ee

The equations of motion are 

\be
\e_{abc} \dot X^b X'^c = 0.
\la{EMM}
\ee

The fastest way to see appearance of non-associativity is to study
Dirac brackets between string coordinates. They are defined by

\be
[X^a (\s), X^b(\s') ]_D = [X^a, X^b]  - \int d \s'' [X^a, \Psi_1 (\s'')] 
 \frac{-1}{2P^2} [\Psi_2(\s''), X^b],
\la{D2mem}
\ee

and explicitly using expression of $P_a$ in terms of $X^a$

\be
[X^a (\s), X^b(\s') ]_D = \frac {X^{[a} P^{b]} }{2P^2} \d (\s -\s').
\la{D3mem}
\ee

Using definition of momenta as in \eq{mP} the Dirac brackets could be rewritten
only in terms of $X$'s as

\be
[X^a (\s), X^b(\s') ]_D = - \e_{abc} \frac {X^c}{h X'^2} \d (\s -\s').
\la{XXmem}
\ee

Last equations leads to non-associativity between membrane end-point
coordinates.

In this example, because we have a mixture of first and second class constraints it 
is more convenient to 
convert them directly in the mixture, i.e. we need to find additional variables which give us all
constraints of the first class. The discussion on conversion in mixture of first and second
class constraints and references therein could be found in \cite{BMRS}.
 To apply conversion covariantly in three dimensions orthogonal to membrane
world-volume let us introduce new variables $\x^a$ and $P^{(\x)}_b$ with 
$[\x^a (\s),P^{(\x)}_b (\s')] = \d^a_b \d (\s - \s')$. Thus, if we want to convert two second
class constraints into the first class, 
counting of degrees of freedom tells us that we need only two extra
variables but
we added six. The resolution of this problem is that not all of the above new variables are
independent and that is why we need to impose two extra first class constraints 
that removes four degrees
of freedom. This procedure gives us  two extra independent variables that we needed.
 It is convenient to proceed as follows. Let us define
 
\be
  X^{a+} = X^a + \x^a, \qquad P^{a+} = P^a + P^{(\x) a} , 
\la{XP+}
\ee
 
\be
  X^{a-} = X^a - \x^a, \qquad P^{a-} = P^a - P^{(\x) a} . 
\la{XP+1}
\ee

Then the modified constraints \eq{mc} could be rewritten as the following first class constraints

\be
\hat \phi_a = P_{a}^{-} + \frac {h}{3} \e_{abc}X^{b+}X'^{c+},
\la{HPhiM}
\ee

and they identically commute.
Now we need two additional constraints that commute with every other constraint .
It is possible to choose them in the form

\be
P^+ X^+ - P^- X^- = 0 ,  \qquad  P'^+ X^+  + P^- X'^-= 0.
\la{NewCM}
\ee

It is not unique but convenient choice. Now all the constraints \eq{HPhiM} and \eq{NewCM}
are of the first class.
The BRST charge is  given by the sum of the first class constraints multiplied by 
the corresponding ghosts
plus contribution from commutators.

This BRST charge describes theory equivalent to initial one not only on the classical but 
also on the quantum level. Fixing gauge $\x=0$ and $P_\x = 0$ gives initial theory.
In some sense one can think that model before conversion was one with fixed gauge symmetries, 
which could be restored on the level of BRST invariant action. By the same way one can proceed
for the first term in \eq{a1-brane}.

In this example we see that even for the nonlinear case of membrane boundary it is possible
to change variables to end up with commutative coordinates. This procedure is close to one in the
end of previous subsection except that for the case of membrane, one of the constraints is of the
first class and we were forced to have conversion directly in the mixture of two types of
constraints.


\subsection{\bf Non-associativity and string in non-constant B-field}

 In this subsection we study string in nonconstant B-field.
The action of the open string in non-constant B-field in linear order
in $H = dB$ where three-form $H$ is covariantly constant is given by
\cite{CS} (and references therein) 
where quantization of this model was considered

\be
S = \frac{1}{2} \int_\S \p X^a \p X^b g_{ab} + \frac{1}{2}
B_{ab} \int_\S \e \p X^a \p X^b + 
\frac{1}{3} H_{abc} \int_\S X^a \dot X^b X'^c .
\la{a}
\ee

In the decoupling limit, using $g_{ab} = 0$, the first kinetic
term disappears and the string action is given by the Lagrangian

\be
L = B_{ab} \dot X^a X'^b + \frac{1}{3} H_{abc} X^a \dot X^b X'^c .
\la{LBfield}
\ee

Momentum is given by

\be
P_a = B_{ab} X'^b - \frac{1}{3} H_{abc} X^b X'^c.
\la{P1}
\ee

 Constraints in this model are

\be
\Phi_a = P_a - B_{ab} X'^b + \frac{1}{3} H_{abc} X^b X'^c.
\la{C1}
\ee

It is more convenient to start from 
the limit when $B_{ab} = 2b \e_{abc} X^c $.
In this case $ H_{abc} = b \e_{abc}$  is constant.
In the Lagrangian \eq{LBfield} two terms could be combined together to produce
action that effectively could be rescaled to give

\be
S = \int d^2 \s \frac {b}{3} \e_{abc} X^a \dot X^b X'^c,
\la{a-bstr}
\ee

The momentum is given by

\be
P_a = - \frac{b}{3} \e_{abc} X^b X'^c.
\la{mPs}
\ee

With constraints

\be
\phi_a = P_a + \frac{b}{3} \e_{abc} X^b X'^c,
\la{mc-new}
\ee

and Poisson brackets

\be
[ \phi_a (\s), \phi_b (\s') ] = {2 \over 3}b \e_{abc} X'^c \d (\s - \s').
\la{mPhPh-new}
\ee

We see that this model is equivalent to the one of membrane in decoupling limit.
It is interesting to discuss this resemblance. It is also straightforward to
show that non-associativity for the case of string in constant B-field is also
could be removed as for the case of membrane. 

Therefore, two models, with commuting and with non-associative coordinates
are equivalent not only on classical but also on quantum level.

The quantization could be done using connection between BRST and
Fedosov quantization as in \cite{BGL} or by applying word-sheet approach
of \cite{CS}.

 For the case of more general $B_{ab}$ and constant $H_{abc}$ it is also
 possible to have a conversion.

It is still not clear how to find spectrum of theory described by
action \eq{a-bstr}.

As we mentioned before, the action of the form \eq{a-bstr} was also 
considered from point of view of
membrane boundary in the case of membrane in decoupling limit in \cite{Berg1}

\bigskip


\subsection{\bf Noncommutative 4-brane: Hamiltonian analysis}

The noncommutativity for higher D-branes was studied in \cite{JXL} and
references therein.

Let us start from  the boundary action of $D4$ brane ending on $D6$ brane, i.e.
 for the case of $p=6$ of D(p-2) brane, i.e. $p-2 = 4$ with 
4-dimensional boundary:

\be
S = \int d^4 \s b \e_{abcde} X^a \dot X^b \p_1 X^c \p_2 X^d \p_3 X^e,
\la{a-64}
\ee

here we chose constant H field in the form $H_{abcde} = b \e_{abcde}$ that could be generalized
for other constant values of H. The coordinates of the boundary are $(\t, \s_1,... \s_3)$ and
$\p_i$ is partial derivative in respect to $i$'s space coordinates of the brane.

The momentum is

\be
P_a = - b \e_{abcde} X^b \p_1 X^c \p_1 X^c \p_2 X^d \p_3 X^e .
\la{m64}
\ee

The constraints are

\be
\phi_a = P_a + b \e_{abcde} X^b \p_1 X^c \p_2 X^d \p_3 X^e .
\la{c64}
\ee

Poisson brackets for those constraints are given by

\be
[ \phi_a (\s), \phi_b (\s') ] = -2b \e_{abcde} \p_1 X^c \p_2 X^d \p_3 X^e  \d^3 (\s - \s').
\la{mPh64}
\ee

Those constraints are the mixture of the first and second class, 
but they could be covariantly separated 
into the  three first class constraints 

\be
P \p_1 X = 0, \qquad P \p_2 X = 0, \qquad P \p_3 X = 0,
\la{123-64}
\ee

and two second class constraints 

\be
PX =\Psi_1 , \qquad P^2 + b \e_{abcde} P^a X^b \p_1 X^c \p_2 X^d \p_3 X^e = \Psi_2 .
\la{2-64}
\ee

Then the noncommutativity appears on the level of Dirac brackets

\be
[X^a (\s), X^b(\s') ]_D = [X^a, X^b]  - \int d \s'' [X^a, \Psi_1 (\s'')] 
 \frac{-1}{2P^2} [\Psi_2(\s''), X^b],
\la{D2S-64}
\ee

and explicitly using \eq{m64}

\be
[X^a (\s), X^b(\s') ]_D = \frac {X^{[a} P^{b]} }{P^2} \d (\s -\s').
\la{D3S-64}
\ee

Using definition of momenta as in \eq{m64} the Dirac brackets could be rewritten
only in terms of $X$'s as

\be
[X^a (\s), X^b(\s') ]_D = - \e_{abcde} \frac {\p_1 X^c \p_2 X^d \p_3 X^e }
{b^2 ( \p_1 X \p_2 X \p_3 X )^2} \d^3 (\s -\s').
\la{XXn-64}
\ee

Last equations leads to higher-order non-associativity between brane boundary 
coordinates. As for the case of membrane in constant C-field, when
Dirac brackets for coordinates produce nonassociativity, it could be 
transfered to the quantum level. Non-associativity could be
also transfered to the quantum level in the framework of
BRST quantization. Here we used Hamiltonian approach to extract nonassociativity
for the 4-brane. On the other hand using properties of Hamiltonian systems it is possible
to find equivalent (on classical and quantum level)  system where all the
constraints will be of the first kind and Dirac brackets will be automatically equal to zero.
It leads to the fact that nonassociativity is not strict but rather removable that
will be showed exactly for the case of p=4 and also could be applied for the p=6.


\subsection{\bf Noncommutativity, non-associativity and WZW theory}

In this section we showed that, by appropriate change of variables and application of 
conversion,
noncommutative string and membrane as well as higher $D$ branes 
 appear to be equivalent to ordinary ones. 
It is possible to argue that noncommutativity of string/membrane endpoints,
 appearing fundamental, 
is rather removable and depends on field content and change of variables.
Not only conversion is doing its job, as for the case of string/membrane in 
decoupling limit, but also
equivalence  of two phase spaces without changing type of constraints is giving the 
same result. It is important to investigate the correspondence between noncommutative
string and ordinary one from worldsheet point of view. It is not quite clear yet
how this equivalence works for the product of vertex operators, which obey
noncommutative algebra.

 Here we presented only analysis of bosonic part of theories. 
It is interesting to consider whole supersymmetric
case extending  approach of \cite{CZ}, \cite{LW}, \cite{LZ}.
For supersymmetric case it is also important to understand situation on the level
of supersymmetric solutions and Dp-Dq brane systems \cite{MPT},\cite{Wit4}.

 Hamiltonian systems with boundaries were also extensively studied in \cite{Z} 
 from the similar point of
view. In \cite{Be1}, \cite{Sol} the different interpretation of 
boundary conditions as constraints was considered.
It is interesting to compare results coming from two approaches.

      It could be noted that membrane boundary string in decoupling limit and 
 critical C-field should be not only
 tensionless but also does not have gravity in its spectrum. This situation
is similar to a Little String 
 Theory concept, see \cite{TH} and references therein. 
 Usual world-sheet formulation of tensionless string \cite{T1}  has kinetic term quadratic
in fields together with null vectors. Worldsheet of such a string is a null surface.
Sometimes in literature it is called null-string.
 Unfortunately  this theory is anomalous and complete spectrum is not known yet.
It could be interesting to assume
 that tensionless string could be described by equations of motion \eq{mc}
and by Lagrangian \eq{a-brane}. But because \eq{a-brane} describes membrane boundary
in decoupling limit the mentioned above tensionless string does not contain graviton
in its spectrum and its low energy limit is described by $(2,0)$ six-dimensional theory.
Therefore there is a connection to Little String Theory.
Modified system of  the first class constraints \eq{HPhiM} and \eq{NewCM}
could be quantized in the BRST framework. It is straightforward to
apply operator BRST quantization. It is interesting to find a spectrum 
of this model as well as build vertex operators and study this model from worldsheet 
point of view. 
 On the other hand string in nonconstant B-field also leads to non-associativity, and in
 decoupling limit action for this string could be parametrized in such a way that it 
 coincides with action of membrane in constant C-field. It is not surpsizing feature
 because  double dimensional reduction of the membrane in C-field results in 
string in non-constant B-field and C-field is still constant.

It is interesting study other nonlinear models which posses the same 
equations of motion 
 as \eq{EMM} and similar to the membrane in the decoupling limit.
 
  First of all, because action \eq{a-brane} coincides with Wess-Zumino term of the
membrane it is useful to start from known nonlinear Wess-Zumino models in two dimensions.
Let us consider Wess-Zumino term of $SU(2)$ WZNW model. It is given by the action:

\be
S_{WZ-WZNW} = \frac{k}{96 \pi} \int dr \int  d^2 \x g^{-1} \dot g  g^{-1}
\e^{\m\n}\p_\m g
g^{-1} \p_\n g.
\la{WZWG}
\ee

Using  parametrization of $SU(2)$ in terms of Euler`s angles $\phi, \theta, \psi$ it
 is possible to rewrite \eq{WZWG} as

\be
S_{WZ-WZNW} = \frac {k}{4 \pi} \int d^2 \x \phi \e^{\m\n}  sin\theta \p_\m \theta 
\p_\n \psi.
\la{WZSU2}
\ee

At least for compact $ X^a$ it is possible to identify $(\phi, -cos\theta, \psi)$ with
 $(X^3, X^4, X^5)$
consequently. In this case WZ part of WZNW $SU(2)$ model could be rewritten as

\be
S_{WZ-WZNW} = \frac {k}{4 \pi} \int d^2 \x X^3(\dot X^4 X'^5 - \dot X^5 X'^4),
\la{WZWX}
\ee

and it is always possible to go to free-field realization of \eq{WZWX}.
This action gives the same equations of motion as \eq{EMM}. Thus, two models, string 
describing membrane endpoints on five-brane  and
Wess-Zumino part of $SU(2)$ WZNW model, are equivalent on the level of equations of motion,
 but  the actions are different. Therefore,
it could be useful to find a connection, if any,  between description of   
evolution of the boundary string on the surface of five-brane and 
 topological part of $SU(2)$ WZNW model at least for compact 
coordinates of membrane end-points.

\bigskip

\newpage




\chapter{\bf Conclusion}

 In this work we have analyzed supersymmetry and its realizations.
The major part of the dissertation has focused on the Poincar\'e superalgebra
generalized by the brane charges. 
We  study the representations of this superalgebra applying the generalization of
the Wigner method and generalized  mass-shell condition. This condition appears
naturally in the model of superparticle with brane charge coordinates which we
have presented in this work. 
Usually, superalgebra with central charges contains
 massive multiplets only, which are called BPS multiplets, and they
  correspond to nonperturbative states in theory. 
One of the remarkable properties of generalized mass-shell condition is 
that it leads to existence of
massive as well as massless representations.
 In this work we mostly
concentrate on massless multiplets of superalgebra with brane charges.

    We also give perturbative 
field-theoretical interpretation of the massless states.
The application of generalized Wigner methods shows that the multiplet shortening
occurs even in the case of massless representations of the superalgebra with brane
charges. We explore these ultra short multiplets first in four dimensions and then
in $D=11$. In eleven dimensions we show that there are other supermultiplets apart
from supergravity multiplet.
The interesting properties of these new supermultiplets in $D=11$ is
that each state is
characterized not only by the value of the usual Casimir operators but also by new Casimir
operators constructed from the two- and five- form charges.  
  
    In this dissertation a superparticle realization of Poincar\'e  
    superalgebra with brane charges is
presented. The superparticle model describes massless as well as
massive particles. The symmetries of the superparticle model with brane charge
coordinates is studied. In addition to global supersymmetry generated by the
superalgebra with brane charges, there are new bosonic gauge symmetries. 
We also show that the superparticle action possesses kappa
symmetry and reduces, in the limit of zero brane charges, to
the ordinary Brink-Schwarz superparticle.  The introduction of
twistor variables in the form of generalized Penrose correspondence in arbitrary
dimension is useful for understanding  the generalized mass-shell condition
as well as the properties of the kappa symmetry. Interestingly, 
we show that different number of spinorial
constituents in Penrose decomposition corresponds to
either massive or massless superparticle models.

     To proceed with quantization of massive/massless
superparticle with brane charge coordinates we have used the 
twistorial formulation of the
superparticle and analyzed the constraint system of this model. We have shown that
the constraint system is a mixture of the first and second class constraints. Those
constraints could not be separated covariantly, and we apply conversion procedure
directly in the mixture of first and  second class constraints.
In this procedure, we extend the phase space of the theory by introducing new
variables which enable us to produce a system of first class constraints only.
We show that the model described by these converted first class constraints
is equivalent to the initial one not only at the classical but also at the quantum
level.
Using Dirac quantization of massive/massless superparticle with brane charge
coordinates we have found the spectrum of a linearized supersymmetric field theory.
We have focused on the quantization in $D=4$, where we have obtained not only
the spectrum but also the linearized equations of motion. However,
 we have also discussed the generalization of the quantization procedure to
eleven dimensional case. In eleven dimensions the Poincar\'e superalgebra with
brane charges coincides with M-theory superalgebra and superparticle with brane
charge coordinates realizes this algebra. The quantization produces a 
spectrum of a eleven dimensional linearized  supersymmetric field theory. 
In this work we have  discussed possible
connection between particle spectrum and spectrum of linearized perturbative
M-theory.

   We also investigate the higher dimensional unification of Type IIA/B and
heterotic superalgebras. We have shown that the unification 
is possible starting
from a superalgebra with brane charges in D$>$11. 
The multi-particle realization 
of the nonlinear deformed superalgebra in dimensions beyond eleven
with more then one timelike directions
is also presented. The introduction of super Yang-Mills background into the 
multi-superparticle action is considered.
We also discuss superstring models in D$>$11.

   In addition to the superparticle, we have also studied
    higher branes using superembedding
approach. In this approach both worldvolume of the p-brane and the target space are 
supermanifolds.
We have applied superembedding approach to construct the equations of motions
and an action for a new class of branes, called L-branes. The L-branes have linear
multiplet on their worldvolume which includes higher rank p-form. The action we have
found for the L-brane generalizes the Born-Infeld type action.

   In this dissertation we have applied BRST/Hamiltonian methods 
to study properties of strings and
membranes in background fields. For the case of string in constant B-field,
the application of the conversion procedure has
given a stringy origin of the Seiberg-Witten map between noncommutative and ordinary
field theory which appears in the low energy limit of string theory.

\bigskip
\newpage





\appendix{Properties of spinors and $\c$-Matrices in arbitrary
dimensions}



Here we collect the properties of spinors and Dirac $\c$-matrices in
$(s,n)$ dimensions where $s(n)$ are the number of space(time)
coordinates. The Clifford algebra is $\{\c_\m,\c_\n\}=2 \eta_{\m\n}$,
where $\eta_{\m\n}$ has the signature in which the time-like directions
are negative and the spacelike directions positive. The possible reality
conditions on spinors are listed in Table 1, where $M, PM, SM, PSM$ stand for
Majorana, pseudo Majorana, symplectic majorana and pseudo symplectic
Majorana, respectively \cite{kt,ss}
\footnote
{Corrections to formulae in p. 5\,\&\,6 of \cite{ss}: 
(a) Change eq. (2) to: $(\C^{d+1})^2=(-1)^{(s-t)(s-t-1)/2}$ (our $n$ here is 
denoted by $t$ in \cite{ss}), 
(b) eq. (3) should read $\C^\dagger_\m=(-1)^t A\C_\m A^{-1}$, 
(c) change the last formula in eq. (5) to $B=CA$, 
(d) multiply eq. (6) with $\e$ on the right hand side, 
(e) the prefactor in eq. (8) should be $2^{-\left[d/2\right]}$, 
(f) interchange the indices $\n_1$ and $\n_2$ in the last term of eq. (9). 
Note that here we have let $ C\ra C^{-1}$ relative to \cite{ss}. }
(see below). An additional chirality condition can be imposed for
$s-n=0\ mod\ 4$. 

The symmetry properties of the charge conjugation matrix $C$ and $(\c^\m
C)_{\a\b}$ are listed in Table 2. The sign factors $\e_0$ and $\e_1$
arise in the relations
\be
C^T=\e_0 C\ , 
\qq  
(\c^\m C)^T = \e_1 (\c^\m C)\ .
\la{sp}
\ee
This information is sufficient to deduce the symmetry of 
$(\c^{\m_1\cdots \m_p} C)_{\a\b}$ for any $p$, since the symmetry
property alternates for $p~{\rm mod}~2$. In any dimension with $n$
times, one finds
\be
(\c^{\m_1\cdots \m_p} C)^T= \e_p\,(\c^{\m_1\cdots \m_p} C)\ , 
\qq\quad
\e_p \equiv \e\,\eta^{p+n}\,(-1)^{(p-n)(p-n-1)/2}\ ,
\ee
where $\eta$ is a sign factor. Note that $\e_n=\e$ and
$\e_{n-1}=-\e\eta$. All possible values of $(n,s,p)$ in which $\e_p=+1$,
i.e. the values of $p$ for which $\c^{\m_1\cdots \m_p} C$ is symmetric
(the antisymmetric ones occur for $mod\ 4$ complements $p$) are listed
in Table 3. Other useful formulae are:
\be
\c_\m^T = (-1)^n\,\eta~C^{-1} \c_\m C\ , 
\qq
\c_\m^*= \eta B\c_\m B^{-1}\ ,
\qq
\c_\m^\dagger = (-1)^n A C_\m A^{-1}\ .
\ee
We can choose $A=\c_0\c_1\cdots \c_{n-1}$. Note that $A=BC$.
The chirality matrix in even dimensions $d$ can be defined as
\be
\c_{d+1}\equiv\pm (-1)^{(s-n)(s-n-1)/4}\, \c_0\c_1\cdots \c_{d-1}\ ,
\qq (\c_{d+1})^2=1\ .
\ee
We can choose the sign in the first equation such that the overall factor in
front is $+1$ or $+i$. Using the matrix $B$, we can express the reality
conditions on a (pseudo) Majorana spinor $\psi^* = B\psi$ and a
(pseudo)symplectic Majorona spinor as $(\psi_i)^* = \Omega^{ij}B\psi_j$,
where $\Omega^{ij}$ is a constant antisymmetric matrix satisfying
$\Omega_{ij}\Omega^{jk}=-\d_i^k$ and the index $i$ labels a pseudo-real
representation of a given Lie algebra which admits such a representation
(e.g. the fundamental representations of $Sp(n)$ and $E_7$).
\\

\begin{tabular}{lr}
\begin{minipage}[t]{4in}
\indent{\bf Table 1:\ }{\it Spinor types in $(s,n)$ dimensions} \\
\vspace{0.5cm}
\begin{tabular}{|c|c|c|c|}
\hline
 {}  &    {}  &      {}      &       {}       \\
$\e$ & $\eta$ & $(s-n)\ mod\ 8$ &   $Spinor\ Type $  \\
\hline\hline
 $+$   &    $+$   &   0,1,2      &       M       \\
\hline
 $+$   &    $-$   &   6,7,8      &       PM      \\
\hline
 $-$   &    $+$  &    4,5,6      &       SM      \\
\hline
 $-$   &    $-$  &    2,3,4      &       PSM      \\
\hline
\end{tabular}
\end{minipage}
\end{tabular}


\vspace{0.5in}


\begin{tabular}{lr}
\begin{minipage}[t]{4in}
\indent{\bf Table 2:\ }{\it Symmetries of $C$ and $(\c^\m C)$ in $(s,n)$ }\\
{ \it dimensions } \\
\vspace{0.5cm}
\hspace{0.25in}
\begin{tabular}{|c|c|c|}
\hline
{} & {}  & {} \\
$n\ mod\ 4$ & $\e_0$ & $\e_1$ \\
\hline\hline
0   & $+\e$      &   $+\e\eta$      \\
\hline
1   & $-\e\eta$  &   $+\e$          \\
\hline
 2   & $-\e$     &   $-\e\eta$      \\
\hline
 3   & $+\e\eta$  &  $-\e$         \\
\hline
\end{tabular}
\end{minipage}
\end{tabular}

\vspace{0.5in}


\begin{tabular}{l}
\begin{minipage}[t]{3.5in}
\indent{\bf Table 3:\ } {\it Symmetric $(\c^{\m_1\cdots \m_p}C)$
in $(s,n)$ dimensions.}
{\it In bold cases, an extra Weyl condition is possible. }\\
\vspace{0.5cm}
\begin{tabular}{||c|c|c||c|c|c||}
\hline\hline
 {}  &    {}  &      {}      &        
 {}  &    {}  &      {}     \\
$n\ mod\ 4$ & $ s\ mod\ 8 $ & $p\ mod\ 4$   & 
$n\ mod\ 4$ & $ s\ mod\ 8 $ & $p\ mod\ 4$    \\
\hline\hline
${\bf 1}$ &  ${\bf 1},2,3$  &  ${\bf 1},2$ & 
${\bf 3}$ &  ${\bf 3},4,5$  &  ${\bf 3},4$  \\
\hline
    & ${\bf 1},7,8$   &  ${\bf 1},4$ &     
    & ${\bf 3},1,2$   &  ${\bf 3},2$  \\    
\hline
    & ${\bf 5},6,7$   &  ${\bf 3},4$ &     
    & ${\bf 7},8,1$   &  ${\bf 1},2$  \\   
\hline
    & ${\bf 5},3,4$   &  ${\bf 3},2$ &     
    & ${\bf 7},5,6$   &  ${\bf 1},4$  \\  
\hline\hline
${\bf 2}$ & ${\bf 2},3,4$ & ${\bf 2},3$ & 
${\bf 4}$ & ${\bf 4},5,6$ & ${\bf 4},1$  \\
\hline
    & ${\bf 2},8,1$   &  ${\bf 2},1$ &  
    & ${\bf 4},2,3$   &  ${\bf 4},3$  \\
\hline
    & ${\bf 6},7,8$   &  ${\bf 4},1$  &
    & ${\bf 8},1,2$   &  ${\bf 2},3$  \\
\hline
    & ${\bf 6},4,5$   &  ${\bf 4},3$  &
    & ${\bf 8},6,7$   &  ${\bf 2},1$  \\
\hline\hline
\end{tabular}
\end{minipage}
\end{tabular}


\appendix{Superspace conventions}

\subsection*{\bf D=9 $\to$ d=6}

The superembeddings studied in this note are taken to break half of the
original target space supersymmetries. This implies that the
worldvolume of a brane has $1 \over 2$ the number of fermionic coordinates of
the target space. In the example of the L5-brane this means that the
target space $\unM$ has $9$ bosonic and $16$ fermionic coordinates while
the worldvolume of the brane $M$ has $6$ bosonic and $8$ fermionic
coordinates. To study the embedding in more detail it is convenient to
split the $9$-dimensional $\c$-matrices in a way that reflects the
embedding, i.e. $SO(1,8) \rightarrow SO(1,5) \xz SU(2)$, as follows:

\begin{eqnarray}
(\C^{a})_{\ua}{}^{\ub} & = & \pmatrix{
\bf 0 &(\c^{a})_{\a \b} \cr
(\c^{a})^{\a \b} &\bf 0}\d_{i}{}^j \nonumber \\
(\C^{a'})_{\ua}{}^{\ub} & = & \pmatrix{
\d_{\a}{}^{\b} &\bf 0 \cr
\bf 0 &-\d^{\a}{}_{\b}}(\c^{a'})_{ij}.
\label{97}
\end{eqnarray}

The $9$-dimensional charge conjugation matrix $C_{\underline{\a \b}}$ is given
by

\be
C_{\underline{\a \b}} = \e_{ij} \pmatrix{
\bf 0 &-\d_\a{}^\b \cr
\d^\a{}_\b &\bf 0}.
\label{98}
\ee

The conventions for the spinors are chosen such that $9$-dimensional
spinors $\psi_{\ua}$ are pseudo-Majorana while $6$-dimensional spinors
are chosen to be symplectic Majorana-Weyl. This implies that

\be
\psi_{\ua} \rightarrow \left\{ \matrix{
\psi_\a  & = & \psi_{\a i} \cr
\psi_{\a'} & = & \psi_i^{\a}.} \right.
\label{99}
\ee

$SU(2)$-indices can be raised and lowered with $\e_{ij}$ using the
convention that $\l_i = \l^j \e_{ji}$. A useful equation for
manipulations of $SU(2)$ $\c$-matrices is

\be
(\c^{c'})_{jk}(\c_{c'})_{il} = \e_{ij}\e_{kl}+\e_{jl}\e_{ik}.
\label{100}
\ee

The conversion between spinor and Lorentz indices is governed by

\be
K_{\b}{}^{\c} = {1 \over 4} \d_j{}^k (\c^{bc})_{\b}{}^{\c} K_{bc}
+ \d_{\b}{}^{\c} K_{jk}
\label{102}
\ee

and

\be
K_{\b}{}^{\c'} = -{1 \over 2}(\c^{b})_{\b \c}(\c^{c'})_j{}^k K_{bc'}.
\label{103}
\ee

\subsection*{\bf D=8 $\to$ d=5}

\be
C_ {\underline{\a \b}}=\e_{ij} \pmatrix{
C_{\a \b}  & \bf 0  \cr
\bf 0  & -C_{\a\b} },
\la{104}
\ee

\begin{eqnarray}
(\C^{a'})_{\ua}{}^{\ub} & = & (\c^{a'})_i{}^j \pmatrix{
\bf  0 & \d_{\a}{}^{\b} \cr
{\d}^{\a}{}_{\b} & \bf  0 }  \nonumber \\
(\C^{a})_{\ua}{}^{\ub} & = & \d_i{}^j \pmatrix {
(\c^a)_{\a}{}^{\b} & \bf 0 \cr
\bf 0 & - (\c^a)^{\a}{}_{\b} } ,
\la{105}
\end{eqnarray}

\be
\psi_{\ua} \rightarrow \left\{ \matrix{
\psi_\a  & = & \psi_{\a i} \cr
\psi_{\a'} & = &\phi_{\a i} .} \right.
\la{106}
\ee

\subsection*{\bf D=7 $\to$ d=4}

\bea
C_ {\underline{\a \b}}=\e_{ij} \pmatrix{
\bf 0  & - C_{\a \b} \cr
C^{\a\b} &\bf 0  },
\la{107}
\eea

\begin{eqnarray}
(\C^{a'})_{\ua}{}^{\ub} & = & -(\c^{a'})_{ij} \pmatrix{
\d_{\a}{}^{\b} & \bf  0 \cr
\bf  0 & -{\d}_{a}{}^{\b} }  \nonumber \\
(\C^{a})_{\ua}{}^{\ub} & = & \d_i{}^j \pmatrix {
\bf 0 &(\c^a)_{\a}{}^{\b} \cr
(\c^a)_{\a}{}^{\b} & \bf 0 } ,
\la{108}
\end{eqnarray}

\be
\psi_{\ua} \rightarrow \{ \matrix{
\psi_\a  & = & \psi^{+}_{\a i} \cr
\psi_{\a'} &  =  & {\psi^{-}_{\a i}} }  \}.
\la{109}
\ee

Our conventions for superforms are as follows. Gauge potentials $A_q$ are
defined by

\be
A_q = {1 \over q!} dz^{M_q} \ldots dz^{M_1} A_{M_1 \ldots M_q}
\label{110}
\ee

and field strengths by

\be
F_q = {1 \over (q+1)!} dz^{M_{(q+1)}} \ldots dz^{m_1} F_{M_1 \ldots M_{(q+1)}}.
\label{111}
\ee

Wedge products between forms are understood. The exterior derivative $d$
acting on a $q$-form $\rho$ we define by

\be
d \rho = dz^{M_{(q+1)}} \ldots dz^{M_1} \partial_{M_1} \rho_{M_2 \ldots
M_{(q+1)}}
\label{112}
\ee

and the interior product $i_v$ by

\be
i_v \rho = q dz^{M_q} \ldots dz^{M_2} v^{M_1} \rho_{M_1 \ldots M_{(q+1)}}.
\label{113}
\ee

Both act from the right on products of forms, i.e.

\be
d( \omega \rho) = \omega (d \rho) + (-1)^q (d \omega) \rho
\label{114}
\ee

and

\be
i_v (\omega \rho) = \omega (i_v \rho) + (-1)^q (i_v \omega ) \rho
\label{115}
\ee

where $\omega$ is a $p$-form and $\rho$ is a $q$-form.

\bigskip




\begin{thebibliography}{177}



\bm{c2} E.Witten, D.Olive Phys.Lett. B { \bf 78} (1978) 97.

\bm{M} P.K.Townsend, {\it M-theory from its superalgebra}, hep-th/9712004.

\bm{pkt1} P.K. Townsend, {\it Four lectures on M-theory}, hep-th/9612121.

\bm{c9} J.A. de Azcarraga, J.P.Gauntlett, J.M.Izquierdo, P.K.Townsend
        Phys.Rev.Lett.  {\bf 63} (1989) 2443. \\
        J.A. de Azc\'arraga, P. K. Townsend , {\it Superspace geometry and     
        classification of supersymmetric extended objects},
        Phys. Rev. Lett. { \bf 62} (1989) 2579.

\bm{ma3} E. Sezgin, {\it The M-algebra}, Phys. Lett. { \bf B392} (1997)
	323, hep-th/9609086. 

\bm{Gstr} M. Green, {\it Supertranslations, superstrings and Chern-Simons forms}, 
       Phys. Lett. { \bf B223} (1989) 157.

\bm{Sstr} W. Siegel, {\it Randomizing the superstring}, Phys. Rev. { \bf D50}
        (1994) 2799, hep-th/9403144.

\bm{b2} I. Bars, {\it S-theory}, Phys. Rev. { \bf D55} (1997) 2373, 
	hep-th/9607112.

\bm{c1} S.Ferrara, C.A.Savoy, B.Zumino Phys.Lett.B { \bf 100} (1981) 393.
	
\bm{c3} P.Fayet, Nucl.Phys.B { \bf 138} (1979) 137.

\bm{c4} P.H.Dondi, M.Sohnius, Nucl.Phys.B { \bf 81} (1974) 317, \\
        M.Sohnius, Nucl.Phys.B { \bf 138} (1978) 109.

\bm{c5} J.T.Lopuszanski, M.Wolf Univ. of Wroclaw preprint No. 482 (1979).

\bm{c6} S.Ferrara, J.Scherk, B.Zumino, Nucl. Phys.B { \bf 121} (1977) 393.

\bm{c7} P.S.Howe, J.M.Izquierdo, G.Papadopoulos, P.K.Townsend,
        Nucl.Phys. { \bf B467} 183-214,1996 

\bm{c8} R.Haag, J.T.Lopuszanski, M.Sohnius, Nucl.Phys.B { \bf 88} (1975)
        257.

\bm{Hull} J.P.Gauntletett, C.M.Hull, {\it BPS states with extra
         supersymmetry}, hep-th/9909098.

\bm{3/4} J.P.Gauntlett, G.W.Gibbons,C.M.Hull,P.K.Townsend {\it BPS states
             of the $D=4$ $N=1$ supersymmetry}, hep-th/0001024.

\bibitem{Ferr}
S.~Ferrara and M.~Porrati,
``AdS(5) superalgebras with brane charges,''
Phys.\ Lett.\ B { \bf 458}, 43 (1999)
hep-th/9903241 \\
R.~D'Auria, S.~Ferrara and M.~A.~Lledo,
``On central charges and Hamiltonians for 0-brane dynamics,''
Phys.\ Rev.\ D { \bf 60}, 084007 (1999)
hep-th/9903089 \\
S.~Ferrara and M.~Porrati,
``Central extensions of supersymmetry in four and three dimensions,''
Phys.\ Lett.\ B { \bf 423}, 255 (1998)
hep-th/9711116 \\
S.~Ferrara and J.~Maldacena,
``Branes, central charges and $U$-duality invariant BPS conditions,''
Class.\ Quant.\ Grav.\ { \bf 15}, 749 (1998)
hep-th/9706097.



\bibitem{dW1}
B.~de Wit and H.~Nicolai,
``Hidden symmetries, central charges and all that,''
hep-th/0011239.\\
B.~de Wit,
``M-theory duality and BPS-extended supergravity,''
hep-th/0010292.

\bibitem{Shif}
A.~Gorsky and M.~Shifman,
``More on the tensorial central charges in N = 1 supersymmetric gauge  theories (BPS wall junctions and strings),''
Phys.\ Rev.\ D { \bf 61}, 085001 (2000)
hep-th/9909015.





\bibitem{RS1}
I.~Rudychev and E.~Sezgin,
``Superparticles, p-form coordinates and the BPS condition,''
Phys.\ Lett.\ B { \bf 424} (1998) 60. \\
I.~Rudychev and E.~Sezgin,
``Superparticles in D $>$ 11 revisited,''
hep-th/9711128.

\bm{BLS} I.Bandos and J.Lukierski, Mod.Phys.Lett. { \bf A14},1257 (1999),\\
         I.Bandos, J.Lukierski and D.Sorokin, hep-th/9904109.

\bm{Pi} P.S.Howe, S.Penati, M.Pernici, P.K.Townsend, Phys.Lett.{ \bf B215}:555,1988. 

\bm{K1}  Piet Claus, Renata Kallosh, J. Rahmfeld, Phys.Lett. { \bf B462}
(1999) 285, hep-th/9906195.

\bibitem{MT}
R.~R.~Metsaev and A.~A.~Tseytlin,
``Superparticle and superstring in AdS(3) x S**3 Ramond-Ramond  background in light-cone gauge,''
hep-th/0011191.

\bm{BAIL}  Igor A. Bandos, Jose A. de Azcarraga, Jose M. Izquierdo, Jerzy Lukierski,
          {\it BPS states in M-theory and twistorial constituents},
          hep-th/0101113.

\bm{RSS-new} I.Rudychev, E.Sezgin, P.Sundell "New representations of super
Poincare algebra with brane charges and Higher Spin Theory" , in preparation.

\bm{W1}O. Barwald, P. West,   {\it Brane Rotating Symmetries and the Fivebrane Equations 
         of Motion}, Phys.Lett. { \bf B476} (2000) 157,

 P. West, {\it Hidden Superconformal Symmetry in M Theory}, JHEP { \bf 0008} (2000) 007.

\bm{B1} Itzhak Bars, Costas Kounnas, Phys.Lett. { \bf B402} (1997) 25, 
                                     Phys.Rev. { \bf D56} (1997) 3664,\\
        I. Bars, C. Deliduman, D. Minic,  Phys.Lett. {\bf B457} (1999) 275,
	                                  Phys.Rev. { \bf D59} (1999) 125004.     
	

I.~Bars and C.~Deliduman,
``High spin gauge fields and two-time physics,''
hep-th/0103042

I.~Bars,
``Survey of two-time physics,''
hep-th/0008164

I.~Bars,
``2T physics formulation of superconformal dynamics relating to twistors  
and supertwistors,''
Phys.\ Lett.\ B { \bf 483}, 248 (2000)
hep-th/0004090 

I.~Bars,
``Two-time physics in field theory,''
Phys.\ Rev.\ D { \bf 62}, 046007 (2000)
hep-th/0003100 

I.~Bars,
``Two-time physics with gravitational and gauge field backgrounds,''
Phys.\ Rev.\ D { \bf 62}, 085015 (2000)
hep-th/0002140 

I.~Bars, C.~Deliduman and D.~Minic,
``Strings, branes and two-time physics,''
Phys.\ Lett.\ B {\bf 466}, 135 (1999)
hep-th/9906223. 







       
\bibitem{RSS}
I.~Rudychev, E.~Sezgin and P.~Sundell,
``Supersymmetry in dimensions beyond eleven,''
Nucl.\ Phys.\ Proc.\ Suppl.\ { \bf 68}, 285 (1998)
hep-th/9711127.

\bm{BL} I.Bandos and J.Likierski {\it New superparticle models outside the
HLS supersymmetry scheme}, hep-th 9812074.

\bibitem{Gun1}
M.~Gunaydin,
``Unitary supermultiplets of OSp(1/32,R) and M-theory,''
Nucl.\ Phys.\ B { \bf 528}, 432 (1998)
hep-th/9803138.

\bibitem{ns}
H.~Nishino and E.~Sezgin,
``Supersymmetric Yang-Mills equations in 10+2 dimensions,''
Phys.\ Lett.\ B { \bf 388}, 569 (1996)
hep-th/9607185.

\bibitem{es}
E.~Sezgin,
``Super Yang-Mills in (11,3) dimensions,''
Phys.\ Lett.\ B { \bf 403}, 265 (1997)
hep-th/9703123.


\bibitem{Nishino:2000cv}
H.~Nishino, ``Supersymmetric Yang-Mills theory in eleven dimensions,''
Phys.\ Lett.\ B { \bf 492}, 201 (2000)
hep-th/0008029.

\bibitem{Nishino:1999ck}
H.~Nishino,
``Twelve-dimensional supersymmetric gauge theory as the large N limit,''
Phys.\ Lett.\ B { \bf 452}, 265 (1999)
hep-th/9901104.

\bibitem{Nishino:1999qn}
H.~Nishino,
``Supergravity theories in D $>$ 12 coupled to super p-branes,''
Nucl.\ Phys.\ B { \bf 542}, 217 (1999)
hep-th/9807199.


\bibitem{Nishino:1998ia}
H.~Nishino,
``Lagrangian and covariant field equations for supersymmetric Yang-Mills  
theory in 12D,''
Phys.\ Lett.\ B { \bf 426}, 64 (1998)
hep-th/9710141.

\bibitem{hn2}
H.~Nishino,
``Supersymmetric Yang-Mills theories in D $>=$ 12,''
Nucl.\ Phys.\ B { \bf 523}, 450 (1998)
hep-th/9708064.

\bibitem{Nishino:1998sw}
H.~Nishino,
``N = 2 chiral supergravity in (10+2)-dimensions as consistent background 
 for super (2+2)-brane,''
Phys.\ Lett.\ B { \bf 437}, 303 (1998)
hep-th/9706148.

\bm{nahm} W. Nahm, {\it Supersymmetries and their representations},
	  Nucl. Phys. { \bf B135} (1978) 149.



\bm{pvn} L. Castellani, P. Fr\'e, F. Giani, K. Pilch and P. van
	Nieuwenhuizen, {\it Beyond D=11 supergravity and Cartan Integrable
	systems}, Phys. Rev. { \bf D26} (1982) 1481.



\bm{duff} M. Blencowe and M.J. Duff, {\it Supermembranes and signature
	of spacetime}, Nucl. Phys. { \bf B310} (1988) 387.
	
	
	
	
	
\bm{hull} C.M. Hull, {\it String dynamics at strong coupling}, Nucl.
	Phys. { \bf B468} (1996) 113, hep-th/9512181.

\bm{vafa} C. Vafa, {\it Evidence for F-theory}, Nucl.
	Phys. { \bf B469} (1996) 403, hep-th/9602022.

\bm{km1} D. Kutasov and E. Martinec, {\it New principles for
	string/membrane unification}, Nucl. Phys. { \bf B477} (1996) 652,
	hep-th/9602049. 

\bm{at} A. Tseytlin, {\it Self-duality of Born-Infeld action and Dirichlet
	3-brane of Type IIB superstring theory}, Nucl. Phys. { \bf B469} 
	(1996) 51, hep-th/9602077.

\bm{pkt2} P.K. Townsend, {\it p-brane democracy}, hep-th/9507048.
          {\it D-branes from M-branes}, Phys. Lett. { \bf B373} 
	(1996) 68, hep-th/9512062. \\
           {\it Membrane tension and manifest IIB
	S-duality}, hep-th/9705160.

\bm{jr} D.P. Jatkar and S.K. Rama, {\it F-theory from Dirichlet 3-branes},
	 Phys. Lett. { \bf B388} (1996) 283, hep-th/9606009.

\bm{b3} I. Bars and C. Kounnas, {\it A new supersymmetry}, hep-th/9612119.


\bm{n1} H. Nishino, {\it Supergravity in 10+2 dimensions as consistent
	background for superstring}, hep-th/9703214.

\bm{b5} I. Bars, {\it A case for 14 dimensions}, Phys. Lett. { \bf B403}
	(1997) 257, hep-th/9704054.

\bibitem{RS}
I.~Rudychev and E.~Sezgin,
``Superparticles in D $>$ 11,''
Phys.\ Lett.\ B {  415}, 363 (1997)
hep-th/9704057.

 
\bm{b6} I. Bars and C. Kounnas, {\it  String and particle with two times},
	Phys. Rev. { \bf D56} (1997) 3664, hep-th/9705205.

\bm{b7} I. Bars and C. Deliduman, {\it Superstrings with new
	  supersymmetry in (9,2) and (10,2) dimensions}, hep-th/9707215.

\bm{n2} H. Nishino, {\it Supersymmetric Yang-Mills theories in 
	$D\ge 12$}, hep-th/9708064.

\bm{b8} I. Bars and C. Deliduman, {\it Gauge principles for
	multi-superparticles}, hep-th/9710066.

\bm{n3}  H. Nishino, {\it Lagrangian and covariant field equations
	for supersymmetric Yang-Mills theory in 12D}, hep-th/9710141. 

\bm{es1} E. Sezgin, {\it Super $p$-from charges and a reformulation of the
        supermembrane action in eleven dimensions}, in {\it Leuven Notes in 
        Mathematical Physics, Series B, Vol. 6}, eds. B. de Wit et al, 
	hep-th/9512082.

\bm{db} J. de Boer, F. Harmsze and T. Tjin, {\it Nonlinear finite W
	symmetries and applications in elementary systems}, Phys. Rep. 
	{ \bf 272} (1996) 139, hep-th/9503161. 
		

\bm{w1} F. Barbarin, E. Ragouchy and P. Sorba, {\it Finite W-algebras
	and intermediate statistics}, hep-th/9410114. 

\bm{w2} K. Schoutens, A. Sevrin and P. van Nieuwenhuizen, 
	{\it Nonlinear Yang-Mills theories}, Phys.Lett.{ \bf B255}(1991) 549.

\bm{w3} M. Rocek, {\it Representation theory of the nonlinear 
	SU(2) algebra}, Phys.Lett. { \bf B255} (1991) 554.

\bm{kt} T. Kugo and P.K. Townsend, {\it Supersymmetry and the division
	algebras}, Nucl. Phys. { \bf B221} (1984) 368.

\bm{pkt3} P.K. Townsend, {\it p-brane democracy}, hep-th/9507048.

\bm{bars1} I. Bars, {\it Supersymmetry, p-brane duality and hidden spacetime
	dimensions}, Phys. Rev. { \bf D54} (1996) 5203, hep-th/9604139.

\bm{km2} D. Kutasov, E. Martinec and M. O'Loughlin, {\it Vacua of M-theory and 
         N=2 strings}, Nucl. Phys. { \bf B477} (1996) 675, hep-th/9603116.

\bm{hull2} C.M. Hull, {\it Actions for (2,1) sigma-models and strings},
	hep-th/9702067.

\bm{ma1} E. Bergshoeff and E. Sezgin, {\it New spacetime algebras and
	their Kac-Moody extension}, Phys. Lett. { \bf B232} (1989) 96.

\bm{ma2} E. Bergshoeff and E. Sezgin, {\it Super $p$-brane theories and
	new spacetime algebras}, Phys. Lett. { \bf B354} (1995) 256, 
	hep-th/9504140.

\bm{green}  Michael B. Green, Michael Gutperle, Hwang-h. Kwon, JHEP { \bf 9908} 
         (1999) 012.
	 
\bm{BLPS} I.Bandos, J.Lukierski, C.Preitschopf and D.Sorokin,{\it OSp
          supergroup manifolds, superparticles and supertwistors},
          hep-th/9907113.

\bm{H1} P.S.Howe,Phys.Lett.{ \bf B258}:141,1991,Addendum-ibid.{ \bf B259}:511,1991 
 
 
\bm{tc} T. Curtright, {\it Are there any superstrings in eleven dimensions?}, 
        Phys. Rev. Lett. { \bf D60} (1988) 393.
                   Phys.Lett. { \bf B273}:90,1991.
	 
	 
\bibitem{AmBa}
R.~Amorim and J.~Barcelos-Neto,
``Superstrings with tensor degrees of freedom,''
Z.\ Phys.\ C { \bf 64} (1994) 345.

\bm{RS-unpub} I.Rudychev, E.Sezgin {\it Strings and Membranes with
central charge coordinates}, unpublished.

\bibitem{LZnew}
A.~A.~Zheltukhin and U.~Lindstrom,
``Strings in a space with tensor central charge coordinates,''
hep-th/0103101.




\bm{Berk1} N.Berkovits, {\it Super-Poincare Covariant Quantization of the
                      Superstring} hep-th/0001035.

\bm{w} E. Witten, {\it Twistor-like transform in ten dimensions}, 
	Nucl.Phys { \bf B266} (1986) 245. 

\bm{bst} E. Bergshoeff, E. Sezgin and P.K. Townsend, {\it Supermembranes and
         eleven-dimensional supergravity}, Phys. Lett. { \bf 189B} (1987) 75;
        {\it Properties of eleven dimensional supermembrane theory}, 
         Ann. Phys. { \bf 185} (1988) 330.

\bm{hull3} C.M. Hull, {\it Actions for (2,1) sigma-models and strings},
	hep-th/9702067.


\bm{deli1} I. Bars and C. Deliduman, {\it Superstrings with new
	supersymmetry in (9,2) and (10,2) dimensions}, hep-th/9707215.



\bm{deli2} I. Bars and C. Deliduman, {\it Gauge principles for
	multi-superparticles}, hep-th/9710066. 

\bm{ov} H. Ooguri and C. Vafa, {\it N=2 heterotic strings}
	Nucl. Phys. { \bf B367} (1991) 83.

\bm{km3} D. Kutasov and E. Martinec, {\it Geometrical structures of M-theory}, 
	hep-th/9608017.

\bm{ff} J.M. Figueroa-O'Farrill, {\it F-Theory and the universal string
	theory}, hep-th/9704009.

\bm{BMRS}  I.Bandos, A.Maznytsia, I.Rudychev, D.Sorokin, Int.J.Mod.Phys.
{ \bf A12} (1997) 3259.

\bibitem{penrose}
R. Penrose and M.A.H. McCallum, {\sl Phys. Rep.} {\bf 6}, 291 (1972); \\
R. Penrose and W. Rindler, Spinors and space--time. 2v. Cambridge University
Press, 1986.

\bibitem{ferber}
A.Ferber, {\sl Nucl. Phys.} { \bf B132}, 55 (1978); \\
T. Shirafuji, {\sl Progr. of Theor.  Phys.} { \bf 70}, 18 (1983).


\bibitem{nispach}
E.R. Nissimov, S.J. Pacheva and S. Solomon,
{\sl Nucl. Phys.} { \bf B296}, 469 (1988);
{\sl Nucl. Phys.} { \bf B297}, 349 (1988);
{\sl Nucl. Phys.} { \bf B299}, 183 (1988);
{\sl Nucl. Phys.} {\bf B317}, 344 (1988);
{\sl Phys. Lett.} { \bf B228}, 181 (1989); \\
E.R. Nissimov and S.J. Pacheva,
{\sl Phys. Lett.} { \bf B221}, 307 (1989); \\
R. Kallosh and M. Rahmanov,
{\sl Phys. Lett.} { \bf B209}, 233 (1988);
{\sl Phys. Lett.} { \bf B214}, 549 (1988).

\bibitem{stvz}
D.P. Sorokin, V.I. Tkach and D.V. Volkov, {\sl Mod. Phys. Lett.} 
{ \bf A4}, 901
(1989); \\
D.P. Sorokin, V.I. Tkach, D.V. Volkov and A.A. Zheltukhin,
{\sl Phys. Lett.} { \bf B216}, 302 (1989); \\
D.V. Volkov and A.A. Zheltukhin, {\sl  Sov. Phys. JETP Lett.} { \bf 48}, 61
(1988); {\sl Lett. in Math. Phys.} { \bf 17}, 141 (1989); {\sl Nucl.
Phys.} { \bf B335}, 723 (1990).\\
N. Berkovits, {\sl Phys. Lett.} { \bf 232B}, 184 (1989).\\
M. Tonin, {\sl Phys. Lett.} { \bf B266}, 312 (1991); 
{\sl Int. J. Mod. Phys} { \bf A7}, 6013 (1992); \\
S. Aoyama, P. Pasti and M. Tonin, {\sl Phys. Lett.} { \bf
B283}, 213 (1992).\\
A. Galperin and E. Sokatchev, {\sl Phys. Rev.} { \bf D46}, 714 (1992).\\
F. Delduc, A. Galperin, P. Howe and E. Sokatchev, {\sl Phys. Rev.}
{ \bf D47}, 587 (1992).

\bibitem{sor-rev}
D.~Sorokin,
``Superbranes and superembeddings,''
Phys.\ Rept.\ { \bf 329}, 1 (2000)
hep-th/9906142.

\bibitem{sfortschr}
D.P. Sorokin, {\sl Fortsch. der Phys.} { \bf 38}, 923 (1990).

\bibitem{sg}
A.I. Gumenchuk and D.P. Sorokin,
{\sl Sov. J. Nucl. Phys.} { \bf 51}, 549 (1990).

\bibitem{benght}
A.K.H. Bengtsson, I. Bengtsson, M. Cederwall and N. Linden, {\sl Phys.
Rev.}
{ \bf D36}, 1766 (1987); \\
I. Bengtsson and M. Cederwall, {\sl Nucl. Phys} { \bf B302}, 81 (1988); \\
M. Cederwall, {\sl Phys. Lett.} { \bf B226}, 45 (1989).

\bibitem{ced}
M. Cederwall, in Proceedings of the Ahrenshoop Symposium, DESY-IFH,
1995, p 352; hep-th/9410014.

\bibitem{harm}
D.V. Volkov and A.A.  Zheltukhin, {\sl Ukr. Fiz. Zhurnal} { \bf 30}, 
809 (1985); \\
A.P.  Balachandran, F. Lizzi and G. Sparano,
{\sl Nucl.  Phys.} { \bf B263}, 608 (1986); \\
E. Sokatchev, {\sl Phys.  Lett.} { \bf B169}, 209 (1987); \\
A.A. Zheltukhin {\sl Theor.  Math. Phys.} { \bf 77}, 377 (1988); \\
P. Wiegmann, {\sl Nucl. Phys.} { \bf B323}, 330 (1989).

\bibitem{bz0}
I.A. Bandos and A.A. Zheltukhin, {\sl JETP Lett.} { \bf 51}, 547 (1990);
{\sl JETP Lett.} {\bf 53}, 7 (1991); {\sl Phys. Lett.} { \bf B261}, 245 
(1991);
{\sl Theor. Math. Phys.} { \bf 88}, 358 (1991).

\bibitem{bh}
I.A. Bandos, {\sl Sov. J. Nucl. Phys.} { \bf 51}, 906 (1990); {\sl JETP Lett.}
{ \bf 52}, 205 (1990).

\bibitem{bfortschr}
I.A.  Bandos and A.A. Zheltukhin, {\sl Fortsch.  der Phys.} {bf 41}, 619
(1993).

\bibitem{bzstr}
I.A. Bandos and A.A. Zheltukhin,{\sl Phys. Part. and Nucl.} { \bf 25}, 453
(1994).

\bibitem{bzoth}
I.A. Bandos and A.A. Zheltukhin, {\sl Class. Quant. Grav.}
{ \bf 12}, 609 (1995) and refs. therein.

\bibitem{zima}
S.A. Fedoruk and V.G. Zima, {\sl JETP Lett.} { \bf 61}, 420 (1995) (in
Russian); {\sl Theor. Math. Phys.} { \bf 102}, 241 (1995) (in Russian).

\bibitem{tomsk}
S. M. Kuzenko, S. L. Lyakhovich and A. Yu. Segal, 
{\sl Int. J. Mod. Phys} { \bf 10}, 1529 (1995);
{\sl Phys. Lett. B} { \bf 348}, 421 (1995).
A. A. Deriglasov, A. V. Galajinsky and S. L. Lyakhovich, hep-th/9512036.

\bibitem{bpstv}
I.A.  Bandos, P. Pasti, D.P. Sorokin, M. Tonin and D.V. Volkov,
{\sl Nucl. Phys.} { \bf B446}, 79 (1995); \\
I.A.  Bandos, D.P. Sorokin and D.V. Volkov, {\sl Phys. Lett.} { \bf B352}, 
269 (1995).

\bibitem{spinsup}
S.J. Gates and H. Nishino, {\sl Class. Quant. Grav.} { \bf 3}, 391 (1986); \\
J. Kovalski-Glikman, {\sl Phys. Lett.} { \bf  B180}, 358 (1986); \\
R. Brooks, F. Muhammed and S.J. Gates, {\sl Class. Quant. Grav.} { \bf 3}, 745
(1986); \\
R. Brooks, {\sl Phys. Lett.} { \bf B186}, 313 (1987); \\
G. Kovalski-Glikman, J.W. van Holten, S. Aoyama and J. Lukierski,
{\sl Phys. Lett.} { \bf B201}, 487 (1988); \\
A. Kavalov and R.L. Mkrtchyan, Spinning superparticles. Preprint Yer.Ph.I.
1068(31)88, Yerevan, 1988 (unpublished); \\
J.M.L. Fisch, {\sl Phys. Lett.} { \bf B219}, 71 (1989).

\bibitem{gsw}
M.B. Green, J.H. Schwarts and E.Witten, Superstring theory. 2v. Cambridge
University Press, 1987.

\bibitem{ks}
J.A. de Azcarraga and J. Lukierski, {\sl Phys. Lett.} { \bf B113}, 
170 (1982); \\
W. Siegel, {\sl Phys. Lett.} { \bf B128}, 397 (1983).

\bibitem{ber}
N. Berkovits, Preprint IFUSP-P-1212, April 1996 and references therein.

\bibitem{teit}
C.  Teitelboim, {\sl Phys. Rev.  Lett.} { \bf 38}, 1106 (1977); \\
E. Gozzi and M. Reuter, {\sl Nucl. Phys.} { \bf B320}, 160 (1989).

\bibitem{polyakov}
A.M. Polyakov, Gauge fields and strings.
Academic Publishers, Harwood, London, 1987.

\bibitem{govaerts}
J. Govaerts, {\sl Int. J. Mod. Phys} { \bf 4}, 4487 (1989).

\bibitem{sf}
S.A. Frolov and A.A. Slavnov, {\sl Phys. Lett} { \bf B208}, 245 (1988).

\bibitem{EiSo}
Y. Eisenberg and S. Solomon, {\sl Nucl. Phys.} { \bf B309}, 709 (1988);
{\sl Phys. Lett.} { \bf B220}, 562 (1989); \\
Y. Eisenberg, {\sl Phys. Lett.} { \bf B225}, 95 (1989);
{\sl Mod. Phys. Lett.}
{ \bf A4}, 195 (1989); {\sl Phys. Lett.}
{ \bf B276}, 325 (1992);\\
M. Pluschay, {\sl Phys. Lett.} { \bf B240}, 133 (1990);
{\sl Mod. Phys. Lett.} { \bf A4}, 1827 (1989).

\bibitem{bf}
E.S. Fradkin and G.A. Vilkovisky, {\sl Phys. Lett.} { \bf B55}, 224
(1975); \\
I.A. Batalin and G.A. Vilkovisky, {\sl Phys. Lett.} { \bf B69}, 309 (1977); \\
I.A. Batalin and E.S. Fradkin, {\sl Phys. Lett.} { \bf B128}, 303 (1983).

\bibitem{mhenn}
M. Henneaux, {\sl Phys. Rep.} { \bf 126}, 1 (1985).\\
M. Henneaux and C. Teitelboim, Quantization of Gauge Systems, Princeton
University Press, Princeton, N.J., 1992.

\bibitem{bff}
I.A. Batalin and E.S. Fradkin, {\sl Nucl. Phys.} { \bf B279}, 514 (1987); \\
I.A. Batalin, E.S. Fradkin and T.E. Fradkina, {\sl Nucl. Phys.} { \bf B314}, 
158 (1989).

\bibitem{fs}
L.D. Faddeev and S.L. Shatashvili, {\sl Phys. Lett.} { \bf B167}, 225
(1986); \\
E. Egorian and R. Manvelyan, {\sl Theor. Math. Phys.} { \bf 94}, 241 (1993)
(in Russian) and references therein.


\bibitem{moshe}
J. Feinberg and M. Moshe, {\sl Phys. Lett.} { \bf B247}, 509 (1990);
{\sl Ann. of Phys.} { \bf 206}, 272 (1991).


\bibitem{6spin}
T. Kugo and P. Townsend, {\sl Nucl. Phys.} { \bf B221}, 357 (1983); \\
P.S. Howe, G. Sierra and P.K. Townsend, {\sl Nucl. Phys.} { \bf B221}, 
331 (1983).

\bibitem{west}
A. Neveu and P. West, {\sl Nucl. Phys.} { \bf B293}, 266 (1987); \\
J. Govaerts, {\sl Int. J. Mod. Phys} { \bf 4}, 173 (1989).

\bibitem{rivelles}
M. Gomes, V.O. Rivelles and A.J. da Silva, {\sl Phys. Lett.} { \bf B218}, 63
(1989), 63; \\
J. Gamboa and V.O. Rivelles, {\sl Phys. Lett.} { \bf B241}, 45 (1990); \\
M. Pierri and V.O. Rivelles, {\sl Phys. Lett.} { \bf B251}, 421 (1990).



\bm{conv1}  M.A. Grigoriev, S.L. Lyakhovich, {\it Fedosov Deformation Quantization 
            as a BRST Theory.} hep-th/0003114. \\
          I.A. Batalin, M.A. Grigoriev, S.L. Lyakhovich,
	  {\it Star Product for Second Class Constraint Systems from a 
	  BRST Theory},
	  hep-th/0101089.

\bibitem{R}
I.~Rudychev,
``From noncommutative string/membrane to ordinary ones,''
hep-th/0101039.


\bibitem{L-brane}
P.~S.~Howe, O.~Raetzel, I.~Rudychev and E.~Sezgin,
``L-branes,''
Class.\ Quant.\ Grav.\ { \bf 16}, 705 (1999)
 hep-th/9810081 .


\bm{se1} D. Sorokin, V. Tkach and D.V. Volkov, {\it Superparticles,
twistors and Siegel symmetry }, Mod. Phys. Lett. { \bf A4 } (1989) 901



\bm{se2} I.A. Bandos, D. Sorokin, M. Tonin, P. Pasti and D. Volkov,
        {\it Superstrings and supermembranes in the doubly supersymmetric
        geometrical approach }, Nucl. Phys. { \bf B446 } (1995) 79

\bm{hs} P.S. Howe and E. Sezgin, {\it Superbranes}, hep-th/9607227.

\bm{chs1} C.S. Chu, P.S. Howe and E. Sezgin, {\it Strings and D-branes
        with Boundaries}, Phys. Lett. { \bf B428} (1998) 59



\bm{bdr} E. Bergshoeff, M. de Roo, E. Eyras, B. Janssen and J. P. van der
        Schaar, {\it Multiple intersections of D-branes and M-branes},
        Nucl.Phys.  {\bf B494} (1997) 119, hep-th/9612095.



\bm{cod1} P.S. Howe, A.Kaya, E. Sezgin and P. Sundell {\it Codimension
          one branes }, Nucl.Phys. {\bf B587} (2000) 481-513



\bm{hos} P.S. Howe, O. Raetzel and E. Sezgin, {\it On brane actions and
        superembeddings }, JHEP { \bf 08} (1998) 011



\bm{bik} S. Bellucci, E. Ivanov and S. Krivinos, {\it Partial breaking $N=4$ to
$N=2$; hypermulitplet as a Goldstone superfield}, hep-th/9809190



\bm{bg} J. Bagger and A. Galperin, {\it The tensor Goldstone multiplet for
partially broken supersymmetry}, Phys. Lett. { \bf B412} (1997) 296-300.


\bm{hsw} P.S. Howe, E. Sezgin and P.C. West, {\it Aspects of
        superembeddings}, in {\it Supersymmetry and Quantum Field Theory}, eds.
        J. Wess and V.P. Akulov (Springer 1998), hep-th/9705093.




\bm{act} A. Ach\'ucarro, J.M. Evans, P.K. Townsend and D.L. Wiltshire, {\it
         Super p-branes}, Phys. Let. { \bf 198B} (1987) 441.


\bm{pkt4} P.K. Townsend, {\it Four lectures in M-theory},
        hep-th/9612121.


\bm{jhs4} M. Aganagic, J. Park, C. Popescu and J.H. Schwarz, {\it Dual
        D-brane actions}, Nucl. Phys. { \bf B496} (1997) 215, hep-th/9702133.



\bm{gns} S.J. Gates, H. Nishino and E. Sezgin, {\it Supergravity in
        $D=9$ and its coupling to the noncompact sigma model}, Class. and
        Quantum Grav. { \bf 3} (1986) 253.


\bm{at-new} M. Awada, P.K. Townsend, M. G\"unaydin and G. Sierra, {\it Convex
  cones, Jordan algebras and geometry of $D=9$ Maxwell-Einstein supergravity},
  Class. and Quantum Grav. { \bf 2} (1985) 801.


\bm{duff1} M.J. Duff, {\it Ani-de Sitter space, branes, singletons,
        superconformal field theories and all that}, hep-th/9808100.


\bm{nst} H. Nicolai, E. Sezgin and Y. Tanii, {\it Conformally invariant
	supersymmetric field theories on $S^p\times S^1$ and super $p$-branes},
	Nucl. Phys. { \bf B305} (1988) 483.

\bm{bps-new} E. Bergshoeff, E. Sezgin and A. van Proeyen, {\it Superconformal
	tensor calculus and matter couplings in six dimension}, Nucl. Phys. 
	{ \bf B264} (1986)653.








\bm{Wit1} Nathan Seiberg, Edward Witten, {\sl String Theory and Noncommutative
 Geometry}, JHEP { \bf 9909} (1999) 032, hep-th/9908142.
 
\bm{OM1} Rajesh Gopakumar, Shiraz Minwalla, Nathan Seiberg, Andrew Strominger,
{\sl OM Theory in Diverse Dimensions}, JHEP { \bf 0008} (2000) 008, hep-th/0006062.

\bm{SST} N. Seiberg, L. Susskind, N. Toumbas, {\sl Strings in Background Electric Field, 
Space/Time Noncommutativity and A New Noncritical String Theory}, JHEP { \bf 0006} (2000) 021,
 hep-th/0005040.

\bm{Berg1} E. Bergshoeff, D. S. Berman, J. P. van der Schaar, P. Sundell,
{\sl A Noncommutative M-Theory Five-brane}, Nucl.Phys. { \bf B590} (2000) 173,
hep-th/0005026.

\bm{Berg2} E. Bergshoeff, D. S. Berman, J. P. van der Schaar, P. Sundell,
{\sl Critical fields on the M5-brane and noncommutative open strings},
Phys.Lett. { \bf B492} (2000) 193, hep-th/0006112

\bm{Sch1} Volker Schomerus, {\sl D-branes and Deformation Quantization},
JHEP { \bf 9906} (1999) 030, hep-th/9903205.

\bm{Chu1} Chong-Sun Chu, Pei-Ming Ho, {\sl Noncommutative Open String and D-brane},
Nucl.Phys. {\bf B550} (1999) 151, hep-th/9812219.

\bm{Chu2} Chong-Sun Chu, Pei-Ming Ho, {\sl Constrained Quantization of Open String in
 Background B Field and Noncommutative D-brane }, Nucl.Phys. { \bf B568} (2000) 447,
 hep-th/9906192.
 
\bm{Jab1}  F. Ardalan, H. Arfaei, M.M. Sheikh-Jabbari, {\sl Noncommutative Geometry 
From Strings and Branes}, JHEP { \bf 9902} (1999) 016, hep-th/9810072.
 
\bm{Jab2} M.M. Sheikh-Jabbari, A. Shirzad, {\sl Boundary Conditions as Dirac Constraints},
hep-th/9907055.


\bm{Oh1}  Won Tae Kim, John J. Oh, {\sl Noncommutative open strings from Dirac
quantization}, Mod.Phys.Lett. { \bf A15} (2000) 1597, hep-th/9911085.

\bm{Lee} Taejin Lee, {\sl Canonical Quantization of Open String and Noncommutative
Geometry}, Phys.Rev. { \bf D62} (2000) 024022, hep-th/9911140.

\bm{Fad} L.Faddeev, Phys.Let. { \bf B145} (1984) 81,\\
 L.Faddeev, S.Shatashvili, Phys.Let. { \bf B167} (1986) 225.
 
\bm{Manv} E.Sh. Egorian, R.P. Manvelian, Theor.Math.Phys.{ \bf 94}:173,1993, 
Teor.Mat.Fiz. { \bf 94}:241,1993 

\bm{BF1} I.A. Batalin, E.S. Fradkin, Nucl.Phys. { \bf B279}:514,1987,\\
 I.A. Batalin, E.S. Fradkin, T.A. Fradkina, Nucl.Phys. { \bf B314}:158,1989.

\bm{Lyakh} M.A. Grigoriev, S.L. Lyakhovich, {\sl Fedosov Deformation Quantization 
as a BRST Theory}, hep-th/0003114.

\bm{CZ} Chong-Sun Chu, Frederic Zamora, {\sl Manifest Supersymmetry in Non-Commutative 
Geometry}, JHEP { \bf 0002} (2000) 022, hep-th/9912153.

\bm{LW} N.D. Lambert, P.C. West, {\sl D-Branes in the Green-Schwarz Formalism}, hep-th/9905031.

\bm{LZ} P.Haggi-Mani, U.Lindstrom, M.Zabzine, {\sl Boundary Conditions, Supersymmetry and
A-field Coupling for an Open String in a B-field Background}, Phys.Lett. { \bf B483}
 (2000) 443-450, hep-th/0004061.

\bm{JS} B. Jurco, P. Schupp, {\sl Noncommutative Yang-Mills from equivalence of star products},
Eur.Phys.J. { \bf C14} (2000) 367, hep-th/0001032.

\bm{BH} L.Brink, M. Hennaux, "Principles of String Theory", Plenum Press, New York 1988. 

\bm{KS} Shoichi Kawamoto, Naoki Sasakura, {\sl Open membranes in a constant C-field background and
noncommutative boundary strings}, JHEP { \bf 0007} (2000) 014, hep-th/0005123.

\bm{ES1} P.S. Howe, E. Sezgin, P.C. West, {\sl The Six Dimensional Self-Dual Tensor},
Phys.Lett. { \bf B400} (1997) 255-259, hep-th/9702111,\\
 {\sl Covariant Field Equations of the M Theory Five-Brane},
Phys.Lett. { \bf B399} (1997) 49-59, hep-th/9702008.

\bibitem{CS}
L.~Cornalba and R.~Schiappa,
``Nonassociative star product deformations for D-brane worldvolumes in  curved backgrounds,''
hep-th/0101219.

\bm{BGL} I.A. Batalin, M.A. Grigoriev, S.L. Lyakhovich,
{\sl Star Product for Second Class Constraint Systems from a BRST Theory},
hep-th/0101089.


\bibitem{JXL}
J.~X.~Lu,
``(1+p)-dimensional open D(p-2) brane theories,''
hep-th/0102056.

\bm{MPT} M. Mihailescu, I.Y. Park, T.A. Tran, {\sl D-branes as Solitons of an N=1, D=10 Non-commutative
Gauge Theory}, hep-th/0011079.

\bm{Wit4}E.Witten, {\sl BPS Bound States Of D0-D6 And D0-D8 Systems In A B-Field}, hep-th/0012054.

\bm{Z} M.Zabzine, JHEP { \bf 0010} (2000) 042, hep-th/0005142.

\bm{Be1}  K.Bering,  J.Math.Phys. { \bf 41} (2000) 7468-7500, Phys.Lett. { \bf B486} (2000) 426-430.

\bm{Sol} V.O.Soloviev, {\sl Bering's proposal for boundary contribution to the Poisson bracket},
hep-th/9901112.

\bm{TH} T. Harmark, {\sl Open Branes and Little Strings}, hep-th/0012142.

\bm{T1} J. Isberg, U. Lindstrom, B. Sundborg, G. Theodoridis, 
{\sl Classical and Quantized Tensionless Strings}, Nucl. Phys. { \bf B411} (1994) 122,
 hep-th/9307108.\\
  U.Lindstrom, M. Zabzine, A.A. Zheltukhin, {\sl Limits of the D-brane action},
JHEP { \bf 9912} (1999) 016, hep-th/9910159.\\
 I.A. Bandos, A.A. Zheltukhin, Fortsch.Phys.   41:619, 1993. 

\bm{ss} A. Salam and E. Sezgin, {\it Supergravities in Diverse
	Dimensions,\ Vol. 1,\ p.5} (World Scientific, 1989).


\end{thebibliography}
\end{document}